\documentclass[a4paper,11pt]{article}

\usepackage{natbib}
\usepackage{amssymb}
\usepackage{graphicx}
\usepackage{rotating}
\usepackage{morefloats}
\usepackage{hyperref}
\hypersetup{colorlinks,%
          citecolor=blue,%
	    filecolor=black,%
            linkcolor=black,%
            urlcolor=blue}
\usepackage{breakurl}
\usepackage[margin=3cm]{geometry}

\begin{document}

\title{Dissolution on Titan and on Earth: Towards the age of Titan's karstic landscapes}
\begin{center}
{\bf\Large Dissolution on Titan and on Earth: Towards the age of Titan's karstic landscapes}
\end{center}

\vspace{0.5cm}
\begin{itemize}
  \item[] {\small \textbf{Thomas Cornet (tcornet@sciops.esa.int)}, European Space Agency (ESA), European Space Astronomy Centre (ESAC), P.O. BOX 78, 28691 Villanueva de la Canada (Madrid), Spain.}
  \item[] {\small \textbf{Daniel Cordier}, Universit{\'e} de Franche-Comt{\'e}, Institut UTINAM, CNRS/INSU, UMR 6213, 25030 Besan\c{c}on Cedex, France.}
  \item[] {\small \textbf{Tangui Le Bahers}, Universit\'e de Lyon, Universit\'e Claude Bernard Lyon 1, ENS Lyon, Laboratoire de Chimie UMR 5182, 46 all\'ee d'Italie, F-69007 Lyon Cedex 07, France.}
  \item[] {\small \textbf{Olivier Bourgeois}, LPG Nantes, UMR 6112, CNRS, OSUNA, Universit\'e de Nantes, 2 rue de la Houssini\`ere, BP92208, F-44322 Nantes Cedex 3, France.}
    \item[] {\small \textbf{Cyril Fleurant}, LETG - UMR CNRS 6554, Universit\'e d'Angers, UFR Sciences, 2 bd Lavoisier, F-49045 Angers Cedex 01, France.}
  \item[] {\small \textbf{St\'ephane Le Mou\'elic}, LPG Nantes, UMR 6112, CNRS, OSUNA, Universit\'e de Nantes, 2 rue de la Houssini\`ere, BP92208, F-44322 Nantes Cedex 3, France.}
  \item[] {\small \textbf{Nicolas Altobelli}, European Space Agency (ESA), European Space Astronomy Centre (ESAC), P.O. BOX 78, 28691 Villanueva de la Canada (Madrid), Spain.}
\end{itemize}
\vspace{0.5cm}

\textbf{Paper published in JGR Planets, April 2015, available at \burl{http://onlinelibrary.wiley.com/doi/10.1002/2014JE004738/full}.} \\

\textbf{Abstract}   
  {Titan's polar surface is dotted with hundreds of lacustrine depressions. 
Based on the hypothesis that they are karstic in origin, we aim at determining
the efficiency of surface dissolution as a landshaping process on Titan, in a comparative planetology 
perspective with the Earth as reference.
Our approach is based on the calculation of solutional denudation rates and allow inference of 
formation timescales for topographic depressions developed by chemical erosion on both planetary bodies. 
The model depends on the solubility of solids in liquids, the density of solids and liquids, and the 
average annual net rainfall rates. We compute and compare
the denudation rates of pure solid organics in liquid hydrocarbons and of minerals in liquid water 
over Titan and Earth timescales. We then investigate the denudation rates of a superficial organic 
layer in liquid methane over one Titan year. At this timescale, such a layer on Titan would behave 
like salts or carbonates on Earth depending on its composition, which means
that dissolution processes would likely occur, but would be 30 times slower on Titan compared 
to the Earth due to the seasonality of precipitation. Assuming an average depth of
100 m for Titan's lacustrine depressions, these could have developed in a few tens 
of millions of years at polar latitudes higher than 70$^\circ$ N and S, 
and a few hundreds of million years at lower polar latitudes.
The ages determined are consistent with the youth of the surface ($< 1$ Gyr) 
and the repartition of dissolution-related landforms on Titan.}
                 
\section{Introduction}

\begin{figure}
\center
\includegraphics[width=\textwidth]{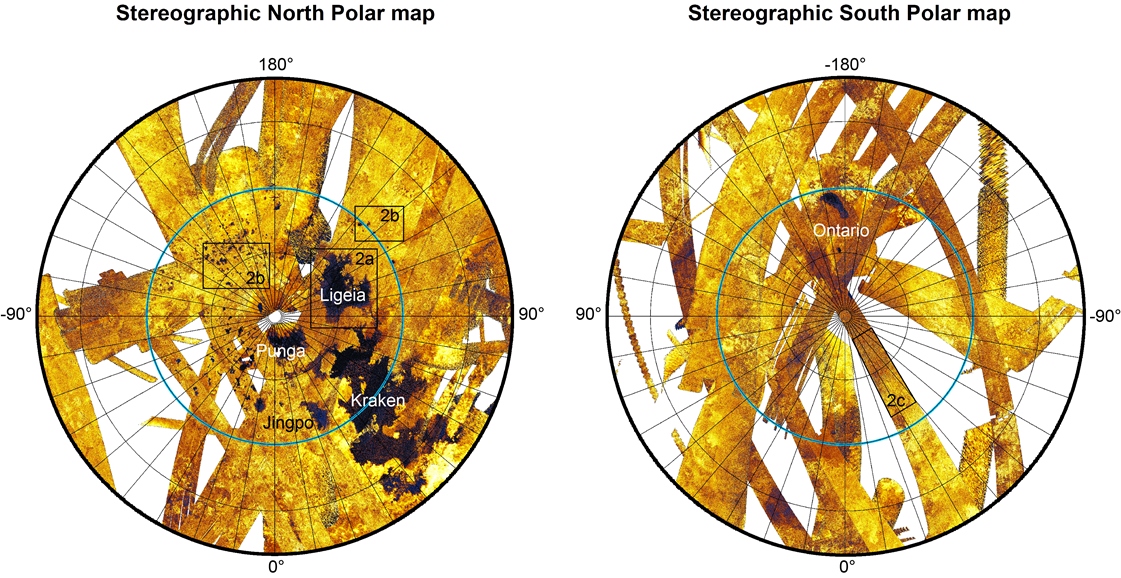} 
\caption{RADAR SAR mosaic of the TA (2004) to T95 (2013) flybys in north and south polar stereographic 
projections (down to 55$^\circ$ latitudes). All current lakes observed to date are located above 70$^\circ$ in latitude (blue circles).}
\label{fig:Titan_lakes}
\end{figure}

Along with the Earth, Saturn's icy moon Titan is the only planetary body of the entire Solar System that possesses lakes and seas 
\citep{Lopes2007,Stofan2007,Hayes2008}. Some of these lakes and seas are currently covered by liquids, while others are not 
\citep{Hayes2008}. Most are located in the polar regions \citep{Hayes2008,Aharonson2009}, 
although a few occurrences have been reported at lower latitudes \citep{Moore2010,Vixie2012lac}. 
The currently-filled lakes and seas are located poleward of 70$^\circ$ of latitude in both hemispheres whereas
most empty depressions are located at lower latitudes (Figure \ref{fig:Titan_lakes})
\citep{Hayes2008,Aharonson2009}.

Altogether, empty depressions, lakes, seas and fluvial channels, argue for the presence of an active 
``hydrological'' cycle on Titan similar to that of the Earth, with exchanges between the 
subsurface (ground liquids), the surface (lakes, seas, fluvial channels) and Titan's methane-rich 
atmosphere, where convective clouds and sporadic intense rainstorms have been imaged by 
the Cassini spacecraft instruments \citep{Turtle2011rains}. Methane, rather than water as on Earth, probably 
dominates the cycle on Titan \citep{Lunine2008} and thus constitutes one of the 
main components of the surface liquid bodies observed in the polar regions \citep{Glein2013,Tan2013}. 
Ethane, the main photodissociation product of methane \citep{Atreya2007}, is also implied in Titan's lakes 
chemistry, as predicted by several thermodynamical models 
\citep{Lunine1983,Raulin1987,Dubouloz1989,Cordier2009,Tan2013,Cordier2013erratum,Glein2013}, 
or as identified in Ontario Lacus thanks to the Cassini/VIMS instrument \citep{Brown2008}. 

Titan's lakes are located in topographic depressions carved into the
ground by geological processes that are poorly understood to date. 
The origin of the liquid would be 
related to precipitation, surface runoff and underground circulation, leading to the accumulation
of liquids in local topographic depressions. 

In the present work, we aim to constrain the origin and the age of these depressions. 
Section \ref{sec:geology} first provides a brief overview of their geology and a discussion of their possible origin based on
their morphological characteristics and from considerations about Titan's surface composition and climate. Based on this
discussion, we propose a new quantitative model, whereby the depressions have formed by the dissolution of a 
surface geological layer over geological timescales, such as in karstic landscapes on Earth. 

In terrestrial karstic landscapes, the maximum quantity of mineral that can be dissolved per year, namely 
the solutional denudation rate, can be
computed using a simple thermodynamics-climatic model presented in Section \ref{sec:denudation_rate_model}.
The denudation rate depends on the nature of the surface material (solubility and density of the minerals) and on 
the climate conditions (precipitation, evaporation, surface temperature). 
Using this simple model, it is possible to determine theoretical timescales for the formation of specific karstic landforms on Earth,
which are compared to relative or absolute age determinations in Section \ref{sec:earth_ages}. 

We apply the same model to Titan's surface in Section \ref{sec:denudation_rates_titan}.
Section \ref{sec:DR_Earth_time} is dedicated to the comparative study of denudation rates of 
pure solid organics in pure liquid methane, ethane and propane, and of common soluble minerals 
(halite, gypsum, anhydrite, calcite and dolomite), cornerstones of karstic landscapes development on Earth,
in liquid water over terrestrial timescales. Section \ref{sec:DR_Titan_time} describes the computation of 
denudation rates of pure solids and of mixed organic surface layers in liquid methane over Titan timescales 
by using the methane precipitation rates extracted from the GCM of \citet{Schneider2012}.
Based on these denudation rates, we compute the timescales needed to develop the typical 
100 m-deep topographic depressions observed in the polar regions of Titan (Section \ref{sec:ages_Titan})
and compare them to timescales estimated from other observations (e.g. crater counting, dune formation).

%
%

\section{Geology of Titan's lacustrine depressions} \label{sec:geology}

\subsection{Geomorphological settings} \label{sec:geomorphology}

\begin{figure}
\center
\noindent \includegraphics[width=\textwidth]{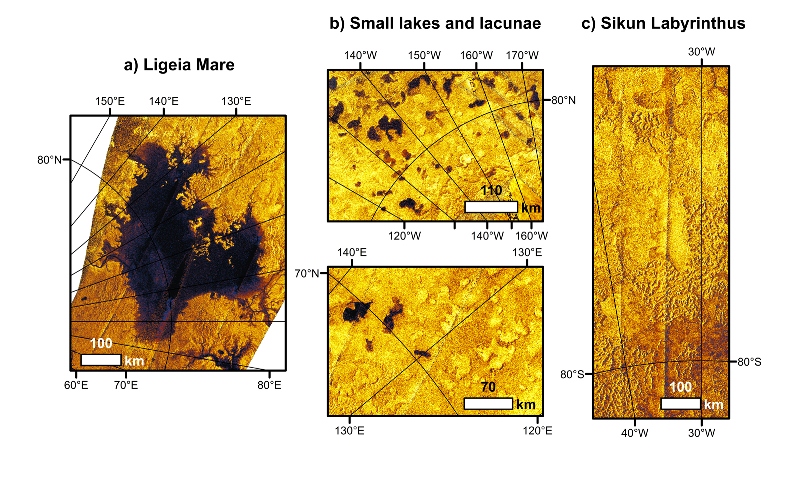} 
\caption{RADAR SAR views of Titan's second largest sea Ligeia Mare (a), empty and liquid-covered 
lacustrine depressions (b), and the heavily dissected terrains of 
Sikun Labyrinthus (c). Ligeia Mare clearly differs in size and shape from the smaller lacustrine depressions, 
which suggests a different formation mechanism. Lacustrine depressions and labyrinthic terrains pertain to the 
class of features potentially related to the dissolution of the surface.}
\label{fig:Titan_lakes2}
\end{figure}

Seas and lacustrine depressions strongly differ in shape (Figure \ref{fig:Titan_lakes2}). 
On one hand, seas are large (several hundred km in width) and deep 
(from 150 to 300 - 400 m in depth) \citep{Lorenz2008organic,Lorenz2014,Mastrogiuseppe2014}. 
They possess dendritic contours and are connected to fluvial channels (e.g. Ligeia Mare, Figure \ref{fig:Titan_lakes2}a)
\citep{Stofan2007,Sotin2012,Wasiak2013}. They seem 
to develop in areas associated with reliefs, which constitute some parts of their coastlines.

On the other hand, Titan's lacustrine depressions (Figure \ref{fig:Titan_lakes2}b) 
develop in relatively flat areas. They lie between 300 and 800 meters above the level of the northern 
seas \citep{Stiles2009,Kirk2012}. They are typically rounded or lobate in shape and some of them 
seem to be interconnected \citep{Bourgeois2008}. Their widths vary from 
a few tens of km, such as for most of Titan's lacunae, up to a few hundred 
km, such as Ontario Lacus or Jingpo Lacus. Their depths have been tentatively estimated to range from
a few meters to 100 - 300 meters \citep{Hayes2008,Kirk2008,Stiles2009,Lorenz2013}, 
with ``steep''-sided walls \citep{Mitchell2007,Bourgeois2008,Kirk2008,Hayes2008}. 
The liquid-covered depressions would lie 250 meters below the floor of the empty depressions \citep{Kirk2007}, which 
could be indicative of the presence of an alkanofer in the sub-surface, analog to terrestrial aquifers, 
filling or not the depressions depending on their base level \citep{Hayes2008,Cornet2012}. 
The depressions sometimes possess a raised rim, ranging from a few hundred meters up to 600 meters in 
height \citep{Kirk2007,Kirk2008}. All these numbers are likely subject to 
modification following future improvements in depth-deriving techniques.

\subsection{Geological origin of the depressions} \label{sec:origins}

The geological origin of the topographic depressions and how they are fed by liquids are 
still debated. The geometric analysis of the lakes by \citet{Sharma2010,Sharma2011} led 
to the conclusion that, unlike on Earth, the formation mechanism of the lacustrine 
depressions cannot be derived from the analysis of their coastline shapes. Recently, \citet{Black2012}
and \citet{Tewelde2013} showed that mechanical erosion due to fluvial activity
would have a minor influence in landscape evolution on Titan. 
Given this context, several hypotheses are being explored to understand how Titan's lakes
have formed. These include:
\begin{enumerate} 
\item cryovolcanic origin \citep{Mitchell2007,Wood2007}, 
forming topographic depressions in which lakes can exist, such as in terrestrial calderas \citep{Acocella2007}
or maars \citep{Lorenz1986};
\item thermokarstic origin \citep{Kargel2007,Mitchell2007,Harrisson2012}, 
where the cyclic destabilization of a methane frozen ground would form topographic depressions, such as in 
periglacial areas on Earth where the permafrost cyclically freezes and thaws and forms thermokarst lakes,
pingos or alases \citep{French2007} ; 
\item solutional origin \citep{Mitchell2007,Bourgeois2008,Mitchell2008,Mitchell2011karst,Malaska2011dissolution,Barnes2011evaporites,Cornet2012}, 
where processes analogous to terrestrial karstic dissolution create topographic depressions, 
such as terrestrial sinkholes/dolines, playas and pans under various climates \citep{Shaw2000,Ford2007}. 
\end{enumerate}

On one hand, the general lack of unequivocal cryovolcanic features on Titan tends to limit the 
likelihood of the cryovolcanic hypothesis \citep{Moore2011}. A methane-based 
permafrost would be difficult to form on Titan due to the presence of nitrogen in the
atmosphere \citep{Lorenz2002,Heintz2009}. Its putative cyclic destabilization would also be challenging,
given the tiny temperature variations between summer and winter, day and night, equator 
and poles \citep{Jennings2009,Lora2011,Cottini2012} over all timescales 
\citep{Aharonson2009,Lora2011}. On the other hand, solid organics have been shown to be quite 
soluble in liquid hydrocabons under Titan's surface conditions 
\citep{Lunine1983,Raulin1987,Dubouloz1989,Cordier2009,Cordier2013erratum,Cordier2013,Glein2013,Tan2013}, provided that they
are available at the surface. The observation of bright terrain around present lakes and inside of 
empty depressions, analogues of terrestrial evaporites produced by the evaporitic crystallization 
of dissolved solids, also strengthens this hypothesis \citep{Barnes2011evaporites, MacKenzie2014}.

On Earth, dissolution-related landforms are not restricted to sinkholes/dolines, pans or 
playa, which characterize relatively young karsts \citep{Ford2007}. Spectacular instances of reliefs 
nibbled by dissolution, known as cone/cockpit karsts, fluvio karsts or tower karsts, exist 
under temperate to tropical/equatorial climates, such as in China \citep{Xuewen2006,Waltham2008}, 
Indonesia \citep{Ford2007} or the Carribeans \citep{Fleurant2008,LyewAyee2010}.
The observation of possible mature karst-like terrains in 
Sikun Labyrinthus (Figure \ref{fig:Titan_lakes2}c) by \citet{Malaska2010}, similar to these terrestrial 
karstic landforms also gives further credence to the hypothesis that lacustrine depressions on Titan 
are karstic in origin.

\subsection{Composition of Titan's solid surface} \label{sec:composition}

Titan's surface can be divided into five main spectral units identified by the Cassini/VIMS instrument: bright terrain, 
dark equatorial dune fields or dark-brown units, blue units, 5 $\mu$m-bright units and the dark lakes \citep{Barnes2007,Stephan2009}.
In the polar regions, the solid surface appears dominantly as bright terrain, lakes, and patches 
identified as the 5 $\mu$m-bright unit in VIMS data \citep{Barnes2011evaporites,Sotin2012, MacKenzie2014}.
The spectral characteristics of the VIMS 5 $\mu$m-bright unit seen inside and around some polar (and equatorial)
lacustrine depressions indicate the presence of various hydrocarbons and nitriles \citep{Clark2010, Moriconi2010}
and are not compatible with the presence of water ice \citep{Barnes2009ontario}. 

The origin of the organic materials is probably linked to the atmospheric photochemistry, which 
results in the formation of various hydrocarbons and nitriles \citep{Lavvas2008a,Lavvas2008b,Krasnopolsky2009}
detected by Cassini \citep{Cui2009,Magee2009,Clark2010,Coustenis2010,Vinatier2010,Cottini2012}.
Table \ref{t:products} gives the estimated fluxes of some produced organics, as derived from several models. 
Most of these compounds could condense as solids and sediment onto the surface over geological timescales 
\citep{Atreya2007, Malaska2011dissolution}. Most of them would be relatively soluble in liquid alkanes
\citep{Raulin1987,Dubouloz1989,Cordier2009,Cordier2013erratum,Glein2013,Tan2013}.

It is therefore reasonable to assume that a superficial soluble layer, composed of organic products, 
exists at the surface of Titan. Episodic dissolution of this layer would be responsible for the development 
of karst-like depressions and labyrinthic terrains
\citep{Bourgeois2008,Malaska2010,Malaska2011dissolution,Mitchell2011karst,Cornet2012}.
Evaporitic crystallization could also occur after episodes of dissolution in the liquids,
forming evaporite-like deposits \citep{Barnes2011evaporites,Cornet2012,Cordier2013,MacKenzie2014}.

%
%

\section{Solutional denudation rates on Earth} \label{sec:denudation_rate_model}

On Earth, karstic landforms develop thanks to the 
dissolution of carbonate (calcite, dolomite) and evaporite (gypsum, anhydrite, halite) 
minerals under the action of groundwater and rainfall percolating through pore space 
and fractures present in rocks. The mineral solubilities vary as a function of the 
environmental conditions (amount of rain and partial pressure of carbon dioxide) \citep{Ford2007}. 
Karstic landforms like dolines or sinkholes are often located under temperate to humid climates 
in carbonates (though gypsum or halite karsts also exist on Earth). They 
reach depths of up to a few hundred meters \citep{Ford2007}.
Karsto-evaporitic landforms like pans are located under semi-arid to arid climates. They reach 
depths of up to a few tens of meters \citep{Goudie1995,Bowen2012}. Evaporitic landforms like playas are 
located under arid climates. They are characterized by their extreme flatness 
and can occur in any kind of topographic depression \citep{Shaw2000}. 

Denudation rates (hereafter $DR$) in terrestrial karstic landscapes are primarily constrained by geological 
(age and physico-chemical nature of rocks) and climate (net precipitation rates evolution) 
analyses. For many limestone-dominated areas (the majority of karst areas), it is commonly
assumed that dissolution features are mainly created in the epikarstic zone located in the
top few meters below the surface \citep{White1984,Ford2007}. Classically, in karstic terrains, the chemical/solutional 
denudation rate $DR$ (in meters per Earth year, or m/Eyr) is related to the rock physico-chemical 
properties and the climate by the following equation \citep{White1984,Tucker2001,Ford2007,Fleurant2008}:
\begin{equation} \label{eqn:DR}
DR = \rho_w \, \frac{M_{\rm calcite}}{\rho_{\rm calcite}} \, \tau \, m_{\rm Ca}, \qquad{\textrm{[m/Eyr]}}
\end{equation}

where $\rho_w$ is the mass density of liquid water ($\simeq 1000$ kg/m$^3$), $M_{\rm calcite}$ is the molar 
mass of calcite (in kg/mol), $\rho_{\rm calcite}$ is the mass density of calcite (in kg/m$^3$), $\tau$ is the 
mean annual net precipitation rate (in m/yr), 
equivalent to the sum of runoff and infiltration or the difference between precipitation and 
evapotranspiration over long timescales \citep{White2012}, and $m_{\rm Ca}$ is the equilibrium
 molality of calcite (Ca$^{2+}$ cations, in mol/kg),
assuming an instantaneous dissolution. Following this equation, denudation rates depend linearly on
the climate precipitation regime and on the molality of dissolved materials.

Molality calculations are provided in Appendix \ref{sec:solubility} for various soluble minerals, based on
two thermodynamic hypotheses: an Ideal Solution Theory (IST, no preferential interactions between molecules) 
and an Electrolyte Solution Theory (EST, preferential interactions between molecules).
We incorporated the effect of CO$_2$ gas dissolved in water, which acidifies water
due to a series of intermediate reactions (one of which producing carbonic acid), 
and increases the dissolution rates of carbonates (calcite and dolomite, 
see Appendix \ref{sec:solubility} for more details). 
The partial pressure of CO$_2$ (hereafter $P_{CO_2}$, in atm) under different climates can be computed 
as follows \citep{Ford2007}:
\begin{equation}
\log P_{CO_2} = -3.47 + 2.09 \, (1-e^{-0.00172 \, AET}),
\end{equation}

where $AET$ is the mean annual evapotranspiration rate in mm/Eyr. 
Applying the above formula gives $P_{CO_2} < 3$ matm
for arid areas, $3 < P_{CO_2} < 12$ matm for temperate areas and  $P_{CO_2} > 12$ matm for tropical or 
equatorial climates. The resulting typical values of denudation rates on Earth for common soluble minerals in liquid water 
are given in Table \ref{t:DRmin}. Depending on the climatic conditions, denudation rates of salts are on the order of
a few hundred $\mu$m to a few cm per Earth year, whereas denudation rates of carbonates are on the 
order of a few $\mu$m to a few hundred $\mu$m per Earth year. 
Note that chemical erosion is just one of the landshaping processes on Earth. Other mechanisms, such as 
mechanical erosion, would tend to increase the total denudation rate of a surface \citep{Fleurant2008}.
The dissolved solids can take part in the crystallization of soluble deposits at the surface (e.g. calcretes or salt 
crusts in arid environments) or in the sub-surface (e.g. cements), when they saturate the liquids. They do not 
necessarily crystallize at the location where they dissolved \citep{Ford2007}.

%
%

\section{Denudation rates to determine ages of terrestrial karstic landscapes} \label{sec:earth_ages}

Using the denudation rates, theoretical timescales for the development of dissolution landforms 
of a given depth can be inferred. We test hereafter this hypothesis on terrestrial instances by comparing
ages determined from the denudation rates and ages determined from relative or absolute
chronology. For carbonates, the partial pressure of CO$_2$ is computed based on the 
MOD16 Global Evapotranspiration Product of the University of Montana/ESRI Mapping Center.
The notation GEyrs, MEyrs and kEyrs will refer to timescales 
expressed respectively in Giga Earth years ($10^9$ years), Million Earth years ($10^6$ years) and kilo Earth years 
($10^3$ years) in the following sections.

\subsection{Example under a hyper-arid climate: the caves of the Mount Sedom diapir (Israel)}
Mount Sedom is a $11\times1.5$ km salt diapir located in Israel, 250 meters above the level of the Dead sea,
under a hyper-arid climate \citep{Frumkin1994,Frumkin1996}.
It is composed of a Pliocene-Pleistocene ($< 7$ MEyrs) halite basis covered by a 5 - 50 m-thick
layer composed of anhydrite and sandstones. 
Several sinkholes and caves exist in Mount Sedom \citep{Frumkin1994}. 

In this part of Israel, the present-day hyper-arid climate leads to average rainfall rates 
of 50 mm/Eyr, of which 10 - 15 mm/Eyr constitutes
the average effective rainfall rate ($\tau$). Gvishim and Mishquafaim Caves are respectively 60 m and 20 m-deep caves and are essentially 
carved into halite in the northern part of Mount Sedom. 
Radioisotopic age measurements in several caves of Mount Sedom give an Holocene age, about 8 kEyrs 
(and around 3.2 - 3.4 kEyrs from age measurements performed in Mishquafaim Cave) \citep{Frumkin1996}.

Assuming that underground dissolution is somewhat connected to the rainfall percolating through anhydrite and reaching halite,
the denudation rates computed with our model (Equation \ref{eqn:DR}) over this region are $DR \simeq 1.69 - 2.53$ mm/Eyr.
The timescales required to form these caves would be between 23.7 and 35.5 kEyrs for Gvishim Cave, and  
between 7.9 and 11.8 kEyrs for Mishquafaim Cave, longer than those estimated from radioisotopic measurements. 

However, it should be noted that Mount Sedom experiences a rather deep dissolution, the load of dissolved
solids in the underground liquid water flowing in its caves being up to 3 times that of surface runoff \citep{Frumkin1994}. 
Therefore, considering that the denudation rate should be 3 times greater, 
we find timescales of development between 7.9 and 11.8 kEyrs for Gvishim Cave and
between 2.6 and 3.9 kEyrs for Mishquafaim Cave, in better agreement with age measurements.

\subsection{Example under a semi-arid climate: The Etosha super-pan (Namibia)}
The Etosha Pan, located in Namibia under a semi-arid climate, is a flat 120$\times60$ km 
karsto-evaporitic depression that has already been suggested as a potential 
analogue for Titan's lacustrine depressions \citep{Bourgeois2008,Cornet2012}.
The depression is about 15 to 20 m-deep at most, and has been carved into the carbonate layer 
essentially composed of calcretes and dolocretes lying ontop a middle 
Tertiary-Quaternary detritical sedimentary sequence that covers the Owambo basin
\citep{Buch1996,Hipondoka2005}. This sedimentary sequence is believed to have 
accumulated under a semi-arid climate, relatively similar to the current one.
The age of the calcrete layer is not known precisely, but its formation is believed 
to have started during the Miocene/Pliocene transition
about 7 MEyrs ago \citep{Buch1997a,Buch1997b}, or even later during the late Pliocene, 4 MEyrs ago \citep{Miller2010}. 
The Etosha Pan is believed to have developed at the expense of the calcrete layer since 2 MEyrs \citep{Miller2010}.

In the Owambo sedimentary basin, $AET \simeq 350 $ mm/Eyr, yielding $P_{CO_2} = 3$ matm. 
Precipitation rates during the summer rainy season reach up to 500 mm/Eyr, which leads to $\tau = 150$ mm/Eyr at most.
These conditions lead to $DR = 6.5$ $\mu$m/Eyr in a substratum composed of calcite (calcrete) and $DR = 5.6$ $\mu$m/Eyr 
in a substratum composed of dolomite (dolocrete). These denudation rates give an approximate age for the Etosha Pan of about  
2.31 - 3.08 MEyrs in calcretes and 2.68 to 3.57 MEyrs in dolocretes. These ages are consistent with ages estimated from
geological observations ($< 4$ MEyrs).

\subsection{Example under a temperate climate: the doline of Crveno Jezero (Croatia)}
Crveno Jezero is a 350 m-wide collapse doline located in the Dinaric 
Karst of Croatia, under a presently mediterranean temperate climate. This part of the 
Dinaric Karst is essentially composed of limestones and dolomites that have been 
deposited during the Mesozoic (Triassic to Cretaceous), when the area was 
covered by shallow marine carbonate shelves \citep{Vlahovic2002,Mihevc2010}. 
The cliffs forming the doline are up to 250 m in height. The bottom
of the depression is covered by a lake, named Red Lake, about 250 m-deep.
The total depth of the doline is therefore about 500 m \citep{Garasic2001,Garasic2012}. 

Although the formation process of the collapse dolines in the area is still not completely understood, 
especially in terms of the relative importance and timing of the collapse compared to the dissolution in the 
development of the structure, the uplift linked to the formation of the Alps (Eocene-Oligocene) led to the exposure 
of these carbonates and their subsequent karstification since at least Oligo-Miocene times (last 30 MEyrs) \citep{Sket2012}.
At some point, dissolution occurred to form an underground cave; then the load of the capping rocks exceeded their 
cohesion, leading to the brutal or progressive collapse of the cap in the empty space beneath. The 
timescale calculated for the doline formation therefore constitutes a higher estimate. Age determinations in the northern Dinaric caves
have been performed and show that karstification would be older than 4 to 5 MEyrs \citep{ZupanHajna2012}.
Crveno Jezero thus developed between 4 and 30 MEyrs ago.

In this region of Croatia, $AET \simeq 500$ mm/Eyr, which leads to $P_{CO_2} = 5.44$ matm.
Annual mean precipitation rates are about 1100 mm/Eyr \citep{Mihevc2010}, leading to $\tau = 600$ mm/Eyr. 
These parameters give $DR = 31.8$ $\mu$m/Eyr in limestones dominated by calcite 
($DR = 27.5$ $\mu$m/Eyr in limestones dominated by dolomite). 
The time required to dissolve 500 m of calcite under these present conditions would be around 
15.7 MEyrs (18.2 MEyrs in dolomite). It is likely that cap rocks fell into the cave, 
leading to a younger age (1 m of calcite and dolomite dissolve in about 31 and 36 kEyrs respectively). 
However, the age determined by our method is still consistent with the geological records.

\subsection{Example under a tropical climate: the Xiaozhai tiankeng (China)}
The Xiaozhai tiankeng is a 600 m-wide collapse doline located in the Chongqing province of China, under a 
tropical climate. The depth of this doline is evaluated between 511 and 662 m depending on the location \citep{Xuewen2006,Ford2007}.
It is organized into two major collapse structures, an upper 320 m-deep structure and a lower 342 m-deep shaft.
The tiankeng developed into Triassic limestones in a karst drainage basin. It 
results from various surface and subsurface processes such as dissolution, fluid flows and collapse.
Tiankengs of China would have formed during the late Pleistocene (last 128,000 Eyrs) 
\citep{Xuewen2006} under an equatorial monsoonal climate similar to nowadays,
emplaced 14 MEyrs ago \citep{Wang2009}. 

In the area, annual precipitation rates are about 1500 mm/Eyr, for an annual
evapotranspiration rate of about 800 mm/Eyr, which leads to $P_{CO_2} = 0.0123$ atm and 
$\tau=700$ mm/Eyr. Under these conditions, $DR =49.6 $ $\mu$m/Eyr (calcite) or 
$DR =42.6 $ $\mu$m/Eyr (dolomite) and the timescale to form the Xiaozhai tiankeng 
would be between 10.3 and 13.3 MEyrs (calcite) and 12.0 and 15.5 MEyrs (dolomite),
considerably longer timescales than those estimated from geological records. However,
such a difference is to be expected, since the formation of gigantic tiankengs are
subject to complex interconnected surface processes not taken into account in these
simple calculations (collapse and underground circulation of water). This illustrates the limits
of our method, generally over-estimating the ages of karstic features when they 
do not result from dissolution only.

%
%

\section{Denudation rates on Titan} \label{sec:denudation_rates_titan}

Equation \ref{eqn:DR} has to be adapted for Titan. From thermodynamics, we compute molar volumes 
($V_{\rm m,i}$ in m$^3$/mol, see Appendix \ref{sec:molar_volumes}) and mole fractions at saturation 
($X_{i,\rm sat}$, see Appendix \ref{sec:solubility}) instead of mass densities and molalities. We considered the Ideal 
and the non-ideal Regular Solutions Theories (IST and RST respectively) and ensured that our results
are consistent with experimentally determined solubilities. The RST has been developed to determine the solubility
of non-polar to slightly polar molecules, such as simple hydrocarbons or carbon dioxide, in non-polar solvents. 
It is less likely to be appropriate for polar molecules such as nitriles/tholins and water ice. 
We therefore always provide a comparison between the IST and the RST in order to assess the uncertainty
of the calculation by using the RST for polar molecules (see also Appendix \ref{sec:solubility}). 
However, we consider the RST, which provides our lowest and more realistic solubilities (at least for hydrocarbons), 
as the most suited model for the study. We assume that dissolution is instantaneous, which is not unreasonable 
given the rapid saturation of solid hydrocarbons in liquid ethane inferred from
recent dissolution experiments \citep{Malaska2014}, compared to the long geological timescales considered in our work.
For a binary solute-solvent system, Equation \ref{eqn:DR} becomes:
\begin{equation} \label{eqn:DR2}
DR_i =\frac{ X_{i, \rm sat}}{1- X_{i, \rm sat}}\, \tau \,  \frac{V_{{\rm m},i}^S }{V_{\rm m, solv}^L}, \qquad \textrm{[m/Eyr]}
\end{equation}

where $V_{{\rm m},i}^S $ and $V_{\rm m, solv}^L$ are the molar volumes of the 
solid (crystallized) phase and of the liquid phase respectively. 

\subsection{Comparison between Titan and the Earth over terrestrial timescales} \label{sec:DR_Earth_time}

\begin{figure}
\center
\includegraphics[width=\textwidth]{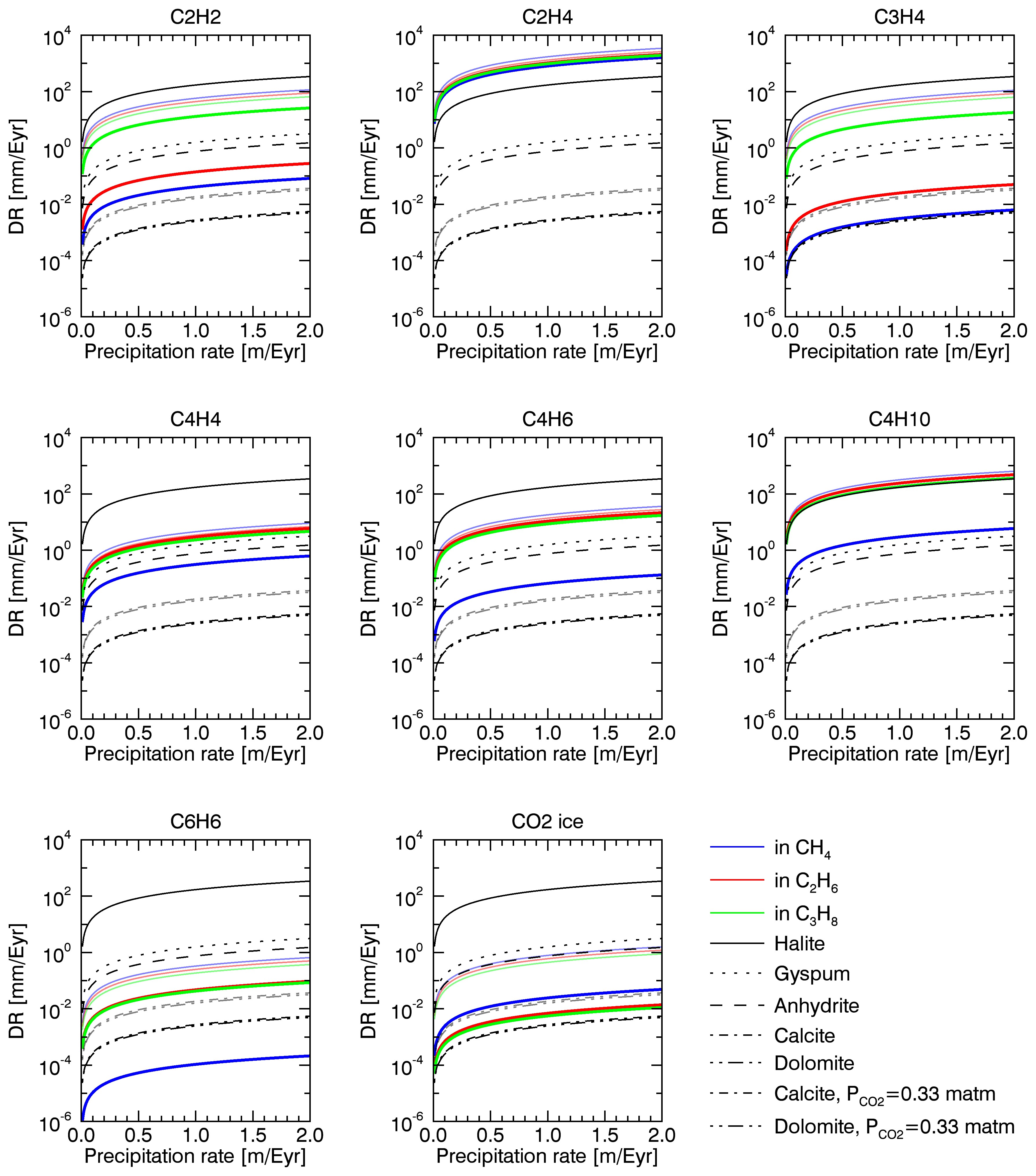} 
\caption{Denudation rates (in mm/Eyr) of non to slightly polar molecules (hydrocarbons and carbon dioxide ice) 
in Titan's liquids at 91.5 K (light curves: IST, bold curves: RST) compared to those of Earth's minerals in liquid water at 
298.15 K (EST). The partial pressure of carbon dioxide for acid dissolution is set to 0.33 matm.}
\label{fig:DR1}
\end{figure}

\begin{figure}
\center
\includegraphics[width=\textwidth]{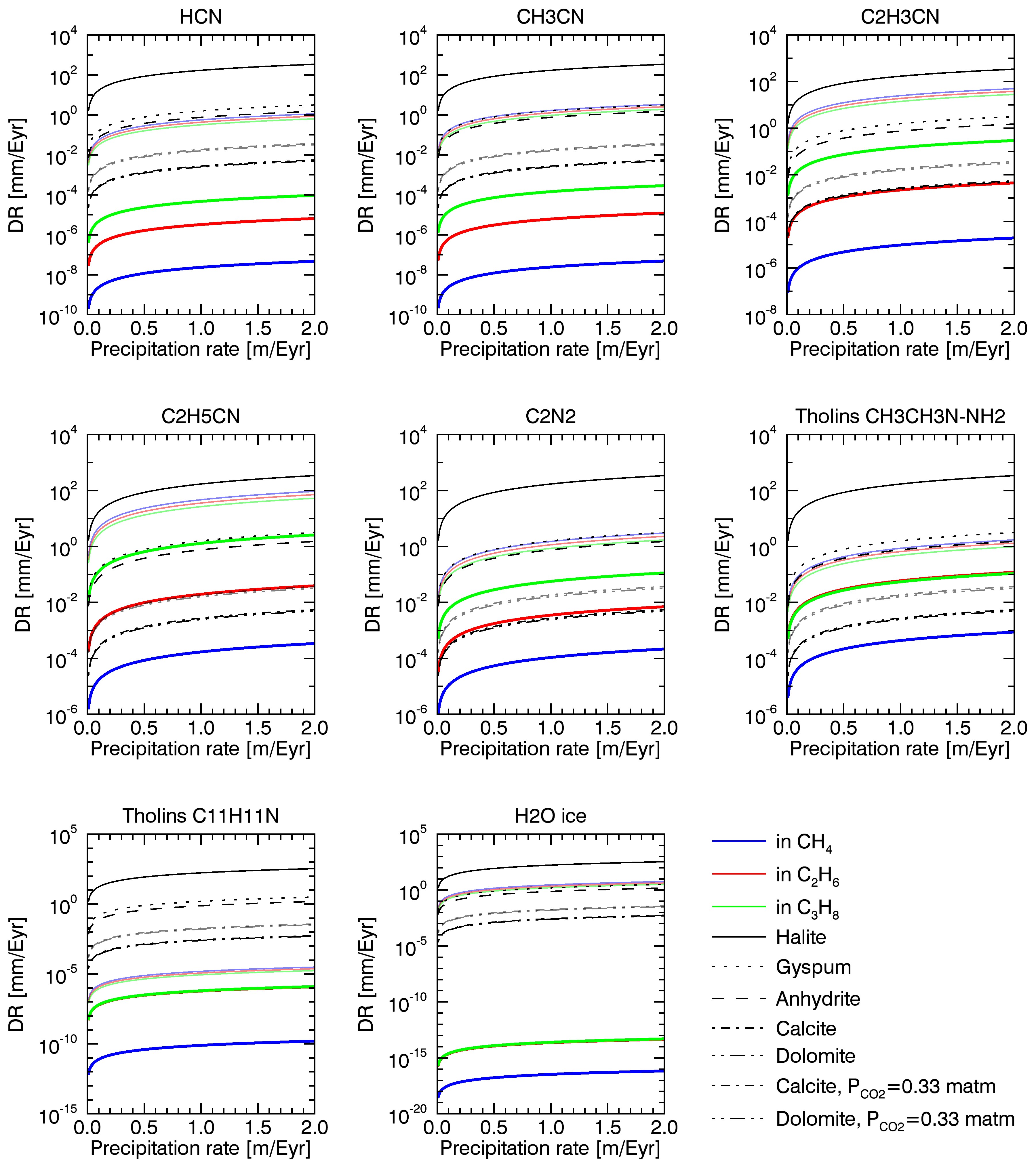} 
\caption{Same as Figure \ref{fig:DR1} but for polar molecules (nitriles and water ice).}
\label{fig:DR2}
\end{figure}

In order to compare the behavior of Titan's solids in Titan's liquids with that of minerals in liquid water on Earth, 
we computed Titan's and Earth's denudation rates over Earth timescales by varying the precipitation rate between 
0 and 2 m/Eyr. This range of precipitation rates encompasses the expected range on Titan (up to 1.2 - 1.3 m/Eyr during the rainy season \citep{Schneider2012}). It also covers the precipitation range between 
arid and tropical climates on Earth. This comparison is illustrated in Figures \ref{fig:DR1} (for hydrocarbons and 
carbon dioxide ice) and \ref{fig:DR2} (for nitriles and water ice), for which we 
fixed Titan's temperature to 91.5 K (the surface temperature during the rainy season according to 
\citet{Schneider2012}) and Earth's temperature to 298.15 K (with a partial pressure of carbon dioxide equal to 0.33 matm).

All organic compounds except C$_{11}$H$_{11}$N would behave like common mineral salts (halite, gypsum or anhydrite) according to the IST, which means that they would experience 
dissolution rates in the range of hundred $\mu$m to a few cm over one Eyr. According to the RST and assuming that the liquid
is methane, C$_2$H$_{4}$ would behave like halite (cm-scale dissolution), C$_4$H$_{10}$ and C$_4$H$_{4}$ 
like gypsum (hundred $\mu$m to mm-scale dissolution), C$_2$H$_{2}$, C$_3$H$_{4}$, C$_4$H$_{6}$ and CO$_{2}$
like carbonates (up to hundred $\mu$m-scale dissolution). C$_6$H$_6$ and nitriles would be less soluble in methane than 
calcite is in pure liquid water, with the least soluble nitriles being C$_{11}$H$_{11}$N, HCN and CH$_3$CN. As expected, 
water ice is completely insoluble according to the RST, developed for non-polar molecules.

The denudation rates of all organic compounds are higher in ethane and propane than in methane. 
Nitriles and C$_6$H$_6$, which are poorly soluble in methane, are quite soluble in ethane and propane
(except C$_{11}$H$_{11}$N, HCN and CH$_3$CN still under the calcite level) 
and reach denudation rates similar to those of carbonates or even gypsum.
C$_2$H$_2$ and C$_3$H$_4$ would behave like carbonates (in ethane) or 
gypsum (in propane). C$_{4}$H$_{4}$ and C$_{4}$H$_{6}$ behave like salts in those liquids and 
are even more soluble than gypsum. C$_4$H$_{10}$ and C$_{2}$H$_{4}$ are halite-like materials, 
extremely soluble in ethane and propane. Therefore, the likelihood 
of developing dissolution landforms in a hydrocarbon dominated substrate, by analogy 
with the Earth, is high.

\subsection{Present denudation rates on Titan over a Titan year} \label{sec:DR_Titan_time}

\subsubsection{Net precipitation rates on Titan}

Computing denudation rates over Titan timescales requires us to define the evolution
of the net precipitation over one Titan year (1 Tyr $\simeq 29.5$ Earth years).
Titan's climate is primarily defined by a rainy warm ``summer'' season and a cold dry ``winter'' season, 
both spanning about 10 Earth years. Southern summers are shorter and more 
intense than those in the north \citep{Aharonson2009}. Precipitation occurs 
as sporadic and intense rainstorms during summer, when cloud formation is observed
\citep{Roe2002,Schaller2006tempete,Schaller2009,Rodriguez2009cloud,Rodriguez2011,Turtle2011rains,Turtle2011grl}.

A few Global Circulation Models (GCM) attempt to describe, at least qualitatively, the
methane cycle on Titan (e.g. \citet{Rannou2006}, \citet{Mitchell2008},
\citet{Tokano2009} and \citet{Schneider2012}). Usually, 
net accumulation of rain is predicted at high latitudes 
$>60^\circ$ during summer, in agreement with the presence of lakes, whereas 
mid-to-low latitudes experience net evaporation, in agreement with the absence of 
lakes and the presence of deserts. Quantitatively, the model predictions are 
subject to debate since they depend on their physics 
(e.g. cloud microphysics, size of the methane reservoir, radiative transfer scheme).
Still, they remain our best estimates about Titan's climate.

During the summer season, the model of \citet{Rannou2006} predicts net precipitation 
rates of methane lower than 1 cm/Eyr at 70$^\circ$ latitude and up to 1 m/Eyr poleward of 70$^\circ$, 
equivalent of a few $\mu$m to 2.7 mm/Earth day (Eday) respectively. The model of \citet{Mitchell2008} predicts 
precipitation rates of about 2 mm/Earth day (Eday) in the polar regions 
and along an ``Inter-Tropical Convergence Zone'' (ITCZ), nearly moving
``pole-to-pole''. The intermediate and moist models of \citet{Mitchell2009} predict precipitation
roughly varying between 2 and 4 mm/Eday at the poles and along the ITCZ. These rates are consistent with
those estimated in \citet{Mitchell2011} in order to reproduce the tropical storms seen in 2010
\citep{Turtle2011rains} and with those estimated by the model of \citet{Schneider2012}. The model of
\citet{Tokano2009} also predicts similar precipitation rates (800 to 1600 kg/m$^2$ in half a Titan year, 
equivalent to precipitations between 3 and 6 mm/Eday).

\begin{figure}
\center
\includegraphics[width=0.5\linewidth]{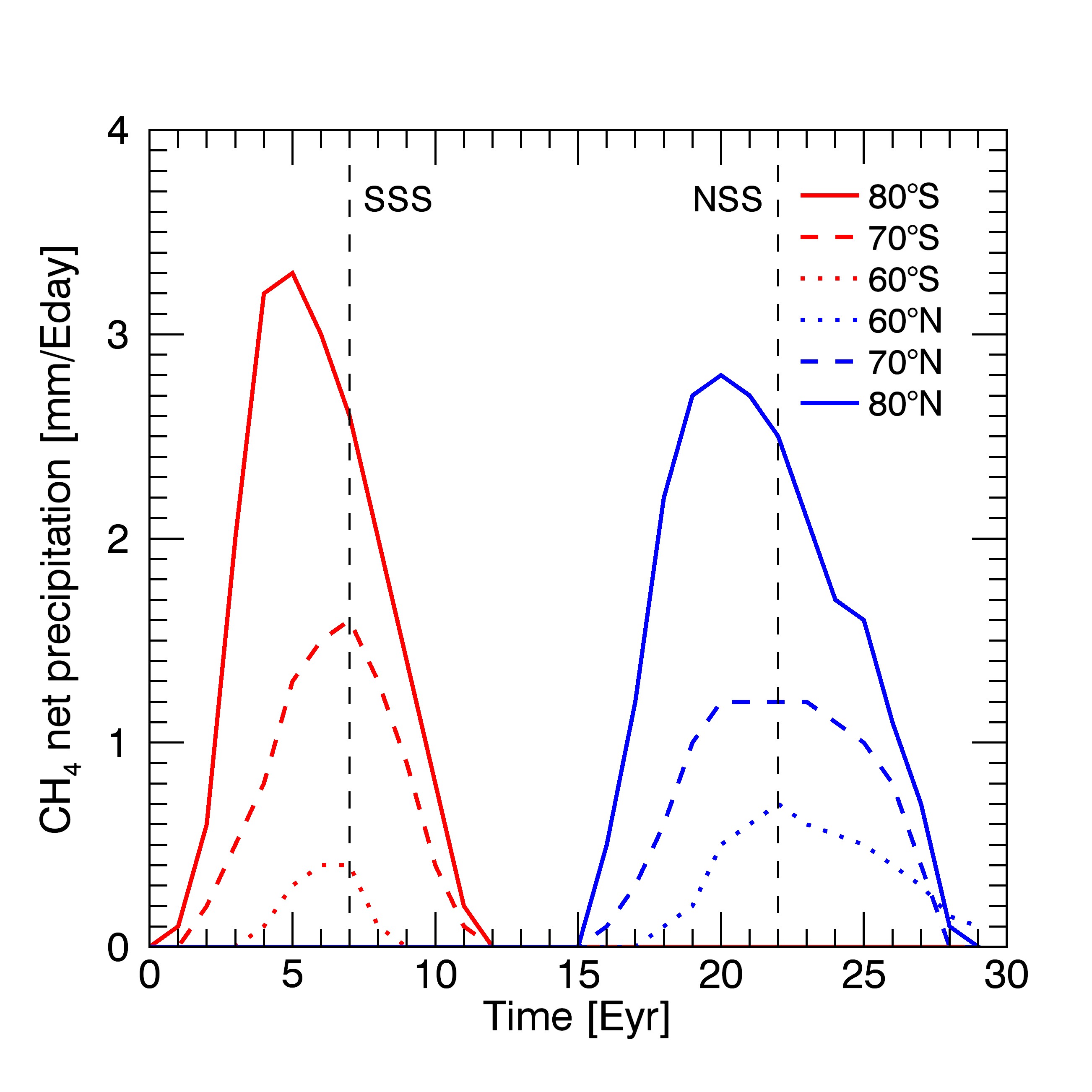} 
\caption{Net precipitation rates in mm/Earth day (Eday) at high southern and northern latitudes from \citet{Schneider2012}
over a Titan year (29.5 Eyrs). SSS and NSS: Southern and Northern Summer Solstices.}
\label{fig:PE_Titan}
\end{figure}

Here, we use the methane net 
precipitation rates extracted from the GCM of 
\citet{Schneider2012} to compute the present-day denudation rate on Titan. 
Figure \ref{fig:PE_Titan} represents the mean net precipitation rates
at various polar latitudes. High latitudes $> 80^\circ$
would be quite humid ($\tau = 7 - 8$ m/Tyr). Lower latitudes would be less humid
($\tau = 3 - 3.6$ m/Tyr at $70^\circ$, decreasing to $\tau = 0.4 - 1.6$ m/Tyr at $60^\circ$).
Southern low latitudes would be much drier than northern latitudes over a Titan year
as a result of sparser but more intense rainstorms.

\subsubsection{Case of pure compounds} \label{sec:DR_Titan_time_pure}
\begin{figure}
\center
\includegraphics[width=\textwidth]{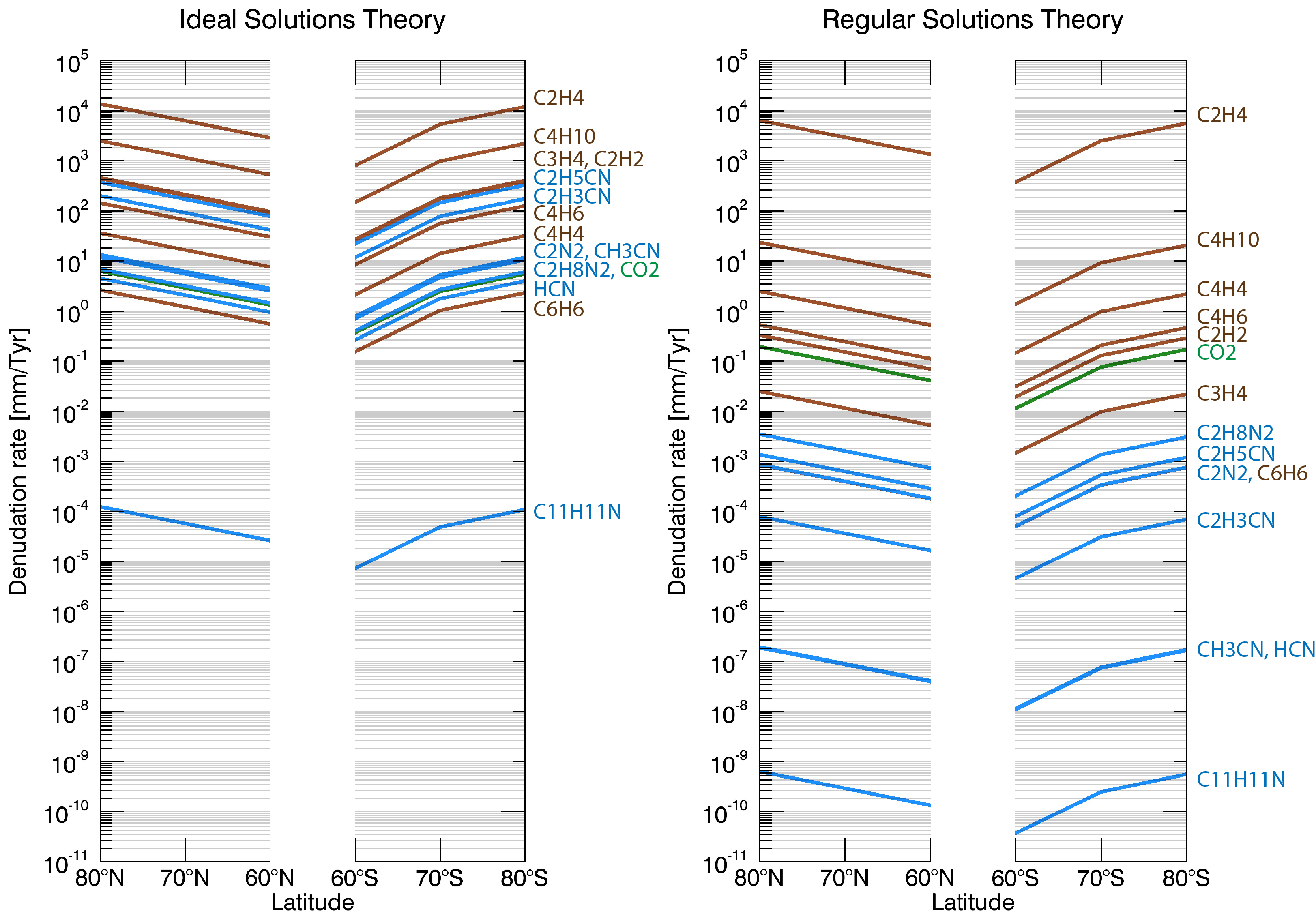} 
\caption{Denudation rates (in mm/Tyr) of Titan's surface solids in liquid methane according to the methane net
precipitation rates of \citet{Schneider2012} over an entire Titan year for the IST and RST approaches. In brown
are represented simple hydrocarbons, in blue the nitriles and in green carbon dioxide.}
\label{fig:DRTyr}
\end{figure}

Figure \ref{fig:DRTyr} illustrates the denudation rates of a surface composed of pure 
organic compounds exposed to methane rains at several southern and northern polar latitudes 
according to the IST and RST hypotheses. Ethane and propane are not shown since 
the model of \citet{Schneider2012} only considers methane, but the behavior
of Titan's solids in these liquids is already discussed in Section \ref{sec:DR_Earth_time}.

Over one Titan year, the denudation rates are the highest at high latitudes
and the lowest at low southern latitudes. According to the IST, all compounds would experience dissolution on the order
of a few mm to a few meters at almost all latitudes (salt-like material), except C$_{11}$H$_{11}$N, the denudation rate of which 
would be a few hundred nm. The dissolution of C$_2$H$_8$N$_2$, CO$_2$, HCN and C$_6$H$_6$ at 
60$^\circ$S however would be on the order of several hundred $\mu$m (carbonate-like material).

According to the RST, C$_2$H$_4$ and C$_4$H$_{10}$ denudation rates are between a few 
mm up to a few meters per Titan year (salt-like materials). The dissolution
of C$_4$H$_4$, C$_4$H$_6$, C$_2$H$_2$, CO$_2$, C$_3$H$_4$ is $\mu$m to hundred $\mu$m-scale over a 
Titan year (carbonate-like materials). The denudation rates of nitriles and C$_6$H$_6$ are
between $10^{-7} - 10^{-10}$ mm/Tyr (for HCN, CH$_3$CN, C$_{11}$H$_{11}$N) and a few $\mu$m/Tyr
(carbonate to siliceous-like materials). 

At high latitudes $> 70^\circ$, we do not see much differences between the northern and the southern 
denudation rates, as expected from the similarities in precipitation rate between the two poles shown in Figure \ref{fig:PE_Titan}. At low southern latitudes, net precipitation rates are too low over one Titan year 
to allow a rapid and significant dissolution. Interestingly, Ontario Lacus and Sikun Labyrinthus, two 
landforms compared with terrestrial karsto-evaporitic and karstic landforms \citep{Malaska2010,Cornet2012}, 
are observed at latitudes greater than 70$^\circ$S, and no other well-developed
dissolution-related landforms are seen at lower southern latitudes.

Therefore, if Titan's surface is composed of pure hydrocarbons, dissolution processes are likely to occur 
but the formation of a karstic-like landscapes would be roughly 30 times slower on Titan than on Earth due to 
Titan's seasonality in precipitation.
Of course, this latter consideration depends on the actual composition of the surface, which is unlikely to be
pure, and of the accuracy of the climate model used. 

\subsubsection{Case of a mixed surface layer} \label{sec:DR_Titan_time_mixed}
We now assume the presence of a surface layer, the composition of which 
is proportional to the accumulation rates at the surface ($h_i$) of solids coming from the atmosphere,
calculated in the same way as \citet{Malaska2011dissolution} did:
\begin{equation}
h_i = p_i \, V^S_{\rm m, i} / N_A, \qquad \textrm{[m/Eyr]}
\end{equation}

where $p_i$ is the production rate of molecules (in molecules/m$^2$/Eyr) listed in Table \ref{t:products}, $V^S_{\rm m, i}$ is the 
molar volume of the solid (or subcooled liquid if the former is not known, in m$^3$/mol) and $N_A$ is the Avogadro 
number ($\simeq 6.022\times10^{23}$ mol$^{-1}$). The composition of the mixed organic layer
is then determined as percentages ($f_i$) of each organic compound in the layer (so that $f_i = h_i / \sum_i h_i$).
This method allows us to consider the volume occupied by each molecule, which is important especially for tholins 
because these contribute up to 20 \% on average to the total production rates of molecules, but they build up to $\simeq 50$ \% 
of the total thickness of the surface deposits due to their higher molar volumes compared to those of simple hydrocarbons or nitriles. 

We consider 3 cases for the surface composition: without tholins and with C$_2$H$_8$N$_2$ or C$_{11}$H$_{11}$N as 
tholins. Tholins production rate is that mentioned in \citet{Cabane1992}. Methane precipitation rates are those of \citet{Schneider2012}. 
The denudation rates for these mixed organic layers ($DR_{mix}$) are computed in a linear mixing model scheme where:
\begin{equation}
DR_{mix} = \sum_i f_i \, DR_i. \qquad \textrm{[m/Eyr]}
\end{equation}

\begin{figure}
\center
\includegraphics[width=\textwidth]{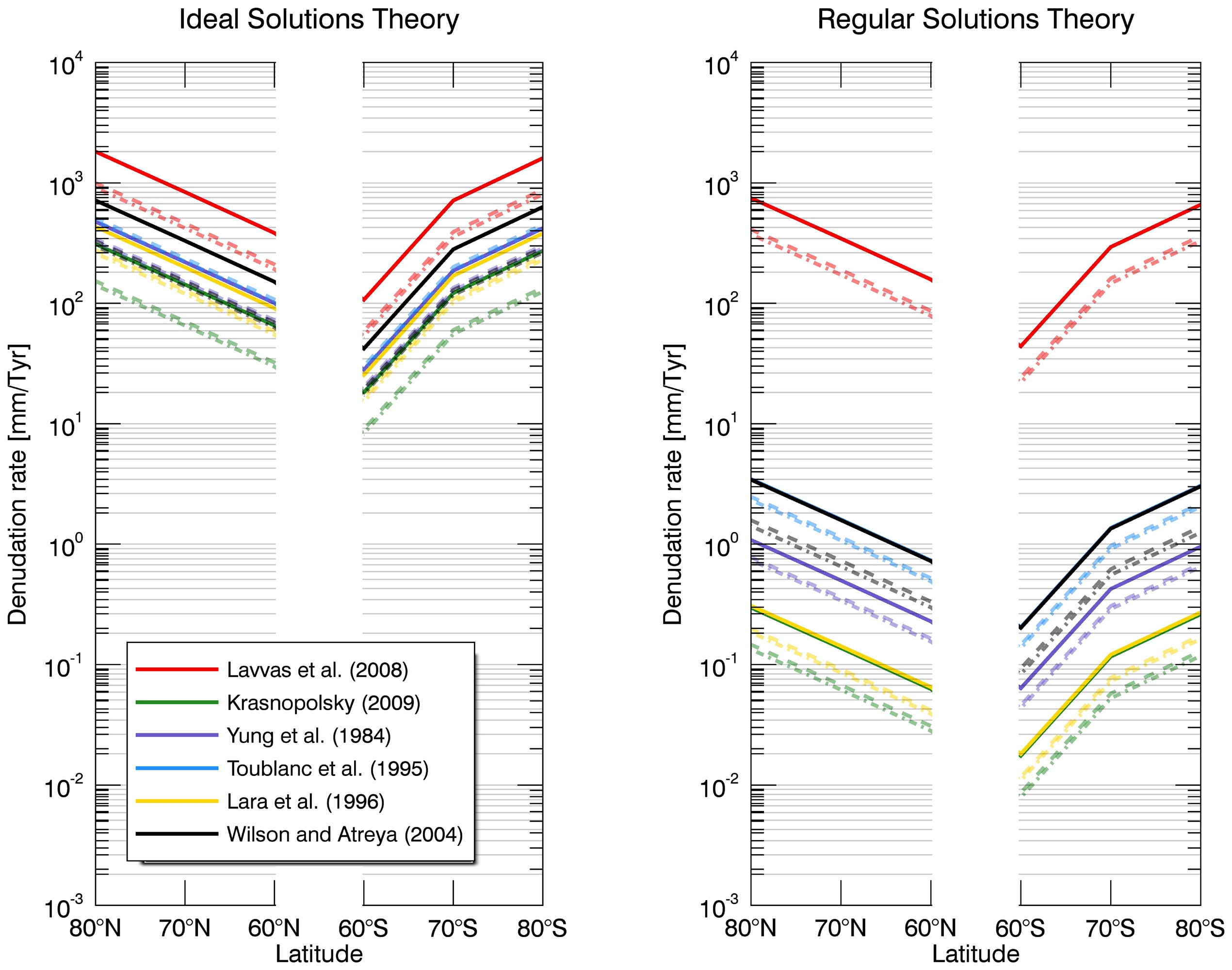} 
\caption{Denudation rates (in mm/Tyr) of Titan's surface solids in liquid methane according to the \citet{Schneider2012} model net 
precipitation rates over an entire Titan year at various latitudes and for different surface layer compositions established from
photochemical models, including (dashed lines) or not (solid lines) the influence of tholins (dashed line: with C$_2$N$_8$N$_2$ ; 
dashed-dot line: with C$_{11}$H$_{11}$N).}
\label{fig:DR_photochemical}
\end{figure}

Figure \ref{fig:DR_photochemical} gives the repartition of denudation as a function of latitude and photochemical models 
at the end of a Titan year. According to the IST, the denudation rate of all mixed layers would
be on the order of a few cm to a few meters over a Titan year (salt-like layers). According to the RST,
the organic layer originating from the \citet{Lavvas2008b} model would be the most soluble (dissolution rates of a 
few cm to a few dm per Titan year, salt-like layer). The organic layers originating from the other models would be 
more carbonate-like layers over a Titan year 
(dissolution rates of a few tens of $\mu$m to a few mm per Titan year), whether tholins are included or not. 
The lowest solubility of all mixed layers is 
reached using a \citet{Krasnopolsky2009}-type composition. Over a Titan year, the likelihood 
of developing dissolution-related landforms is therefore non-negligible, even if the surface is not composed of 
pure soluble simple solids. These mixed organic layers would behave like carbonate or salty terrestrial layers over a 
Titan year.

\section{Discussion: How old are Titan's karstic landscapes?} \label{sec:ages_Titan}
Despite the strong assumptions of the method described in Section \ref{sec:earth_ages} to infer timescales of 
formation of terrestrial karstic landforms (we consider only chemical erosion at equilibrium without significant 
climate changes over the past few MEyrs), the resulting ages are consistent with ages determined by relative or
absolute chronology or constitute upper limits. Therefore, the determination of the age of the lacustrine depressions
on Titan probably result in maximum timescales of development. Titan's climate is believed to have 
remained quite stable over the recent past, with a small periodic insolation variation of $\pm 2$ W/m$^2$ at the poles 
during the last MEyr, for the current low insolation at the North pole \citep{Aharonson2009}. This probably brings some
stability to the calculations.

\begin{figure}
\center
\includegraphics[width=\textwidth]{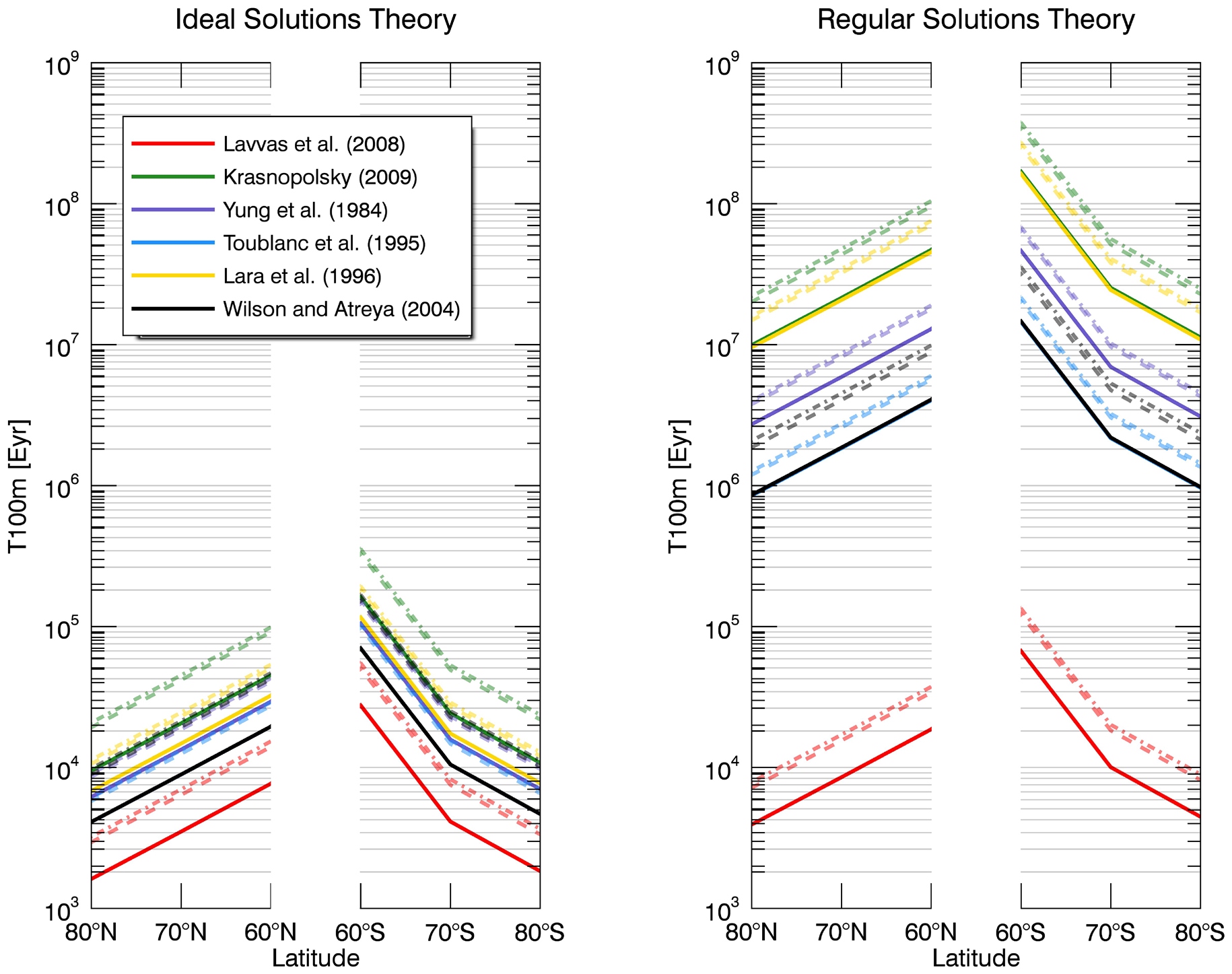} 
\caption{Same as Figure \ref{fig:DR_photochemical} but for the timescales of formation (in Eyr) of 100 m-deep depressions on Titan.}
\label{fig:T100m}
\end{figure}

By applying our simple model, we compute the timescales needed to form a 100 m-deep depression by 
dissolution of a superficial mixed organic layer under the current climate conditions evaluated by \citet{Schneider2012}. 
These are shown in Table \ref{t:Titan_ages} and Figure \ref{fig:T100m}. We compute our formation timescales
using both the IST and the RST since the solubility of polar molecules is not well constrained by the RST 
and could get closer to that computed using the IST. However, we consider timescales evaluated using the RST 
as our references since they present the most conservative values for the age of the lacustrine depressions.

Independent of the thermodynamic theory considered,
Titan's lacustrine depressions would be young. Among all the compositions tested, the time needed to carve 
a 100 m-deep depression by dissolution under current climate conditions at latitudes poleward of 70$^\circ$ would be 
between a few kEyrs (IST) and 56 MEyrs (RST). 
At 60$^\circ$N, a 100 m-deep depression would be created in 7.7 kEyrs (IST) to 104.6 MEyrs (RST) 
while the same depression would be created in 27.6 kEyrs (IST) to 375.1 MEyrs (RST) at 60$^\circ$S. 
This strong difference between the two hemispheres could explain why Titan's south polar 
regions are deprived of well developed lacustrine depressions 
compared to the North. 

It should be noted that the hypothesized timescale difference 
between the northern and the southern low latitudes results from an extrapolation
of Titan's current climate to the past. \citet{Aharonson2009} showed that
Croll-Milankovich-like cycle with periods of 45 (and 270) kEyrs could exist on Titan, resulting in 
a N-S reversal in insolation and likely subsequent climate conditions.
Over geological timescales, the N-S differences in denudation rates and timescales 
estimated at these latitudes could be smoothed by these cycles. However,
as noted earlier, we made the same extrapolation for the Earth, whose climate 
dramatically changed over time, without obtaining unreasonable formation 
timescales. 

In any case, all these timescales are consistent with the youth of Titan's surface as determined from:
\begin{enumerate}
\item crater counting ($0.3 - 1.2$ GEyrs, \citet{Neish2012}), 
\item dune sediment inventory ($50 - 730$ MEyrs, \citet{Sotin2012} and \citet{Rodriguez2014}), 
\item the flattening of the poles due to the substitution of methane by ethane in clathrates (500 MEyrs if restricted 
to the poles or $0.3 - 1.7$ GEyrs if not, \citet{Choukroun2012})
\item the possible methane outgassing event ($1.7 - 2.7$ GEyrs, \citet{Tobie2006}). 
\end{enumerate}

In summary, the morphology of Titan's lacustrine depressions suggests that dissolution occurs on Titan.
The denudation rates of pure organic compounds and a mixed organic layer as compared to those of soluble minerals 
on Earth also supports this hypothesis. The timescales needed to 
dissolve various amounts of material as compared to the timescales of development of karstic 
landforms on Earth are also quite consistent in the sense that karstic landscapes are usually relatively young landscapes. 
Finally, the latitudinal repartition of denudation rates and timescales of dissolution is consistent with the latitudinal 
repartition of the possible dissolution-related landforms at the surface of 
Titan. The surface dissolution scenario for the origin of Titan's lakes
appears very likely and Titan's lakes could be among the youngest features of the moon.

\section{Conclusion}

Titan's lakes result from the filling of topographic depressions by surface or subsurface liquids.
Their morphology led to analogies with terrestrial landforms of various
origins (volcanic, thermokarstic, karstic, evaporitic or karsto-evaporitic). The karstic/karsto-evaporitic
dissolution scenario seems to be the most relevant, given the nature of surface materials on Titan and its climate.
We constrained the timescales needed for the formation of Titan's depressions by dissolution, 
on the basis of the current knowledge on the development of terrestrial karsts.

We computed solutional denudation rates from the theory
developed by \citet{White1984}. This simple theory needs three parameters: the solubilities 
and the densities of solids and liquids at a given temperature and a climatic parameter linked 
to the net precipitation rates onto the surface. We computed the solubilities of terrestrial minerals 
in liquid water at 25$^\circ$C and tested the model by computing the denudation rates and timescales of formation
of several terrestrial examples of karstic landforms.

We then applied the same model to Titan. We computed the solubilities of Titan's
surface organic compounds in pure liquid methane, ethane and propane at 91.5 K
using different thermodynamic theories. We evaluated the molar volumes of
liquid and solid Titan's surface compounds at 91.5 K
and we used the results of the recent GCM of \citet{Schneider2012} as input for the precipitation rates 
of methane on Titan, which allowed us to compute denudation rates at several latitudes. 
Denudation rates have then been computed for pure organic compounds at Earth and Titan timescales 
and have been compared to those determined for soluble minerals on Earth. 
We also computed denudation rates for three different compositions of the surface organic layer. 
Over one Titan year, these mixed layers of organic compounds behave like terrestrial salts or carbonates, which indicates
their high susceptibility to dissolution, though these processes would be 30 times slower on Titan than on Earth due to 
the seasonality (rainfall occurs only during Titan's summer). 

We computed theoretical timescales for the formation of 100 m-deep depressions in mixed organic layers under present
climatic conditions. As with dissolution landforms on Earth, Titan's depressions
would be young. At high polar latitudes, we found that the timescales of development 
for depressions are relatively short (on the order of 50 MEyrs at maximum to carve 100 m) and consistent with the young age
of Titan's surface. These timescales are consistent with the existence of numerous lacustrine depressions and 
dissected landscapes at these latitudes. At southern low latitudes, the computed timescales are as long as 375 MEyrs
due to the low precipitation rates. This low propensity to develop depressions by dissolution is consistent with their relative
absence/bare formation at low latitudes. Over geological timescales greater than those of
Titan's Croll-Milankovitch cycles (45 and 270 kEyrs), this difference would probably be strongly attenuated. 
However, climate model predictions are not presently available over geologic timescales, and the present-day seasonal climate 
variations are the best that can be currently constrained.

The results of these simple calculations are consistent with the hypothesis that Titan's depressions most likely originate
from surface dissolution. Theoretical timescales for the formation of these landforms are consistent with the other age estimates of Titan's surface. 
Future works could include the effects of rain in equilibrium with the nitrogen, ethane and propane atmospheric gases 
in the raindrop composition (e.g. \citet{Graves2008} or \citet{Glein2013}). Experimental 
constraints on the solubility of gases and solids in liquids thanks to recent technical developments for Titan 
experiments \citep{LuspayKuti2012grl,LuspayKuti2014,Malaska2014,Chevrier2014,Leitner2014,Singh2014} would also be of extreme 
importance for such work. 
Finally, the influence of other landshaping mechanisms such as collapse or subsurface fluid flows, 
which play a significant role in the development of some karstic landforms on Earth, could also be implemented.

\section*{Acknowledgements}
The Cassini/RADAR SAR imaging datasets are provided through the NASA Planetary Data System Imaging Node portal 
($http://pds-imaging.jpl.nasa.gov/volumes/radar.html$). Terrestrial annual evapotranspiration
data taken from the Numerical Terradynamic Simulation Group (NTSG) database ($http://ntsg.umt.edu/project/mod16$) 
and precipitation data taken from the WordClim database ($http://www.worldclim.org/current$) have been used in this study.
The authors want to thank Fran\c cois Raulin and S\'ebastien Rodriguez for helpful discussions, Axel Lef\`evre and 
Manuel Giraud for their contribution in the Cassini SAR data processing, and Tim Rawle for the careful proofreading of the 
manuscript. The authors would like to acknowledge two anonymous reviewers for their work on the preliminary version of the 
manuscript, as well as the editor and associate editor for useful comments. TC is funded by the ESA Postdoctoral Research 
Fellowship Programme in Space Science.

\appendix 

\section{List of compound names for Titan} \label{sec:appendix_names}

\noindent CH$_4$: methane (liquid) \\ 
C$_2$H$_6$: ethane (liquid)\\ 
C$_3$H$_8$: propane (liquid)\\ 
C$_4$H$_{10}$: n-butane \\ 
C$_2$H$_2$: acetylene \\ 
C$_2$H$_4$: ethylene \\ 
C$_3$H$_4$: methyl-acetylene \\ 
C$_4$H$_4$: vinyl-acetylene \\ 
C$_4$H$_6$: 1-3 butadiene\\ 
C$_6$H$_6$: benzene \\ 
HCN: hydrogen cyanide \\ 
C$_2$N$_2$: cyanogen \\ 
CH$_3$CN: acetonitrile \\ 
C$_2$H$_3$CN: acrylonitrile \\ 
C$_2$H$_5$CN: propionitrile \\ 
C$_2$H$_8$N$_2$: 1-1 dimethyl-hydrazine (tholins-like)\\ 
C$_{11}$H$_{11}$N: quinoline (tholins-like)\\ 
H$_2$O: water ice Ih\\ 
CO$_2$: carbon dioxide ice

%
%

\section{Solubility of solids on Earth and Titan}  \label{sec:solubility}

%
%

\subsection{Dissolution processes on Earth} \label{sec:dissol_earth}

Dissolution occurs in karstic to evaporitic areas on Earth, where the dominant minerals
are calcite, dolomite, halite, anhydrite or gypsum.
Basic aqueous reactions of congruent dissolution for these common soluble minerals 
(i.e., all components of the minerals dissolve entirely) are summarized below:\\

\noindent CaCO$_{3}$ + H$_{2}$O+ CO$_{2}$ $\iff$ Ca$^{2+}$ + 2 HCO$_3^-$ \\
\noindent CaMg(CO$_3)_{2}$ + 2 H$_2$O + 2 CO$_{2}$ $\iff$ Ca$^{2+}$ + Mg$^{2+}$ + 4 HCO$_3^-$ \\
\noindent CaSO$_4.2$H$_2$O $\iff$ Ca$^{2+}$ + SO$_4^{2-}$ + 2 H$_2$O \\
\noindent CaSO$_4$+ H$_2$O $\iff$ Ca$^{2+}$ + SO$_4^{2-}$ + H$_2$O \\
\noindent NaCl+ H$_2$O $\iff$ Na$^+$ + Cl$^-$ + H$_2$O \\

Halite (NaCl), gypsum (CaSO$_4$.2H$_2$O) and anhydrite (CaSO$_4$) dissolve by pure 
dissociation, and calcite (CaCO$_3$) and dolomite (CaMg(CO$_3$)$_2$) dissolve by 
dissociation and acid dissolution, i.e. with the help of the CO$_2$ gas dissolved in 
water \citep{Langmuir1997,Ford2007,Brezonik2011}. 

The physical and thermodynamic parameters needed to compute solubilities
are given in Table \ref{t:parameters_minerals_ions}. Ideal and non-ideal 
electrolyte theories are considered. 
We also consider the acid dissolution of carbonate minerals. 

\subsubsection{Pure dissociation of minerals: halite, anhydrite and gypsum} \label{sec:xsat_earth}
For a given chemical reaction implying $A$ and $B$ and producing $C$ and $D$, with their respective stoichiometric numbers $a$, $b$, $c$ and $d$, the law of mass action gives:
\begin{equation}
aA + bB \iff cC + dD,
\end{equation}

where one can define a thermodynamical equilibrium constant (or thermodynamic solubility 
product) for the reaction $K_{eq}$ 
(often written ${\rm p}K_{eq}$, with ${\rm p}K_{eq}=-\log K_{eq}$), so that:
\begin{equation}
K_{eq} = \frac{(C)^c \, (D)^d}{(A)^a \, (B)^b} = \frac{(m_C \Gamma_C)^c \, (m_D \Gamma_D)^d}{(m_A \Gamma_A)^a \, (m_B \Gamma_B)^b},
\end{equation}

where $(i)$ denotes the activity of the \textit{ith} compound, being the product of an activity 
coefficient $\Gamma_i$ and a molality $m_i$ (in mol/kg of solvent). 
For liquids and solids reacting together (or being produced) during a congruent 
dissolution reaction, the activity is set to unity. 
The equilibrium constant can be derived from thermodynamics at standard state 
($T=25^\circ$C and $P=1$ atm) using the Gibbs-Helmholtz equation:
\begin{equation}
\Delta_{\rm r} G^\circ = - R \, T \, \ln K_{eq} = - 2.303 \, R \, T \, \log K_{eq}, \qquad \textrm{[J/mol]}
\end{equation}

where $\Delta_{\rm r} G^\circ$ is the standard Gibbs free energy of reaction (in J/mol), calculated 
using the Hess' law:
\begin{equation}
\Delta_{\rm r} G^\circ = \sum_{\rm products} n_i \, {\Delta_{\rm f} G_i^\circ} - \sum_{\rm reactants} n_j \,{\Delta_{\rm f} G_j^\circ}, \qquad \textrm{[J/mol]}
\end{equation}

with $\Delta_{\rm f} G^\circ$ being the standard Gibbs free energy of formation of each 
species (in J/mol) and $n$ their stoichiometric numbers. Following the same principle, 
we compute the standard enthalpy of reaction $\Delta_{\rm r} H^\circ$ 
(J/mol) from the individual enthalpies of formation $\Delta_{\rm f} H^\circ$ (in J/mol). The 
standard enthalpy of reaction is then used to compute the equilibrium constants at 
temperatures that differ from standard conditions using the Van't Hoff equation 
\citep{Langmuir1997} as follows:
\begin{equation} \label{eqn:vant_hoff}
\ln \frac{K_{T2}}{K_{T1}} =  \frac{\Delta_{\rm r} H_i^\circ}{R} \, \left(\frac{1}{T_1} -\frac{1}{T_2} \right),
\end{equation} 

where $K_{T1}$ and $K_{T2}$ are the equilibrium constant of a given chemical reaction 
at two different temperatures $T_1$ and $T_2$ (in K) (the subscript 1 is used to indicate 
the reference state, 25$^\circ$C in this case).
Table \ref{t:thermo_constants_minerals} gathers the thermodynamic parameters for all 
the considered dissolution reactions. It also displays each equilibrium constant equation 
for common terrestrial minerals.

\subsubsection{Acid dissolution of carbonates: calcite and dolomite} \label{sec:xsat_earth_acid}

For calcium and magnesium carbonates, a set of reactions happens, not only involving the mineral 
dissociation itself ($K_{cd}$ or $K_{dd}$), but also including the carbon dioxide gas dissolution in 
water ($K_{CO_2}$), which produces carbonic acid that acidifies water. Then, carbonic acid rapidly 
dissociates into bicarbonate ions ($K_{1}$), which are also created by the association of 
protons H$^+$ and carbonate ions ($K_{2}$). The thermodynamic properties of these intermediate 
reactions to the dissolution of calcite and dolomite are summarized in Table \ref{t:thermo_constants_minerals}. 
The activity of carbon dioxide is approximately equal to its partial pressure ($P_{CO_2}$, 
see \citet{Langmuir1997} and \citet{Ford2007}) so that, for calcite acid dissolution, one can write:
\begin{equation}
K_{cal} = \frac{(Ca^{2+}) \, (HCO_3^-)^2}{P_{CO_2}} = \frac{K_1 \, K_{CO_2} \, K_{cd}}{K_2},
\end{equation}

and for dolomite acid dissolution:
\begin{equation}
K_{dol} = \frac{(Ca^{2+}) \, (Mg^{2+}) \, (HCO_3^-)^4}{P_{CO_2}^2} = \frac{K_1^2 \, K_{CO_2}^2 \, K_{dd}}{K_2^2}.
\end{equation}

Assuming a given partial pressure of carbon dioxide, one can compute the activity, molality and mole 
fraction at saturation of calcite and dolomite in water. The molality of CO$_2$ in the system is 
calculated thanks to the Henry law ($m_{CO_2}= k_H(T) \, P_{CO_2}$, $K_H(T)$ being Henry's law 
constant, varying with temperature). We performed the 
calculations for 3 values of $P_{CO_2}$ at 25$^\circ$C: $P_{CO_2} = 0.33$ matm, which represents 
the normal dry air \citep{Langmuir1997} that could be quite analogous to atmospheric conditions in 
arid/semi-arid areas, and $P_{CO_2} = 0.01 - 0.11$ atm, which represents values 
encountered in more humid areas such as under equatorial/tropical climates 
\citep{Ford2007,Fleurant2008}. 

\subsubsection{Activity coefficients for electrolyte solutions} \label{sec:EST}

We infer the molalities $m_i$ and activities $(i)$ of ions in solution from $K_{eq}$ 
at different temperatures under an Ideal (IST, $\Gamma_i = 1$) and a non-ideal Electrolyte 
(EST, $\Gamma_i \ne 1$) Solution Theory. For the EST,  
the activity coefficients $\Gamma_i$ are calculated by iteration using the extended Debye-H\"uckel 
equation as modified by \citet{Truesdell1974}:
\begin{equation}
\log \Gamma_i = -A \, z_i^2 . \left(\frac{\sqrt{I}}{1+B \, a_i \, \sqrt{I}} \right) + b_i \, I,   \label{eqn:truesdelljones}
\end{equation}

where $I$ is the molal ionic strength of the solution, defined as follows:
\begin{equation}
I = \frac{1}{2} \, \sum_{i=1}^{n} \, m_i \, z_i^2.  \qquad \textrm{[mol/kg]}
\end{equation}

$z_i$ is the charge of \textit{i}, $a_i$ and $b_i$ are the 2 parameters of the \citet{Truesdell1974} equation 
(a modified hydrated radius, in \AA, and a purely empirical parameter respectively). 
$A$ and $B$ are two variables depending on the temperature $T$ (in $^\circ$C),
defined as follows \citep{Ford2007}:
\begin{eqnarray}
A&=&0.4883+8.074\times10^{-4} \, T, \\
B&=&0.3241+1.6\times10^{-4} \, T.
\end{eqnarray}

Results from these calculations are given in Table \ref{t:minerals_mole_frac}.
The calculation of activity coefficients has a relatively minor impact 
in the change of molality for carbonates, while a clear difference can be 
seen for salts (anhydrite and gypsum). 
The presence of CO$_2$ in the carbonate-water system greatly increases the 
amount of dissolved carbonates, until reaching solubilities roughly similar to those of gypsum and anhydrite. 
The EST molalities has been used for all minerals except for halite, which possesses
a ionic strength too strong to be computed using our Debye-H\"uckel equation.

\subsection{Solubility of Titan's solids} \label{sec:dissol_titan}

The solubility of Titan's solids in pure liquid hydrocarbons is calculated according to the Van't Hoff equation:
\begin{equation} \label{eqn:vanthoff_Titan}
\ln \Gamma_i \, X_{i, \rm sat} = \frac{\Delta_{\rm m} H_i^\circ}{R} \, \left(\frac{1}{T_{{\rm m},i}} -\frac{1}{T} \right),
\end{equation} 

where $\Gamma_i$ is the activity coefficient of the solute, $X_{i, \rm sat}$ its mole fraction at saturation, 
$\Delta_{\rm m} H_i^\circ$ its enthalpy of melting (in J/mol) evaluated at $T_{{\rm m},i}$, its melting 
temperature (in K), which differs from the temperature of the solution $T$ (K). $R$ is the ideal gas 
constant. Melting properties of Titan compounds are given in Table \ref{t:hydrocarbons_constants}.

We consider the case of the Ideal Solutions Theory (IST), for which $X_{i, \rm sat}$
is calculated by assuming $\Gamma_i = 1$ (all the molecules share the same affinity
with each other), and the case of the Regular
Solutions Theory (RST) of \citet{Preston1970} for which $\Gamma_i \ne 1$ 
(molecules have preferential affinity with one or the other). 
The latter approach has already been used in several publications dealing with the lakes and seas composition
\citep{Raulin1987,Dubouloz1989,Cordier2009,Cordier2013}.

The RST requires to compute $\Gamma_i$'s. 
Assuming the subscripts 1 for the solvent and 2 for the solute, the RST gives:
\begin{equation}
\ln \Gamma_2 = \frac{(\delta_1 - \delta_2)^2 + 2 \, l_{12} \delta_1 \delta_2}{R \, T} \, V_{{\rm m}, 2}^L \, \phi_1^2,
\end{equation}

where $V_{{\rm m}, 2}^L$ is the subcooled liquid molar volume of the solute 
(in m$^3$/mol), $ l_{12}$ is the empirical interaction parameter between 
the solvent and the solute taken from various sources (e.g. \citet{Preston1970} and \citet{SzczepaniecCieciak1978}), 
$\phi_1$ is the volume fraction of the liquid, computed as follows:
\begin{equation}
\phi_1 = \frac{(1 - X_2) \, V_{{\rm m},1}^L}{(1 - X_2) \, V_{{\rm m},1}^L + X_2 \, V_{{\rm m},2}^L}, 
\end{equation}

and $\delta_1$ and $\delta_2$ are the Hildebrand solubility parameters 
of the solvent and the solute respectively (in (J/m$^3$)$^{1/2}$ 
computed from their enthalpy of vaporization $\Delta_{\rm v} H_i$ (J/mol), as follows:
\begin{equation}
\delta_i^2 = \frac{\Delta_{\rm v} H_i - R \, T}{V_{{\rm m}, i}^L}, \qquad \textrm{[(J/m$^3)^{1/2}$]}
\end{equation} 

where $\Delta_{\rm v} H_i$ is taken in the literature at the boiling point temperature $T_b$ (Table 
\ref{t:hydrocarbons_constants}) for each compound 
and extrapolated down to lower temperatures using the Watson equation \citep{Poling2007}:
\begin{equation}
\Delta_{\rm v} H_{i,T} = \Delta_{\rm v} H_{i,Tb} \, \left( \frac{ 1 - T/T_c}{1 - T_b/T_c} \right)^{0.38}. \qquad \textrm{[J/mol]}
\end{equation} 

\begin{figure}
\center
\includegraphics[width=\textwidth]{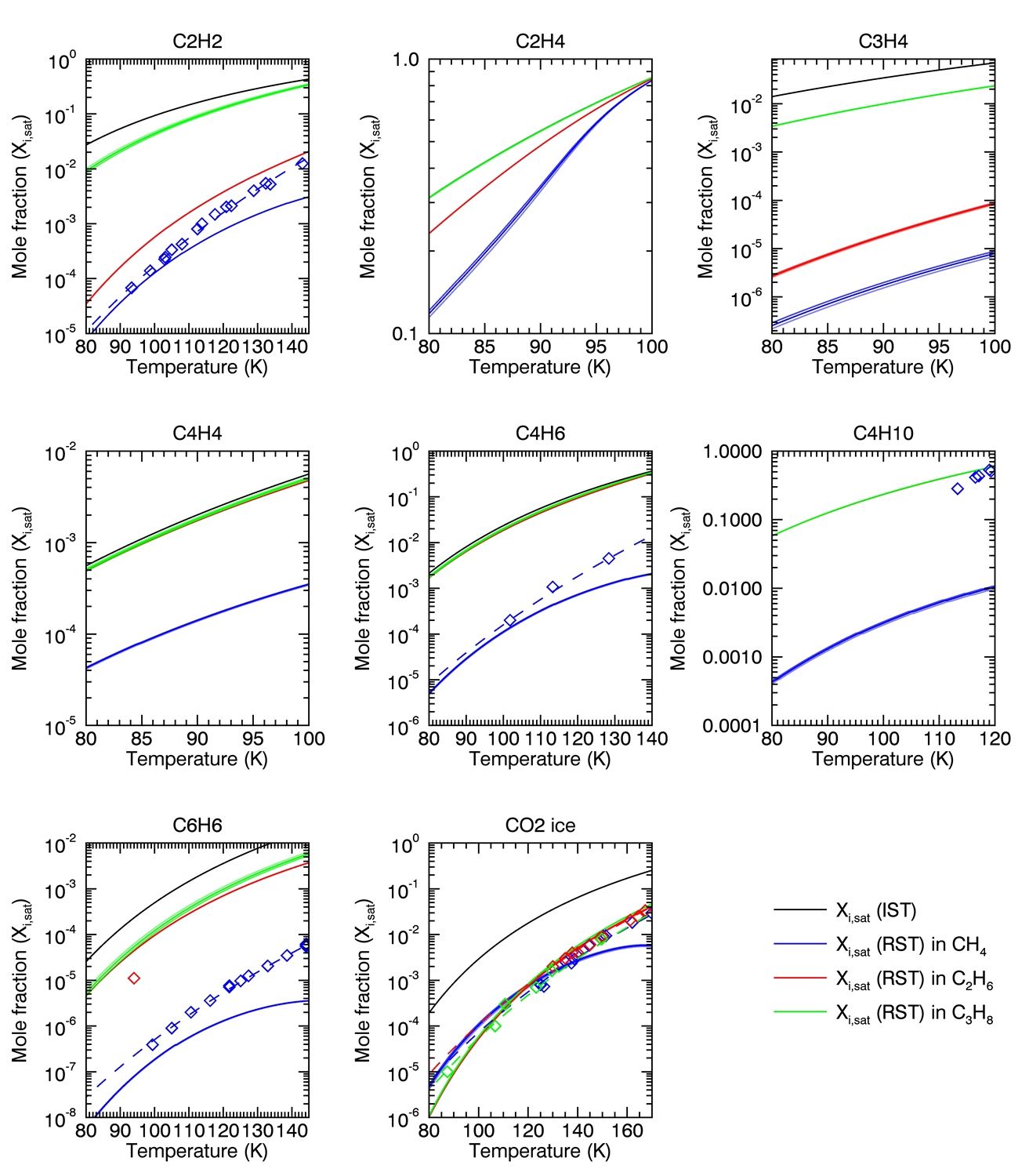} 
\caption{Saturation mole fraction of Titan's ``non-polar solids'' (hydrocarbons and carbon dioxide ice) in Titan's liquids 
computed using the RST and IST, and different Rackett equations (see Section \ref{sec:rackett}). The darkest curves represent 
$X_{i,sat}$ evaluated by Equation \ref{eqn:rackett2} with the $Z_{\rm RA}$ parameter. Diamonds are experimental values 
collected for C$_4$H$_{10}$ \citep{Brew1977,Kuebler1975}, C$_2$H$_2$ \citep{Neumann1969}, 
C$_4$H$_6$ \citep{Preston1971}, C$_6$H$_6$ \citep{Szalghary1972,Luks1981,Kuebler1995,Malaska2014} and CO$_2$ 
\citep{Clark1953,Davis1962,Cheung1968,Preston1971} in the corresponding liquids. Dashed lines are empirical power fits 
of the experimental points (with $R^2 > 0.98$).}
\label{fig:xsat1}
\end{figure}

\begin{figure}
\center
\includegraphics[width=\textwidth]{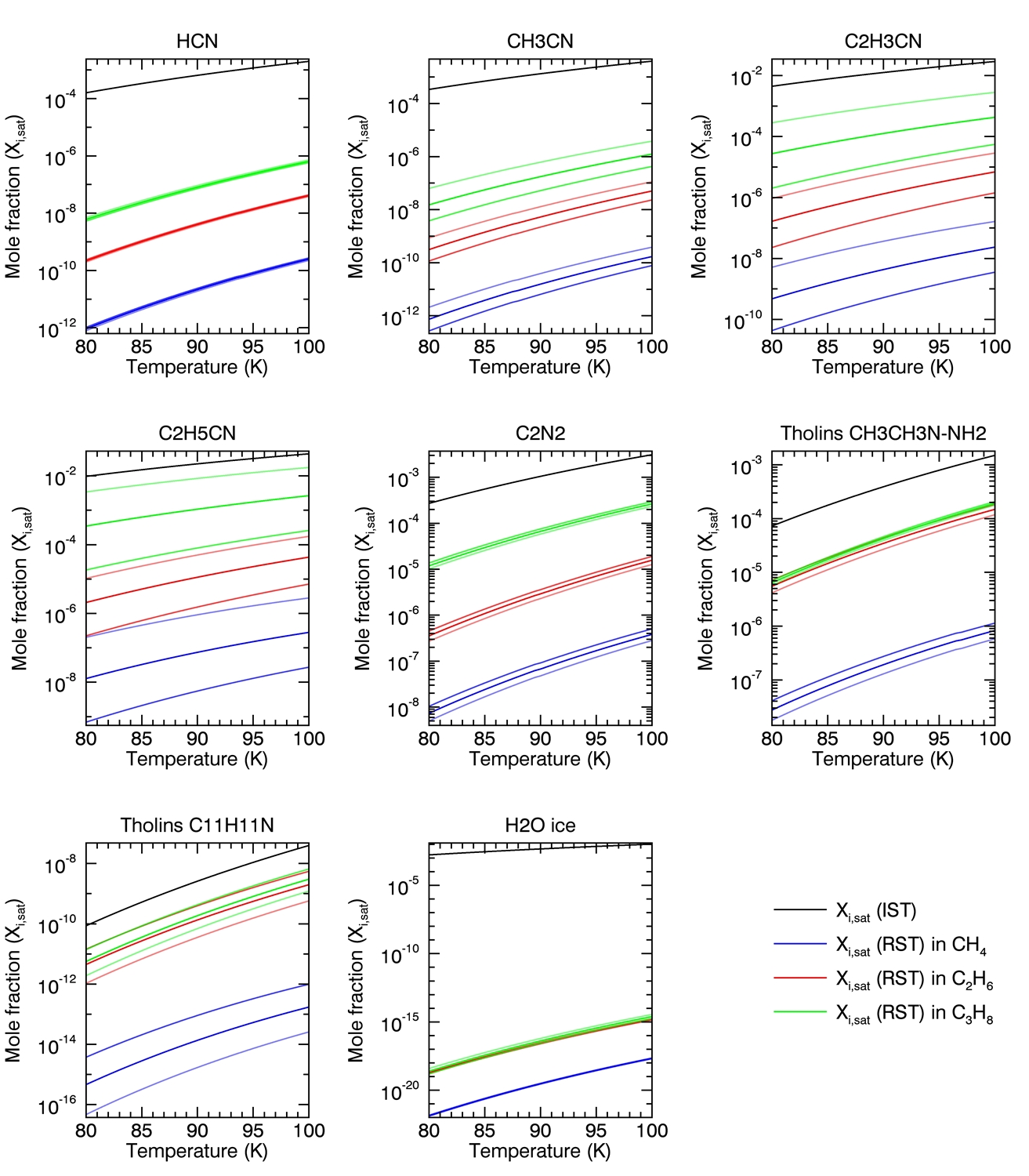} 
\caption{Same as Figure \ref{fig:xsat1} but for Titan's polar solids (nitriles and water ice).}
\label{fig:xsat2}
\end{figure}

The solubilities are given for a wide range of temperatures in Figures \ref{fig:xsat1} (non-polar molecules)
and \ref{fig:xsat2} (polar molecules), ontop of which we also reported
experimental data points gathered from the literature, fitted by an empirical power law ($R^2 > 0.98$) in order 
to extrapolate the experimental values at low temperatures for comparison.

The IST hypothesis provides a first estimate of the solubility, as seen in Figures \ref{fig:xsat1} and \ref{fig:xsat2}.
It does not consider the different behavior of the solutes in the different solvents, but presents
the advantage of not depending on approximate calculations of the activity coefficients of each species. 
According to this theory, all simple hydrocarbons would thus be rather solubles at low temperatures, the least soluble being
nitriles and C$_6$H$_6$.

The RST hypothesis provides a second estimate and shows that solids are less soluble in methane than in ethane and propane. 
Whatever the liquid, simple nitriles generally show an increase in solubility with a higher number of C atoms in their carbon chain.
C$_2$H$_4$, C$_4$H$_{10}$ and C$_2$H$_2$ are the most soluble compounds again, joined by C$_4$H$_4$, C$_4$H$_6$, CO$_2$, C$_3$H$_4$ 
and finally C$_6$H$_6$. Nitriles are the least soluble compounds. At cryogenic temperatures ($<100$ K), the simulated 
solubilities are in good agreement with the power fits of experimental data (with the exception of n-butane, which 
shows a higher solubility in experiments). It is also worth noting 
that recent solubility experiments of benzene in liquid ethane have been performed at 94 K by
\citet{Malaska2014}. The $X_{i,\rm sat}$ for benzene in ethane they determined experimentally 
($1.11 \pm 0.09 \times 10^{-5}$) is also quite close to our theoretical value computed using the RST
at 94 K ($4.42 \times 10^{-5}$), though slightly lower. This difference is probably due to the dissolution
of nitrogen into liquid ethane during the experiment, which tends to lower the solubility of benzene. 
Benzene also reached saturation so quickly in liquid ethane (in less than 2h) \citep{Malaska2014}, compared to the
timescales considered in the present work, that dissolution can be assumed instantaneous.
Finally, as expected, the solubility of water ice is unconstrained 
given the considerable range between IST and RST estimates (about a factor 10$^{17}$), but is probably low. 

Solubilities calculated using the IST and RST are also given at 91.5 K in Table \ref{t:xsat}
along with those gathered from our empirical fits of experimental data and the literature.
The fits of experimental data are quite consistent with our RST values. Previous studies document the solubility
of pure solids in liquid mixtures in equilibrium with the atmosphere (thus composed of methane, ethane, propane and/or nitrogen)
at various temperatures. Direct comparisons are therefore tentative, since the liquid composition, temperatures and the
thermodynamic parameters sources are heterogeneous. Nitrogen tends to decrease the solubility of solids in liquids
while ethane and propane tend to increase it. It should be noted that the solubility of hydrocarbons is quite in agreement 
between our study and others \citep{Raulin1987,Dubouloz1989,Glein2013} but the solubility of nitriles appears lower in our 
simulations than in previous ones \citep{Raulin1987,Dubouloz1989,Cordier2013erratum}, for which they lie between our IST and
RST estimates.

%
%

\section{Molar volumes of solvents and solutes} \label{sec:molar_volumes}

If the density of a compound at 
a given temperature ($T$) is known from experimental data, it is straightforward to derive its molar volume 
$V_{{\rm m}, i}(T)$, as follows:
\begin{equation} \label{eqn:density}
 V_{{\rm m}, i}(T)= M_i /  \rho_i(T) \qquad \textrm{[m$^3$/mol]}
\end{equation}
 
This is the case for water ice \citep{Loerting2011} and minerals \citep{CRC}.
However, experimentally determined densities for solids and liquids at the very low 
temperatures relevant to Titan are rather rare.
We describe two complimentary techniques to evaluate the molar volumes of solids and
liquids.

\subsection{Subcooled liquid molar volumes} \label{sec:rackett}
We first evaluate molar volumes at a given temperature $T$ 
using the Rackett equation \citep{Spencer1972,Poling2007}. This method
is designed to estimate the saturated subcooled liquid molar volumes of pure hydrocarbons and organic 
solvents ($V_{{\rm m}, i}^L$). 
\begin{equation} \label{eqn:rackett}
V_{{\rm m}, i}^L = \frac{R \, T_{{\rm c},i}}{P_{{\rm c}, i}} Z_{c,i} \,^{\left[1+(1-T_{\rm r,i})^{2/7} \right]} \qquad \textrm{[m$^3$/mol]}
\end{equation}

$R$ is the ideal gas constant ($=8.3144621$ J/mol/K), $T_{\rm c}$ and $P_{\rm c}$ are the critical 
temperature (in K) and pressure (in Pa or J/m$^3$) respectively, $Z_c$ is the critical compressibility 
factor, similar to the Rackett parameter ($Z_{\rm RA} = 0.29056-0.08775 \omega$, $\omega$ being the accentric 
factor, see \citet{Poling2007} for more details) and $T_{\rm r}$ the reduced temperature ($T_{\rm r,i} = T/T_{\rm c,i}$). 
All these parameters are given in Table \ref{t:critical_properties}.
An alternative form of this equation is given by:
\begin{equation} \label{eqn:rackett2}
V_{{\rm m}, i}^L = V_{c,i }\, Z_{c,i} \,^{\left(1-T_{\rm r,i}\right)^{2/7}} \qquad \textrm{[m$^3$/mol]}
\end{equation}

\begin{figure}
\center
\includegraphics[width=\textwidth]{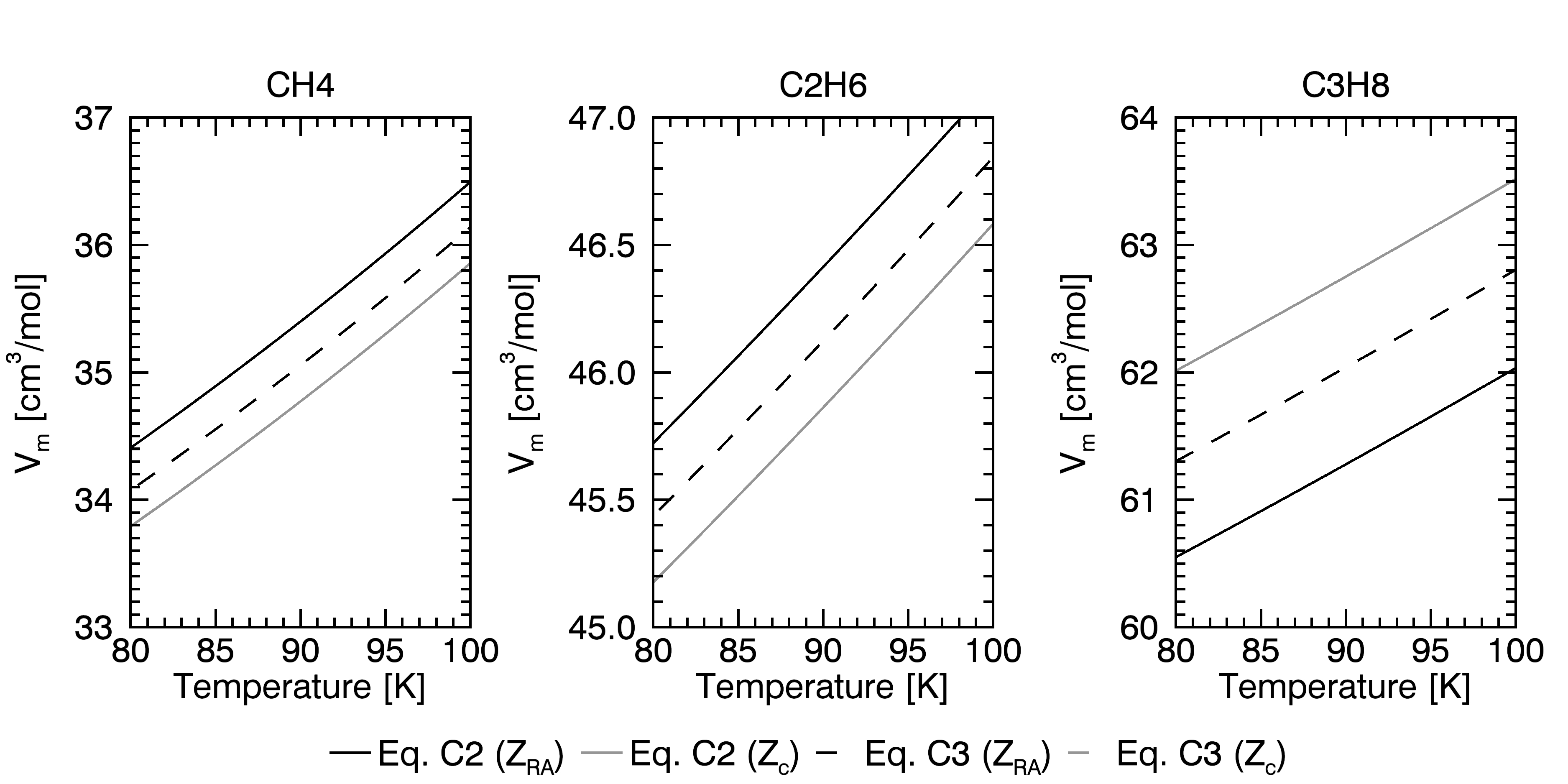} 
\caption{Differences in the calculated molar volumes of Titan's liquids due to the use of different Rackett 
equations and compressibility factors.}
\label{fig:Vm1}
\end{figure}

We investigated uncertainties in the molar volumes computed using the Rackett
equations (\ref{eqn:rackett} and \ref{eqn:rackett2}), with the two different parameters 
($Z_{\rm RA}$, taken from database or computed using the accentric factor, and 
$Z_c$, computed from the ideal gas equation at the critical point). 
The results are reported in Table \ref{t:liquid_molar_volumes} and Figures \ref{fig:Vm1} (liquids), \ref{fig:Vm2} (non-polar solids) and \ref{fig:Vm3} (polar solids). 
Differences in molar volumes are only significant for nitriles, going up to $2\sigma < 24 $ cm$^3$/mol
for propionitrile. To reduce possible over or underestimates for nitriles molar volumes and 
solubilities, we use Equation \ref{eqn:rackett2} with $Z_{\rm RA}$, which is our ``intermediate'' case
and also the most accurate estimates of $V_{{\rm m}, i}^L$ \citep{Poling2007}.

\subsection{Titan's solids molar volumes} \label{sec:abinitio}

The solid molar volumes were determined from crystal structures available in the 
literature \citep{Dulmage1951,Dietrich1975,Simon1980,Refson1986,Antson1987,Etters1989,McMullan1992,Craven1993}.
The CH$_3$CN, HCN, C$_4$H$_{10}$ and C$_2$H$_2$ have a temperature dependent polymorphism. For each of these compounds, 
the selected crystal structure corresponds to the one observed at the temperature of Titan's surface (i.e. 90 - 95 K). 
Except for CH$_3$CN, the cell volume was experimentally determined for different temperatures. A second order polynomial was
used to fit the cell volume as a function of the temperature to extrapolate them at Titan's temperatures 
(90 - 95 K).

Since the pressure on Titan (around 1.5 bar) is not the same as on Earth (1 bar) where the crystal cell volume were determined, 
we checked the influence pressure on molar volumes by calculations based on the Density Functional Theory (DFT) using the 
VASP 5.3 package. Cell volumes of each compound were optimized using vdW-DF2 \citep{Lee2010,Klimes2011} functional 
involving a non-local kernel for the electronic correlation energy calculation. This functional allows to reproduce dispersion 
interactions (such as Van der Waals interactions) which are major inter-molecular interaction encountered in molecular crystals. 
The $E_{\rm ncut}$ value, defining the basis set size, was fixed to 800 eV in order to suppress Pulay's stress. The k-points 
sampling was done with a $12\times12\times12$ Monkhorst-Pack grid. The core electrons were described with the 
projector-augmented plane wave (PAW) approach. Calculations were performed with and without static pressure (set to 100 bar) 
to compare cell volumes. The larger variation was observed for C$_4$H$_{10}$, with a volume decrease of 0.44 \% upon pressure. 
This is reasonably low to conclude that cell volumes measured on Earth (P=1 bar) should be very similar to the ones on Titan (P=1.5 bar).

\begin{figure}
\center
\includegraphics[width=\textwidth]{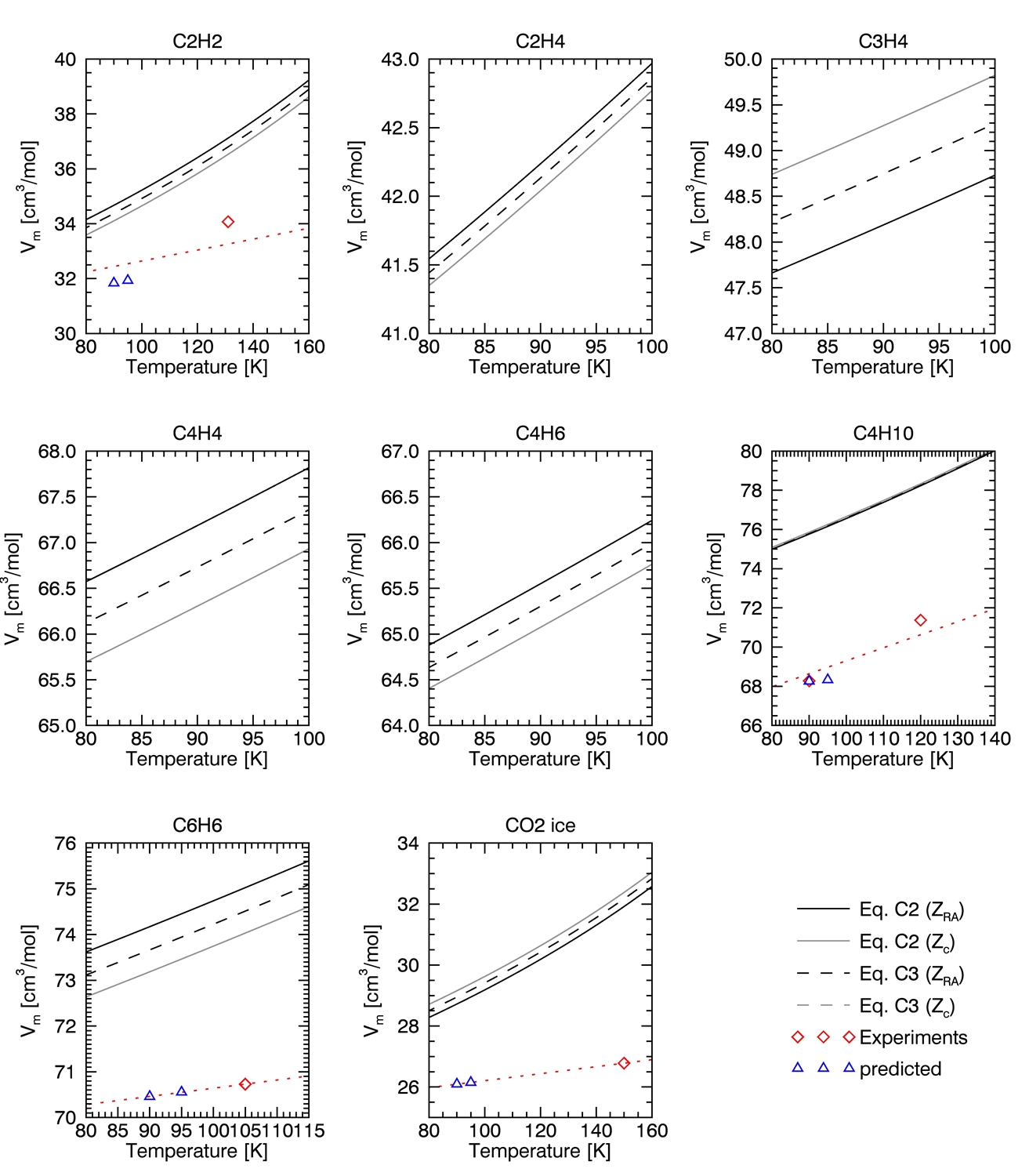} 
\caption{Same as Figure \ref{fig:Vm1} but for ``non-polar'' solids (hydrocarbons and carbon dioxide ice). 
Red and blue diamonds are experimental and predicted values respectively.
Sources: C$_4$H$_{10}$ \citep{Refson1986}, C$_2$H$_2$ \citep{McMullan1992}, C$_6$H$_6$ \citep{Craven1993}
and CO$_2$ \citep{Simon1980}.}
\label{fig:Vm2}
\end{figure}

\begin{figure}
\center
\includegraphics[width=\textwidth]{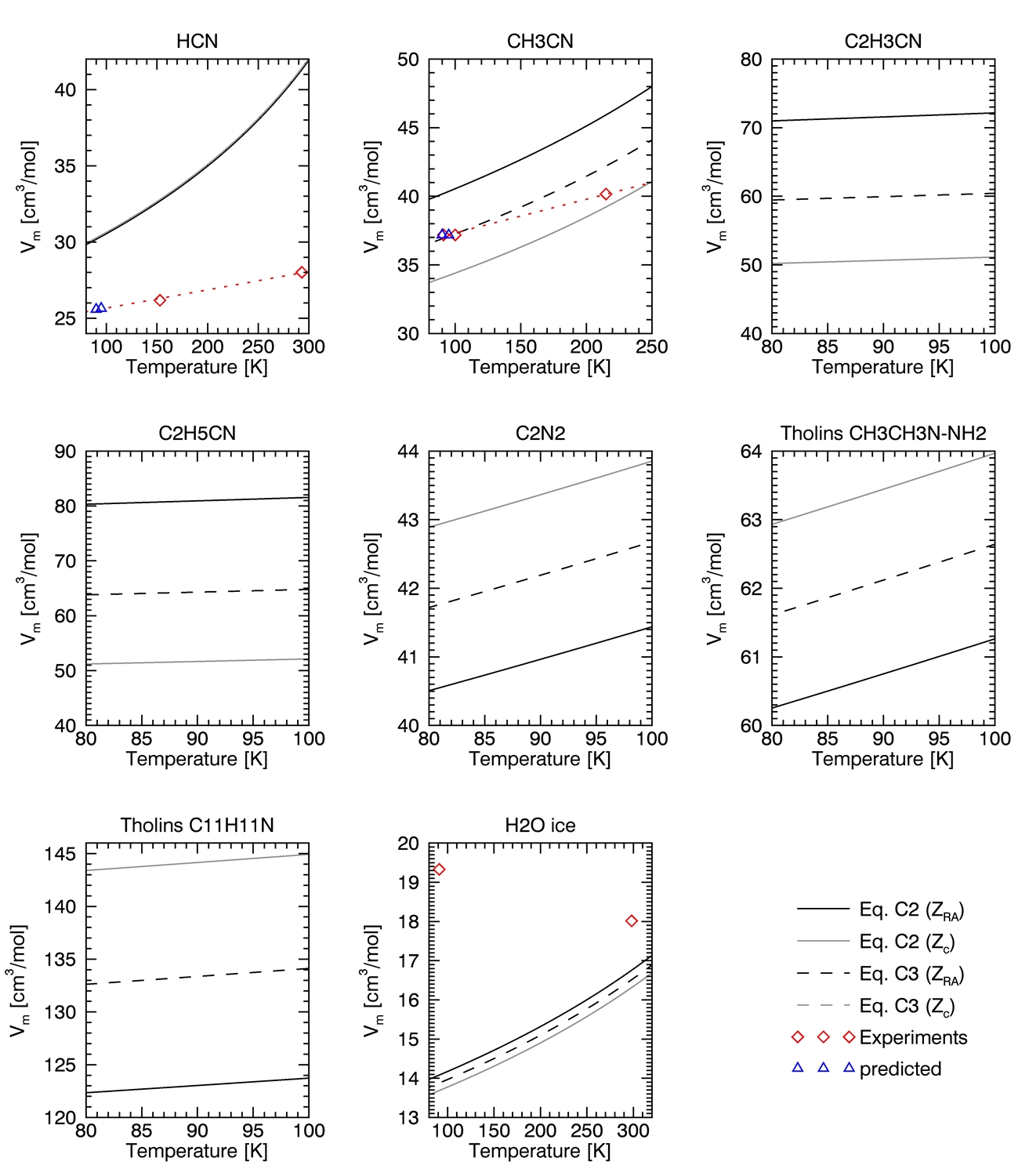} 
\caption{Same as Figure \ref{fig:Vm2} but for polar solids (nitriles and water ice). 
Sources: HCN \citep{Dulmage1951}, CH$_3$CN \citep{Barrow1981,Antson1987} and H$_2$O \citep{Loerting2011}.}
\label{fig:Vm3}
\end{figure}

We reported on Figures \ref{fig:Vm2} and \ref{fig:Vm3} the predicted crystalline 
molar volumes ontop of the experimental and Rackett values (individual values 
also given in Table \ref{t:solid_molar_volumes}). Overall, we get a 
quite good agreement between all these estimates. We estimate 
the solid molar volumes at 91.5 K using a simple empirical linear fit between experimental and 
predicted points. The Rackett method is used to give an approximate molar volume 
for the solids for which we do not have experimental data.

\newpage


\begin{table}
\centering
\caption{ Production rates (in molecules/cm$^ 2$/s) of simple solid organics in Titan's atmosphere derived from several photochemical models. 
Not reported here is the production rate of tholins, estimated by \citet{Cabane1992}, \citet{Rannou2003} and \citet{Krasnopolsky2009} 
to be between 0.2 and 7 kg/cm$^2$/GEyr (Giga Earth year). The list of compound names is available in Appendix \ref{sec:appendix_names}. 
Y84: \citet{Yung1984}, T95: \citet{Toublanc1995}, L96: \citet{Lara1996}, W04: \citet{Wilson2004}, L08: \citet{Lavvas2008a,Lavvas2008b}, 
K09: \citet{Krasnopolsky2009}.}
 \footnotesize{
\begin{tabular}{llllllll}
\hline
            & Y84 & T95 & L96 & W04 & L08 & K09 \\
\hline
\hline
C$_2$H$_6$  &  5.80$\times 10^{9}$ & 1.50$\times 10^{10}$ & 1.41$\times 10^{9}$ & 1.60$\times 10^{9}$ & 4.10$\times 10^{9}$ & 1.16$\times 10^{9}$ \\
C$_3$H$_8$ &  1.40$\times 10^{8}$ & 5.40$\times 10^{8}$ & 3.14$\times 10^{7}$ & 3.30$\times 10^{7}$ & 3.70$\times 10^{8}$ & 2.17$\times 10^{8}$ \\
C$_4$H$_{10}$ &  2.30$\times 10^{7}$ & 8.80$\times 10^{7}$ &       & 3.00$\times 10^{7}$ &       & 7.02$\times 10^{5}$ \\
C$_2$H$_2$ &  1.20$\times 10^{9}$ & 1.10$\times 10^{9}$ & 9.28$\times 10^{8}$ & 3.70$\times 10^{8}$ & 3.20$\times 10^{8}$ & 3.22$\times 10^{8}$\\
C$_2$H$_4$  &       &       &       & 9.30$\times 10^{7}$ & 6.00$\times 10^{7}$ &  \\
C$_3$H$_4$  &       &       &       &       &       & 2.64$\times 10^{6}$ \\
C$_4$H$_4$ &       &       &       &       &       & 1.57$\times 10^{4}$ \\
C$_4$H$_6$  &       &       &       & 6.50$\times 10^{4}$ &       & 1.71$\times 10^{7}$ \\
C$_6$H$_6$  &        &       &       & 4.60$\times 10^{5}$ & 2.10$\times 10^{5}$ & 1.08$\times 10^{6}$ \\
HCN   &  2.00$\times 10^{8}$ & 1.00$\times 10^{8}$ & 8.00$\times 10^{7}$ & 2.10$\times 10^{6}$ & 3.20$\times 10^{8}$ & 1.54$\times 10^{8}$ \\
HC$_3$N  & 1.70$\times 10^{7}$ &       & 3.73$\times 10^{7}$ & 1.30$\times 10^{7}$ & 9.20$\times 10^{6}$ & 1.26$\times 10^{7}$ \\
C$_2$N$_2$  & 1.20$\times 10^{7}$ &       & 2.89$\times 10^{4}$ & 2.00$\times 10^{6}$ & 2.60$\times 10^{4}$ & 3.52$\times 10^{6}$ \\
C$_4$N$_2$  &        &       & 7.20$\times 10^{6}$ & 9.70$\times 10^{5}$ & 1.90$\times 10^{6}$ & \\
CH$_3$CN &       &       & 6.61$\times 10^{5}$ & 3.40$\times 10^{4}$ & 1.70$\times 10^{7}$ & 1.27$\times 10^{7}$ \\
C$_2$H$_3$CN &        &       &       & 1.60$\times 10^{7}$ & 2.80$\times 10^{5}$ & 1.62$\times 10^{7}$ \\
C$_2$H$_5$CN &        &       &       &       &       & 8.00$\times 10^{5}$ \\
\hline
\end{tabular}}%
  \label{t:products}%
\end{table}%

\begin{table}
  \centering
  \caption{Solutional denudation rates in mm/Eyr for terrestrial minerals at 298.15 K 
and under various climatic conditions (CO$_2$ content and precipitation rates). All values 
correspond to the EST hypothesis except those for halite (IST).}
\footnotesize{
\begin{tabular}{lll}
\hline
Name  & $\tau=30$ cm/Eyr & $\tau=1$ m/Eyr\\
\hline
\hline
      & \multicolumn{2}{c}{Dissolution by dissociation}\\
\hline
\hline
Halite & 50.643 & 167.122 \\
Gyspum & 0.504 & 1.662 \\
Anhydrite & 0.224 & 0.740 \\
Calcite & 8.55$\times10^{-4}$ & 2.82$\times10^{-3}$ \\
Dolomite & 7.84$\times10^{-4}$ & 2.59$\times10^{-3}$\\
\hline
\hline
      & \multicolumn{2}{c}{Acid dissolution} \\
\hline
\hline
Calcite, $PCO_2$ = 0.33 matm  & 0.006 & 0.020 \\
Dolomite, $PCO_2$ = 0.33 matm  & 0.005 & 0.017 \\
\hline
Calcite, $PCO_2$ = 0.01 atm  & 0.020 & 0.066 \\
Dolomite, $PCO_2$ = 0.01 atm  & 0.017 & 0.057 \\
\hline
Calcite, $PCO_2$ = 0.11 atm  & 0.047 & 0.156\\
Dolomite, $PCO_2$ = 0.11 atm  & 0.040 & 0.132\\
\hline
\end{tabular}}
  \label{t:DRmin}%
\end{table}%


\begin{table}
  \centering
  \caption{Timescales required to dissolve 100 meters of a surface organic layer (composition given by photochemistry
in the atmosphere) in liquid methane according to the precipitation rates of \citet{Schneider2012}.}
\footnotesize{
\begin{tabular}{lllll}
\hline
Latitude & \multicolumn{2}{c}{IST} & \multicolumn{2}{c}{RST} \\
      & min (kEyrs) & max (kEyrs) & min (kEyrs) & max (MEyrs) \\
\hline
\hline
80$^\circ$N   & 1.6   & 20.9  & 3.9   & 22.1 \\
70$^\circ$N   & 3.5   & 45.2  & 8.5   & 47.8 \\
60$^\circ$N   & 7.7   & 99.0  & 18.7  & 104.6 \\
60$^\circ$S   & 27.6  & 354.9 & 67.1  & 375.1 \\
70$^\circ$S   & 4.1   & 53.2  & 10.1  & 56.2 \\
80$^\circ$S   & 1.8   & 23.7  & 4.5   & 25.0 \\
\hline
\end{tabular}}
  \label{t:Titan_ages}%
\end{table}%

\begin{table}[!t]
  \centering
  \caption{Thermodynamic and physical parameters of minerals, water and ions considered in 
the dissolution reactions, collected from the literature \citep{CRC,Langmuir1997,Ford2007,Brezonik2011,White2013}.}
\footnotesize{
    \begin{tabular}{lllllllll}
    \hline
    Name  & Formula & $M_i$ & $z_i$ & $a_i$  & $b_i$  & $\Delta_{\rm f} G_i^\circ$  & $\Delta_{\rm f} H_i^\circ$ \\
	    &               & [g/mol] & & [\AA] & [kg/mol] & [kJ/mol] & [kJ/mol] \\
\hline
\hline
\multicolumn{8}{c}{\textit{minerals}}\\
\hline
calcite & CaCO$_3$ & 100.087 & 0     & -     & -     & -1129.1 & -1207.6 \\
dolomite & CaMg(CO$_3$)$_2$ & 184.401 & 0     & -     & -     & -2161.7 & -2324.5 \\
gypsum & CaSO$_4$.2H$_2$O & 172.171 & 0     & -     & -     & -1797.2 & -2022.6 \\
anhydrite & CaSO$_4$ & 136.141 & 0     & -     & -     & -1321.7 & -1434.1 \\
halite & NaCl  & 58.443 & 0     & -     & -     & -384.1 & -411.2\\
\hline
\multicolumn{8}{c}{\textit{aqueous species}} \\
\hline
chlore & Cl$^-$ & 35.460 & $-1$  & 3.71  & 0.01  & -131.3 & -167.2 \\
sodium & Na$^+$ & 22.990 & $+1$  & 4.32  & 0.06  & -261.9 & -240.1 \\
hydroxyl & OH$^-$ & 17.007 & $-1$  & 10.65 & 0.21  & -157.3 & -230 \\
hydrogen & H$^+$ & 1.008 & $+1$  & 4.78  & 0.24  & 0     & 0 \\
calcium & Ca$^{2+}$ & 40.080 & $+2$  & 4.86  & 0.15  & -553.54 & -542.83 \\
magnesium & Mg$^{2+}$ & 24.320 & $+2$  & 5.46  & 0.22  & -454.8 & -466.8 \\
sulfate & SO$_4^{2-}$ & 96.060 & $-2$  & 5     & -0.04 & -744.6 & -909.2 \\
carbonate & CO$_3^{2-}$ & 60.010 & $-2$  & 5.4   & 0     & -527.9 & -677.1 \\
bicarbonate & HCO3$^-$ & 61.020 & $-1$  & 5.4   & 0     & -586.8 & -692 \\
carbonic acid & H$_2$CO$_3$ & 62.025 & 0     & -     & -     & -623.2 & -699.7\\
\hline
\multicolumn{8}{c}{\textit{solvent and gas}} \\
\hline
water  & H$_2$O & 18.015 & 0     & -     & -     & -237.18 & -285.83\\
carbon dioxide & CO$_2$ & 44.010 & 0     & -     & -     & -394.4 & -393.51 \\
\hline
    \end{tabular}}%
  \label{t:parameters_minerals_ions}%
\end{table}%

\begin{sidewaystable}
  \centering
  \caption{Thermodynamic constants of dissociation and dissolution reactions in standard conditions 
(at 25$^\circ$C and 1 atm).}
\footnotesize{
    \begin{tabular}{lllllll}
    \hline
    Reaction & Equations & $\Delta_{\rm r} G^\circ$ & $\Delta_{\rm r} H^\circ$ & $K_{eq}$ & ${\rm p}K_{eq}$ & Equilibrium constant expression \\
 	        &                  &        [kJ/mol]                    &              [kJ/mol]               & 		&		\\
    \hline
    \hline
    \multicolumn{7}{c}{\textit{Pure dissociation reactions for minerals}}\\
    \hline
    gypsum dissociation & CaSO$_4$.2H$_2$O   $\iff$ Ca$^{2+}$  + SO$_4^{2-}$  + 2H$_2$O  & 24.70 & -1.09 & 4.71$\times 10^{-5} $ & 4.327 & $K_{gd} = (Ca^{2+}) (SO_4^{2-})$ \\
    anhydrite dissociation & CaSO$_4$   $\iff$ Ca$^{2+}$   + SO$_4^{2-}$   & 23.56 & -17.93 & 7.46$\times 10^{-5} $ & 4.128 & $K_{ad} = (Ca^{2+}) (SO_4^{2-})$ \\
    halite dissociation & NaCl   $\iff$ Na$^{+}$   + Cl$^{-}$   & -9.10 & 3.90  & 39.29 & -1.594 & $K_{hd} = (Na^{+}) (Cl^{-})$ \\    
    calcite dissociation & CaCO$_3$   $\iff$ Ca$^{2+}$   + CO$_3^{2-}$   & 47.36 & -12.53 & 5.05$\times 10^{-9} $ & 8.297 & $K_{cd} = (Ca^{2+}) (CO_3^{2-})$ \\
    dolomite dissociation & CaMg(CO$_3$)$_2$   $\iff$ Ca$^{2+}$   + Mg$^{2+}$   + 2CO$_3^{2-}$   & 97.56 & -39.33 & 8.10$\times 10^{-18} $ & 17.092 & $K_{dd} = (Ca^{2+}) (Mg^{2+}) (CO_3^{2-})^2$  \\
    \hline
    \multicolumn{7}{c}{\textit{Secondary dissociation reactions and dissolution reactions for carbonates dissolution}}\\
    \hline
    CO$_2$ dissolution & CO$_2$   + H$_2$O   $\iff$ H$_2$CO$_3^\circ$   & 8.38  & -20.36 & 3.40$\times 10^{-2} $ &   1.468 & $K_{CO_2} = (H_2CO_3^\circ) / P_{CO_2}$  \\
    carbonic acid dissociation & H$_2$CO$_3^\circ$   $\iff$ H$^+$   + HCO$_3^-$   & 36.40 & 7.70  & 4.20$\times 10^{-7} $ &     6.377 & $K_1 = (H^+) (HCO_3^-) / (H_2CO_3^\circ) $\\
    bicarbonate dissociation & H$^+$   + CO$_3^{2-}$   $\iff$ HCO$_3^-$   & 58.90 & 14.90 & 4.80$\times 10^{-11}$ &  10.319 & $K_2 =  (HCO_3^-) / (H^+) (CO_3^{2-})$ \\
    calcite dissolution & CaCO$_3$   + H$_2$O   + CO$_2$   $\iff$ Ca$^{2+}$   + 2 HCO$_3^-$   & 33.24 & -40.09 & 1.50 $\times 10^{-6} $ & 5.823 & $K_{cal} = (Ca^{2+}) (HCO3^- )^2 / P_{CO_2}$\\ 
    dolomite dissolution & CaMg(CO$_3)_2$   + 2H$_2$O   + 2CO$_2$   $\iff$ Ca$^{2+}$   + Mg$^{2+}$   + 4 HCO$_3^-$   & 69.32 & -94.45 & 7.17 $\times 10^{-13}$ &12.144 & $ K_{dol} = (Ca^{2+}) (Mg^{2+}) (HCO3^- )^4 / P_{CO_2}^2$ \\ 
   \hline
    \end{tabular}}%
  \label{t:thermo_constants_minerals}%
\end{sidewaystable}

\begin{table}[htbp]
  \centering
  \caption{Molalities in mol/kg of solvent of minerals in water at 25$^\circ$C according to the Ideal and the Electrolyte Solutions 
Theories. $P_{CO_2} = 0.33$ matm refers to normal dry air \citep{Langmuir1997} while $P_{CO_2} = 0.01$ atm more refers to tropical climate 
environments \citep{Fleurant2008}, which actually could even reach considerable CO$_2$ content in the soils of up to $P_{CO_2}=0.11$ 
atm \citep{Ford2007}.}
  \footnotesize{
   \begin{tabular}{lll}
\hline
Name  & IST   & EST \\
\hline
\hline
      & \multicolumn{2}{c}{Dissolution by dissociation} \\
\hline
\hline
Halite & 6.27 & 6.27 \\
Gyspum & 8.63$\times10^{-3}$ & 2.26$\times10^{-2}$ \\
Anhydrite & 6.86$\times10^{-3}$ & 1.62$\times10^{-2}$ \\
Calcite & 7.10$\times10^{-5}$ & 7.72$\times10^{-5}$ \\
Dolomite & 3.77$\times10^{-5}$ & 4.11$\times10^{-5}$ \\
\hline
\hline
      & \multicolumn{2}{c}{Acid dissolution} \\
\hline
\hline
Calcite, PCO$_2$ = 0.33 matm  & 4.99$\times10^{-4}$ & 5.42$\times10^{-4}$ \\
Dolomite, PCO$_2$ = 0.33 matm  & 2.59$\times10^{-4}$ & 2.76$\times10^{-4}$ \\
\hline
Calcite, PCO$_2$ = 0.01 atm  & 1.55$\times10^{-3}$ & 1.80$\times10^{-3}$ \\
Dolomite, PCO$_2$ = 0.01 atm  & 8.09$\times10^{-4}$ & 8.99$\times10^{-4}$ \\
\hline
Calcite, PCO$_2$ = 0.11 atm  & 3.46$\times10^{-3}$ & 4.28$\times10^{-3}$ \\
Dolomite, PCO$_2$ = 0.11 atm  & 1.80$\times10^{-3}$ & 2.09$\times10^{-3}$ \\
\hline
\end{tabular}}%
  \label{t:minerals_mole_frac}%
\end{table}%


\begin{table}[htbp]
  \centering
  \caption{Melting and boiling points thermodynamic constants of Titanian materials. Data gathered from \citet{CRC} and \citet{Yaws1996}.
The melting properties of C$_{11}$H$_{11}$N have been taken in \citet{Chirico2007}.}
\footnotesize{
\begin{tabular}{lllll}
\hline
 Formula & $T_{\rm m}$ & $\Delta_{\rm m} H^\circ$ & $T_b$ & $\Delta_{\rm v} H^\circ$ \\
            & [K]   & [kJ/mol] & [K]   & [kJ/mol] \\
\hline
\hline
CH$_4$ & 90.680 & 0.940 & 111.670 & 8.190\\
C$_2$H$_6$ & 90.360 & 0.582 & 184.570 & 14.690 \\
C$_3$H$_8$ & 85.460 & 3.500 & 232.040 & 19.040 \\
C$_4$H$_{10}$ & 134.850 & 4.660 & 272.660 & 22.440 \\
C$_2$H$_2$ & 192.450 & 4.105 & 188.450 & 16.674 \\
C$_2$H$_4$ & 104.000 & 3.350 & 169.380 & 13.530 \\
C$_3$H$_4$ & 170.450 & 5.349 & 249.950 & 22.185 \\
C$_4$H$_4$ & 227.600 & 7.687 & 278.250 & 22.470 \\
C$_4$H$_6$ & 164.240 & 7.980 & 268.740 & 22.470 \\
C$_6$H$_6$ & 278.640 & 9.870 & 353.240 & 30.720 \\
HCN   & 259.860 & 8.406 & 298.850 & 26.896 \\
C$_2$N$_2$ & 245.320 & 8.110 & 252.100 & 23.330 \\
CH$_3$CN & 229.330 & 8.160 & 354.800 & 29.750 \\
C$_2$H$_3$CN & 189.670 & 6.230 & 350.450 & 32.600 \\
C$_2$H$_5$CN & 207.150 & 5.030 & 350.500 & 31.810 \\
C$_2$H$_8$N$_2$ & 215.950 & 10.070 & 337.050 & 28.477 \\
C$_{11}$H$_{11}$N & 326.650 & 20.418 & 538.855 & 50.005\\
H$_2$O & 273.150 & 6.010 & 373.200 & 40.650 \\
CO$_2$ & 216.592 & 9.020 & 194.600 & 15.326 \\
\hline
\end{tabular}}
\label{t:hydrocarbons_constants}
\end{table}

\begin{sidewaystable}
  \centering
  \caption{Saturation mole fractions of Titan's solids in pure liquid methane, ethane and propane at 
91.5 K evaluated using the Ideal Solutions and Regular Solutions Theories compared with 
our fits at 91.5 K of experimental data and with values taken from the literature. 
R87: \citet{Raulin1987} (graphical reading), D89: \citet{Dubouloz1989}, 
C13: \citet{Cordier2013}, G13: \citet{Glein2013}, T13: \citet{Tan2013}.}
\footnotesize{
    \begin{tabular}{lllllllllllllll}
\hline
Formula & $X_i$ (IST) & \multicolumn{3}{c}{$X_i$ (RST)} & Experiments & \multicolumn{2}{c}{R87 (RST)} & D89 (RST) & C13 (IST) & G13 (MVL) & \multicolumn{2}{c}{T13 (PC-SAFT)} \\
      & 91.5 K & \multicolumn{3}{c}{91.5 K} & 91.5 K & \multicolumn{2}{c}{94 K} & 92.5 K & 90 K  & 90.7 K & \multicolumn{1}{c}{90 K} & \multicolumn{1}{c}{93.7 K} \\
\hline
\hline
CH$_4$ & -     & 1.00 & -     & -     & 1.00 & 8.60$\times 10^{-1}$ & -     & 7.30$\times 10^{-2}$ & -     & 6.81$\times 10^{-1}$ & 6.84$\times 10^{-1}$ & 3.18$\times 10^{-1}$ \\
C$_2$H$_6$ & -     & -     & 1.00 & -     & -     & -     & 9.70$\times 10^{-1}$ & 9.09$\times 10^{-1}$ & -     & 1.55$\times 10^{-1}$ & 8.30$\times 10^{-2}$ & 5.32$\times 10^{-1}$ \\
C$_3$H$_8$ & -     & -     & -     & 1.00 & -     & -     & -     &       & -     & 1.55$\times 10^{-2}$ & 7.80$\times 10^{-3}$ & 7.20$\times 10^{-2}$ \\
N$_2$ & -     & -     & -     & -     & -     & 1.40$\times 10^{-1}$ & 3.00$\times 10^{-2}$ & 1.80$\times 10^{-2}$ & -     & 1.48$\times 10^{-1}$ & 2.21$\times 10^{-1}$ & 6.89$\times 10^{-2}$ \\
\hline
C$_4$H$_{10}$ & 1.40$\times 10^{-1}$ & 1.51$\times 10^{-3}$ & 1.39$\times 10^{-1}$ & 1.40$\times 10^{-1}$ & -     & -     & -     & -     & 1.22$\times 10^{-1}$ & -     & -     & - \\
C$_2$H$_2$ & 5.90$\times 10^{-2}$ & 4.48$\times 10^{-5}$ & 2.00$\times 10^{-4}$ & 2.45$\times 10^{-2}$ & 5.72$\times 10^{-5}$ & 1.81$\times 10^{-4}$ & 6.93$\times 10^{-4}$ & 4.10$\times 10^{-4}$ & 5.40$\times 10^{-2}$ & 2.20$\times 10^{-4}$ & 3.40$\times 10^{-3}$ & 7.70$\times 10^{-3}$ \\
C$_2$H$_4$ & 5.89$\times 10^{-1}$ & 4.02$\times 10^{-1}$ & 5.33$\times 10^{-1}$ & 5.88$\times 10^{-1}$ & -     & 1.81$\times 10^{-3}$ & 4.62$\times 10^{-2}$ & -     & -     & -     & -     & - \\
C$_3$H$_4$ & 3.85$\times 10^{-2}$ & 2.25$\times 10^{-6}$ & 2.36$\times 10^{-5}$ & 1.14$\times 10^{-2}$ & -     & 7.26$\times 10^{-6}$ & 1.39$\times 10^{-4}$ & 5.00$\times 10^{-5}$ & -     & -     & -     & - \\
C$_4$H$_4$ & 2.38$\times 10^{-3}$ & 1.64$\times 10^{-4}$ & 2.05$\times 10^{-3}$ & 2.16$\times 10^{-3}$ & -     & -     & -     & -     & -     & -     & -     & - \\
C$_4$H$_6$ & 9.60$\times 10^{-3}$ & 3.57$\times 10^{-5}$ & 7.57$\times 10^{-3}$ & 8.06$\times 10^{-3}$ & 4.85$\times 10^{-5}$ & 3.27$\times 10^{-5}$ & 2.77$\times 10^{-4}$ & -     & -     & -     & -     & - \\
C$_6$H$_6$ & 1.64$\times 10^{-4}$ & 5.32$\times 10^{-8}$ & 3.11$\times 10^{-5}$ & 3.76$\times 10^{-5}$ & 1.59$\times 10^{-7}$ & 3.63$\times 10^{-7}$ & 9.24$\times 10^{-6}$ & -     & 2.20$\times 10^{-4}$ & -     & -     & - \\
HCN   & 7.78$\times 10^{-4}$ & 3.25$\times 10^{-11}$ & 5.90$\times 10^{-9}$ & 1.14$\times 10^{-7}$ & -     & 7.26$\times 10^{-7}$ & 2.31$\times 10^{-5}$ & 8.50$\times 10^{-6}$ & 6.46$\times 10^{-4}$ & -     & -     & - \\
C$_2$N$_2$ & 1.25$\times 10^{-3}$ & 8.96$\times 10^{-8}$ & 3.83$\times 10^{-6}$ & 8.30$\times 10^{-5}$ & -     & 1.09$\times 10^{-7}$ & 2.77$\times 10^{-6}$ & 1.40$\times 10^{-6}$ & -     & -     & -     & - \\
CH$_3$CN & 1.59$\times 10^{-3}$ & 2.31$\times 10^{-11}$ & 7.64$\times 10^{-9}$ & 2.42$\times 10^{-7}$ & -     & 2.18$\times 10^{-6}$ & 2.77$\times 10^{-5}$ & 6.30$\times 10^{-5}$ & 3.73$\times 10^{-3}$ & -     & -     & - \\
C$_2$H$_3$CN & 1.44$\times 10^{-2}$ & 5.75$\times 10^{-9}$ & 1.74$\times 10^{-6}$ & 1.54$\times 10^{-4}$ & -     & -     & -     & 2.20$\times 10^{-5}$ & -     & -     & -     & - \\
C$_2$H$_5$CN & 2.49$\times 10^{-2}$ & 9.28$\times 10^{-8}$ & 1.41$\times 10^{-5}$ & 1.25$\times 10^{-3}$ & -     & 2.18$\times 10^{-6}$ & 3.70$\times 10^{-5}$ & 8.90$\times 10^{-5}$ & -     & -     & -     & - \\
C$_2$H$_8$N$_2$ & 4.87$\times 10^{-4}$ & 2.46$\times 10^{-7}$ & 4.41$\times 10^{-5}$ & 5.44$\times 10^{-5}$ & -     & -     & -     & -     & -     & -     & -     & - \\
C$_{11}$H$_{11}$N & 4.07$\times 10^{-9}$ & 2.08$\times 10^{-14}$ & 2.09$\times 10^{-10}$ & 2.94$\times 10^{-10}$ & -     & -     & -     & -     & -     & -     & -     & - \\
H$_2$O & 5.23$\times 10^{-3}$ & 6.17$\times 10^{-20}$ & 5.50$\times 10^{-17}$ & 7.98$\times 10^{-17}$ & -     & 1.81$\times 10^{-15}$ & 3.24$\times 10^{-12}$ & -     & -     & -     & -     & - \\
CO$_2$ & 1.06$\times 10^{-3}$ & 3.29$\times 10^{-5}$ & 1.23$\times 10^{-5}$ & 1.32$\times 10^{-5}$ & 2.68$\times 10^{-5}$ & 7.26$\times 10^{-5}$ & 3.24$\times 10^{-5}$ & 1.20$\times 10^{-5}$ & 8.72$\times 10^{-4}$ & -     & -     & - \\
\hline
    \end{tabular}}%
  \label{t:xsat}%
\end{sidewaystable}%

\begin{table}[htbp]
  \centering
  \caption{ Critical properties of Titanian materials \citep{CRC}. The Rackett parameters 
($Z_{\rm RA}$) have been gathered from the \textit{CHemical and Engineering Research Information Center} 
(CHERIC) database (\burl{http://www.cheric.org/research/kdb/hcprop/cmpsrch.php}) 
and from \textit{Infotherm} (\burl{http://www.infotherm.com/servlet/infothermSearch}). 
The compressibility factors ($Z_c$) have been calculated for comparison with $Z_{\rm RA}$. The 
critical properties of cyanogen and our tholins analogues (1-1 dimethyl-hydrazine C$_2$H$_8$N$_2$, 
as suggested in \citet{Cordier2013}, based on the experiments of \citet{Quirico2008}, and quinoline C$_{11}$H$_{11}$N, as sugested by \citet{Coll1995}) 
have been collected in \citet{Yaws1996}.}
\label{t:critical_properties} 
\footnotesize{
\begin{tabular}{lllllll}
\hline
Formula & $M$   & $T_c$ & $P_c$ & $V_c$ & $Z_{\rm RA}$ & $Z_c$ \\
           & [g/mol] & [K]   & [MPa] & [cm$^3$/mol] &       &  \\
\hline
\hline
CH$_4$ & 16.043 & 190.564 & 4.599 & 98.600 & 0.289 & 0.286\\
C$_2$H$_6$ & 30.069 & 305.320 & 4.872 & 145.500 & 0.281 & 0.279 \\
C$_3$H$_8$ & 44.096 & 369.830 & 4.248 & 203.000 & 0.277 & 0.280 \\
C$_4$H$_{10}$ & 58.122 & 425.160 & 3.787 & 255.000 & 0.273 & 0.273 \\
C$_2$H$_2$ & 26.037 & 308.300 & 6.138 & 112.200 & 0.271 & 0.269 \\
C$_2$H$_4$ & 28.053 & 282.340 & 5.041 & 131.000 & 0.282 & 0.281 \\
C$_3$H$_4$ & 40.064 & 402.000 & 5.630 & 163.500 & 0.272 & 0.275 \\
C$_4$H$_4$ & 52.075 & 455.600 & 4.860 & 218.710 & 0.282 & 0.281 \\
C$_4$H$_6$ & 54.091 & 425.000 & 4.320 & 221.000 & 0.271 & 0.270 \\
C$_6$H$_6$ & 78.112 & 562.050 & 4.895 & 256.000 & 0.270 & 0.268 \\
HCN   & 27.026 & 456.700 & 5.390 & 139.000 & 0.197 & 0.197 \\
C$_2$N$_2$ & 53.034 & 399.900 & 6.303 & 144.520 & 0.266 & 0.274 \\
CH$_3$CN & 41.042 & 545.500 & 4.850 & 171.000 & 0.199 & 0.183 \\
C$_2$H$_3$CN & 53.063 & 536.000 & 4.559 & 214.061 & 0.261 & 0.219 \\
C$_2$H$_5$CN & 55.079 & 561.300 & 4.260 & 229.000 & 0.263 & 0.209 \\
C$_2$H$_8$N$_2$ & 60.098 & 522.200 & 5.400 & 218.241 & 0.265 & 0.271 \\
C$_{11}$H$_{11}$N & 157.212 & 775.100 & 3.353 & 514.500 & 0.247 & 0.268\\
H$_2$O & 18.015 & 647.140 & 22.064 & 56.000 & 0.233 & 0.230 \\
CO$_2$ & 40.010 & 304.130 & 7.375 & 94.000 & 0.272 & 0.274 \\
\hline
\end{tabular}}
\end{table}

\begin{table}[htbp]
  \centering
  \caption{Estimated subcooled liquid molar volumes (in cm$^3$/mol) and associated mass densities for our best molar volume estimates 
(in g/cm$^3$, for Eq. \ref{eqn:rackett2} with $Z_{\rm RA}$) for Titan's solids and liquids, evaluated using the Rackett equations and parameters 
at 91.5 K.}
  \label{t:liquid_molar_volumes}
  \footnotesize{
    \begin{tabular}{lllllll}
    \hline
    Formula & Eq. \ref{eqn:rackett}& Eq. \ref{eqn:rackett} & Eq. \ref{eqn:rackett2} & Eq. \ref{eqn:rackett2}& $2\sigma$ & $\rho$ \\
  		   &  ($Z_{\rm RA}$) & ($Z_c$) &  ($Z_{\rm RA}$)  &  ($Z_c$) &  ($Z_{\rm RA}$) \\
    \hline
    \hline
CH$_4$ & 35.557 & 34.928 & 35.212 & 34.928 & 0.517 & 0.456 \\
C$_2$H$_6$ & 46.520 & 45.968 & 46.229 & 45.968 & 0.456 & 0.650 \\
C$_3$H$_8$ & 61.390 & 62.865 & 62.153 & 62.865 & 1.219 & 0.710 \\
C$_4$H$_{10}$ & 75.882 & 75.978 & 75.932 & 75.978 & 0.080 & 0.766 \\
C$_2$H$_2$ & 34.753 & 34.185 & 34.453 & 34.185 & 0.469 & 0.756 \\
C$_2$H$_4$ & 42.344 & 42.147 & 42.240 & 42.147 & 0.162 & 0.664 \\
C$_3$H$_4$ & 48.267 & 49.354 & 48.828 & 49.354 & 0.898 & 0.821\\
C$_4$H$_4$ & 67.280 & 66.398 & 66.824 & 66.398 & 0.729 & 0.779\\
C$_4$H$_6$ & 65.652 & 65.175 & 65.405 & 65.175 & 0.394 & 0.827\\
C$_6$H$_6$ & 74.256 & 73.269 & 73.748 & 73.269 & 0.817 & 1.059\\
HCN   & 30.232 & 30.323 & 30.279 & 30.323 & 0.075 & 0.893 \\
C$_2$N$_2$ & 41.034 & 43.436 & 42.263 & 43.436 & 1.987 & 1.255\\
CH$_3$CN & 40.219 & 34.105 & 36.956 & 34.105 & 5.041 & 1.111\\
C$_2$H$_3$CN & 71.655 & 50.733 & 60.016 & 50.733 & 17.199 & 0.884\\
C$_2$H$_5$CN & 81.023 & 51.731 & 64.373 & 51.731 & 24.046 & 0.856\\
C$_2$H$_8$N$_2$ & 60.828 & 63.522 & 62.197 & 63.522 & 2.230 & 0.966\\
C$_{11}$H$_{11}$N & 123.135 & 144.275 & 133.476 & 144.275 & 17.564 & 1.178\\
H$_2$O & 14.087 & 13.692 & 13.884 & 13.692 & 0.327 & 1.298\\
CO$_2$ & 28.789 & 29.225 & 29.017 & 29.225 & 0.360 & 1.379\\
\hline
\end{tabular}}
\end{table}


\begin{table}[htbp]
\centering
\caption{Subcooled liquid molar volumes (in cm$^3$/mol) and associated mass densities (in g/cm$^3$) of Titan's solids estimated 
using the Rackett equation \ref{eqn:rackett2} with $Z_{\rm RA}$ at 91.5 K
compared to the solid molar volumes (in cm$^3$/mol) and associated mass densities (in g/cm$^3$) computed from an empirical linear fit between predictions of the model and experimental data for some Titanian materials.}
\label{t:solid_molar_volumes}
\footnotesize{  
  \begin{tabular}{lllllll}
    \hline
    Name  & Formula & $V_{\rm m}^L$ (Eq. \ref{eqn:rackett2}, $Z_{\rm RA}$) & $V_{\rm m}^S$ (fit) & $2\sigma$ & $\rho^L$ & $\rho^S$ \\
    \hline
    \hline
n-butane & C$_4$H$_{10}$ & 75.932 & 68.737 & 7.195 & 0.766 & 0.846\\
acetylene & C$_2$H$_2$ & 34.453 & 32.471 & 1.982 & 0.756 & 0.802\\
benzene & C$_6$H$_6$ & 73.748 & 70.487 & 3.261 & 1.059 & 1.108\\
hydrogen cyanide & HCN   & 30.279 & 25.562 & 4.717 & 0.893 & 1.057 \\
acetonitrile & CH$_3$CN & 36.956 & 37.122 & 0.166 & 1.111 & 1.106\\
water ice Ih & H$_2$O & 13.884 & 19.329 & 5.446 & 1.298 & 0.932\\
carbon dioxide ice & CO$_2$ & 29.017 & 26.109 & 2.909 & 1.379 & 1.532\\
\hline
\end{tabular}}
\end{table}


\bibliographystyle{elsarticle-harv}

\begin{thebibliography}{153}
\providecommand{\natexlab}[1]{#1}
\expandafter\ifx\csname urlstyle\endcsname\relax
  \providecommand{\doi}[1]{doi:\discretionary{}{}{}#1}\else
  \providecommand{\doi}{doi:\discretionary{}{}{}\begingroup
  \urlstyle{rm}\Url}\fi

\bibitem[{\textit{Acocella}(2007)}]{Acocella2007}
Acocella, V. (2007), {Understanding caldera structure and development: An
  overview of analogue models compared to natural calderas},
  \textit{Earth-Science Reviews}, \textit{85}(3-4), 125 -- 160,
  \doi{10.1016/j.earscirev.2007.08.004}.

\bibitem[{\textit{Aharonson et~al.}(2009)\textit{Aharonson, Hayes, Lunine,
  Lorenz, Allison, and Elachi}}]{Aharonson2009}
Aharonson, O., A.~G. Hayes, J.~I. Lunine, R.~D. Lorenz, M.~D. Allison, and
  C.~Elachi (2009), {An asymmetric distribution of lakes on Titan as a possible
  consequence of orbital forcing}, \textit{Nature Geoscience}, \textit{2},
  851--854, \doi{10.1038/ngeo698}.

\bibitem[{\textit{Antson et~al.}(1987)\textit{Antson, Tilli, and
  Andersen}}]{Antson1987}
Antson, O., K.~Tilli, and N.~Andersen (1987), {Neutron Powder Diffraction Study
  of Deuterated {$\beta$}-Acetonitrile}, \textit{Acta Crystallographica Section
  B}, \textit{43}, 296--301, \doi{10.1107/S0108768187097866}.

\bibitem[{\textit{Atreya}(2007)}]{Atreya2007}
Atreya, S.~K. (2007), {Titan's organic factory}, \textit{Science},
  \textit{316}(5826), 843--845, \doi{10.1126/science.1141869}.

\bibitem[{\textit{Barnes et~al.}(2007)\textit{Barnes, Brown, Soderblom,
  Buratti, Sotin, Rodriguez, {Le Mou\'elic}, Baines, Clark, and
  Nicholson}}]{Barnes2007}
Barnes, J.~W., R.~H. Brown, L.~Soderblom, B.~J. Buratti, C.~Sotin,
  S.~Rodriguez, S.~{Le Mou\'elic}, K.~H. Baines, R.~Clark, and P.~Nicholson
  (2007), {Global-scale surface variations on Titan seen from Cassini/VIMS},
  \textit{Icarus}, \textit{186}(1), 242 -- 258,
  \doi{10.1016/icarus.2006.08.021}.

\bibitem[{\textit{Barnes et~al.}(2009)\textit{Barnes, Brown, Soderblom,
  Soderblom, Jaumann, Jackson, {Le Mou\'elic}, Sotin, Buratti, Pitman, Baines,
  Clark, Nicholson, Turtle, and Perry}}]{Barnes2009ontario}
Barnes, J.~W., R.~H. Brown, J.~M. Soderblom, L.~A. Soderblom, R.~Jaumann,
  B.~Jackson, S.~{Le Mou\'elic}, C.~Sotin, B.~J. Buratti, K.~M. Pitman, K.~H.
  Baines, R.~N. Clark, P.~D. Nicholson, E.~P. Turtle, and J.~Perry (2009),
  {Shoreline features of Titan's Ontario Lacus from Cassini/VIMS observations},
  \textit{Icarus}, \textit{201}(1), 217 -- 225,
  \doi{10.1016/j.icarus.2008.12.028}.

\bibitem[{\textit{Barnes et~al.}(2011)\textit{Barnes, Bow, Schwartz, Brown,
  Soderblom, Hayes, Vixie, {Le Mou\'elic}, Rodriguez, Sotin, Jaumann, Stephan,
  Soderblom, Clark, Buratti, Baines, and Nicholson}}]{Barnes2011evaporites}
Barnes, J.~W., J.~Bow, J.~Schwartz, R.~H. Brown, J.~M. Soderblom, A.~G. Hayes,
  G.~Vixie, S.~{Le Mou\'elic}, S.~Rodriguez, C.~Sotin, R.~Jaumann, K.~Stephan,
  L.~A. Soderblom, R.~N. Clark, B.~J. Buratti, K.~H. Baines, and P.~D.
  Nicholson (2011), {Organic sedimentary deposits in Titan's dry lakebeds:
  Probable evaporite}, \textit{Icarus}, \textit{216}(1), 136 -- 140,
  \doi{10.1016/j.icarus.2011.08.022}.

\bibitem[{\textit{Barrow}(1981)}]{Barrow1981}
Barrow, M. (1981), {$\alpha$-acetonitrile at 215 K}, \textit{Acta
  Crystallographica Section B}, \textit{37}, 2239--2242,
  \doi{10.1107/S0567740881008510}.

\bibitem[{\textit{Black et~al.}(2012)\textit{Black, Perron, Burr, and
  Drummond}}]{Black2012}
Black, B.~A., J.~T. Perron, D.~M. Burr, and S.~A. Drummond (2012), {Estimating
  erosional exhumation on Titan from drainage network morphology},
  \textit{Journal of Geophysical Research}, \textit{117}, E08,006,
  \doi{10.1029/2012JE004085}.

\bibitem[{\textit{Bourgeois et~al.}(2008)\textit{Bourgeois, Lopez, {Le
  Mou\'elic}, Fleurant, Tobie, Le~Corre, Le~Deit, Sotin, and
  Bodeur}}]{Bourgeois2008}
Bourgeois, O., T.~Lopez, S.~{Le Mou\'elic}, C.~Fleurant, G.~Tobie, L.~Le~Corre,
  L.~Le~Deit, C.~Sotin, and Y.~Bodeur (2008), {A surface
  dissolution/precipitation model for the development of lakes on Titan, based
  on an arid terrestrial analogue: The pans and calcretes of Etosha}, in
  \textit{Lunar and Planetary Science XXXIX}, p. 1733.

\bibitem[{\textit{Bowen and Johnson}(2012)}]{Bowen2012}
Bowen, M.~W., and W.~C. Johnson (2012), {Late quaternary environmental
  reconstructions of playa-lunette system evolution on the central High Plains
  of Kansas, United States}, \textit{Geological Society of America Bulletin},
  \textit{124}(1), 146--161, \doi{10.1130/B30382.1}.

\bibitem[{\textit{Brew}(1977)}]{Brew1977}
Brew, T. C.~L. (1977), {A study on the solubility of heavy hydrocarbons in
  liquid methane and methane containing mixtures}, Master's thesis, University
  of Ottawa, Canada.

\bibitem[{\textit{Brezonik and Arnold}(2011)}]{Brezonik2011}
Brezonik, P., and W.~Arnold (2011), \textit{{Water chemistry: An introduction
  to the chemistry of natural and engineered aquatic systems}}, 809 pp., Oxford
  University Press, Inc., New York.

\bibitem[{\textit{Brown et~al.}(2008)\textit{Brown, Soderblom, Soderblom,
  Clark, Jaumann, Barnes, Sotin, Buratti, Baines, and Nicholson}}]{Brown2008}
Brown, R.~H., L.~A. Soderblom, J.~M. Soderblom, R.~N. Clark, R.~Jaumann, J.~W.
  Barnes, C.~Sotin, B.~Buratti, K.~H. Baines, and P.~D. Nicholson (2008), {The
  identification of liquid ethane in Titan's Ontario Lacus}, \textit{Nature},
  \textit{454}, 607 -- 610, \doi{10.1038/nature07100}.

\bibitem[{\textit{Buch}(1997)}]{Buch1997a}
Buch, M.~W. (1997), {Etosha Pan - The third largest lake in the world ?},
  \textit{Madoqua}, \textit{20}(1), 49 -- 64.

\bibitem[{\textit{Buch and Rose}(1996)}]{Buch1996}
Buch, M.~W., and D.~Rose (1996), {Mineralogy and geochemistry of the sediments
  of the Etosha Pan Region in northern Namibia: A reconstruction of the
  depositional environment}, \textit{Journal of African Earth Sciences},
  \textit{22}(3), 355 -- 378, \doi{10.1016/0899-5362(96)00020-6}.

\bibitem[{\textit{Buch and Trippner}(1997)}]{Buch1997b}
Buch, M.~W., and C.~Trippner (1997), {Overview of the geological and
  geomorphological evolution of the Etosha region, Northern Namibia},
  \textit{Madoqua}, \textit{20}(1), 65 -- 74.

\bibitem[{\textit{Cabane et~al.}(1992)\textit{Cabane, Chassefi\`ere, and
  Israel}}]{Cabane1992}
Cabane, M., E.~Chassefi\`ere, and G.~Israel (1992), {Formation and growth of
  photochemical aerosols in Titan's atmosphere}, \textit{Icarus},
  \textit{96}(2), 176 -- 189, \doi{10.1016/0019-1035(92)90071-E}.

\bibitem[{\textit{Cheung and Zander}(1968)}]{Cheung1968}
Cheung, H., and E.~H. Zander (1968), {Solubility of carbon dioxide and hydrogen
  sulfide in liquid hydrocarbons at cryogenic temperatures}, \textit{Chemical
  Engineering Progress}, \textit{64}(88), 34.

\bibitem[{\textit{Chevrier et~al.}(2014)\textit{Chevrier, Singh, Nna-Mvondo,
  M\`ege, Leitner, and Wagner}}]{Chevrier2014}
Chevrier, V., S.~Singh, D.~Nna-Mvondo, D.~M\`ege, M.~Leitner, and A.~Wagner
  (2014), {Solubility and detection of simple and complex organics in Titan's
  liquid hydrocarbons}, in \textit{Titan Through Time - 3rd Workshop}, pp.
  14--15.

\bibitem[{\textit{Chirico et~al.}(2007)\textit{Chirico, III, and
  Steele}}]{Chirico2007}
Chirico, R.~D., R.~D.~J. III, and W.~V. Steele (2007), {Thermodynamic
  properties of methylquinolines: Experimental results for
  2,6-dimethylquinoline and mutual validation between experiments and
  computational methods for methylquinolines}, \textit{The Journal of Chemical
  Thermodynamics}, \textit{39}(5), 698 -- 711, \doi{10.1016/j.jct.2006.10.012}.

\bibitem[{\textit{Choukroun and Sotin}(2012)}]{Choukroun2012}
Choukroun, M., and C.~Sotin (2012), {Is Titan's shape caused by its meteorology
  and carbon cycle ?}, \textit{Geophysical Research Letters}, \textit{39}(4),
  L04,201, \doi{10.1029/2011GL050747}.

\bibitem[{\textit{Clark and Din}(1953)}]{Clark1953}
Clark, A.~M., and F.~Din (1953), {Equilibria between solid, liquid and gaseous
  phases at low temperatures. The system carbon dioxide + ethane + ethylene},
  \textit{Discussions of the Faraday Society}, \textit{15}, 202--207.

\bibitem[{\textit{Clark et~al.}(2010)\textit{Clark, Curchin, Barnes, Jaumann,
  Soderblom, Cruikshank, Lunine, Stephan, Hoefen, {Le Mou\'elic}, Sotin,
  Baines, Buratti, and Nicholson}}]{Clark2010}
Clark, R.~N., J.~M. Curchin, J.~W. Barnes, R.~Jaumann, L.~Soderblom, D.~P.
  Cruikshank, J.~Lunine, K.~Stephan, T.~M. Hoefen, S.~{Le Mou\'elic}, C.~Sotin,
  K.~H. Baines, B.~Buratti, and P.~Nicholson (2010), {Detection and mapping of
  hydrocarbon deposits on Titan}, \textit{Journal of Geophysical Research},
  \textit{115}, E10,005, \doi{10.1029/2009WR008896}.

\bibitem[{\textit{Coll et~al.}(1995)\textit{Coll, Cosia, Gazeau, {de Vanssay},
  Guillemin, and Raulin}}]{Coll1995}
Coll, P., D.~Cosia, M.-C. Gazeau, E.~{de Vanssay}, J.-C. Guillemin, and
  F.~Raulin (1995), {Organic chemistry in Titan's atmosphere: New data from
  laboratory simulations at low temperature}, \textit{Advances in Space
  Research}, \textit{16}(2), 93--103, \doi{10.1016/0273-1177(95)00197-M}.

\bibitem[{\textit{Cordier et~al.}(2009)\textit{Cordier, Mousis, Lunine, Lavvas,
  and Vuitton}}]{Cordier2009}
Cordier, D., O.~Mousis, J.~I. Lunine, P.~Lavvas, and V.~Vuitton (2009), {An
  estimate of the chemical composition of Titan's lakes}, \textit{The
  Astrophysical Journal}, \textit{707}, L128 -- L131,
  \doi{10.1088/0004-637X/707/2/L128}.

\bibitem[{\textit{Cordier et~al.}(2013{\natexlab{a}})\textit{Cordier, Mousis,
  Lunine, Lavvas, and Vuitton}}]{Cordier2013erratum}
Cordier, D., O.~Mousis, J.~I. Lunine, P.~Lavvas, and V.~Vuitton
  (2013{\natexlab{a}}), {Erratum: An estimate of the chemical composition of
  Titan's lakes}, \textit{The Astrophysical Journal Letters}, \textit{768},
  L23, \doi{10.1088/2041-8205/768/1/L23}.

\bibitem[{\textit{Cordier et~al.}(2013{\natexlab{b}})\textit{Cordier, Barnes,
  and Ferreira}}]{Cordier2013}
Cordier, D., J.~Barnes, and A.~Ferreira (2013{\natexlab{b}}), {On the chemical
  composition of Titan's dry lakebed evaporites}, \textit{Icarus},
  \textit{226}(2), 1431 -- 1437, \doi{10.1016/j.icarus.2013.07.026}.

\bibitem[{\textit{Cornet et~al.}(2012)\textit{Cornet, Bourgeois, {Le
  Mou\'elic}, Rodriguez, {Lopez Gonzalez}, Sotin, Tobie, Fleurant, Barnes,
  Brown, Baines, Buratti, Clark, and Nicholson}}]{Cornet2012}
Cornet, T., O.~Bourgeois, S.~{Le Mou\'elic}, S.~Rodriguez, T.~{Lopez Gonzalez},
  C.~Sotin, G.~Tobie, C.~Fleurant, J.~W. Barnes, R.~H. Brown, K.~H. Baines,
  B.~J. Buratti, R.~N. Clark, and P.~D. Nicholson (2012), {Geomorphological
  significance of Ontario Lacus on Titan: Integrated interpretation of Cassini
  VIMS, ISS and RADAR data and comparison with the Etosha Pan (Namibia)},
  \textit{Icarus}, \textit{218}(2), 788 -- 806,
  \doi{10.1016/j.icarus.2012.01.013}.

\bibitem[{\textit{Cottini et~al.}(2012)\textit{Cottini, Nixon, Jennings,
  Anderson, Gorius, Bjoraker, Coustenis, Teanby, Achterberg, B\'ezard, de~Kok,
  Lellouch, Irwin, Flasar, and Bampasidis}}]{Cottini2012}
Cottini, V., C.~A. Nixon, D.~E. Jennings, C.~M. Anderson, N.~Gorius, G.~L.
  Bjoraker, A.~Coustenis, N.~A. Teanby, R.~K. Achterberg, B.~B\'ezard,
  R.~de~Kok, E.~Lellouch, P.~G.~J. Irwin, F.~M. Flasar, and G.~Bampasidis
  (2012), {Water vapor in Titan's stratosphere from Cassini CIRS far-infrared
  spectra}, \textit{Icarus}, \textit{220}(2), 855 -- 862,
  \doi{10.1016/j.icarus.2012.06.014}.

\bibitem[{\textit{Coustenis et~al.}(2010)\textit{Coustenis, Jennings, Nixon,
  Achterberg, Lavvas, Vinatier, Teanby, Bjoraker, Carlson, Piani, Bampasidis,
  Flasar, and Romani}}]{Coustenis2010}
Coustenis, A., D.~E. Jennings, C.~A. Nixon, R.~K. Achterberg, P.~Lavvas,
  S.~Vinatier, N.~A. Teanby, G.~L. Bjoraker, R.~C. Carlson, L.~Piani,
  G.~Bampasidis, F.~M. Flasar, and P.~N. Romani (2010), {Titan trace gaseous
  composition from CIRS at the end of the Cassini-Huygens prime mission},
  \textit{Icarus}, \textit{207}(1), 461 -- 476,
  \doi{10.1016/j.icarus.2009.11.027}.

\bibitem[{\textit{Craven et~al.}(1993)\textit{Craven, Hatton, Howard, and
  Pawley}}]{Craven1993}
Craven, C., P.~Hatton, C.~Howard, and G.~Pawley (1993), {The structure and
  dynamics of solid benzene. I. A neutron powder diffraction study of
  deuterated benzene from 4 K to the melting point}, \textit{Journal of
  Chemical Physics}, \textit{98}, 8236.

\bibitem[{\textit{Cui et~al.}(2009)\textit{Cui, Yelle, Vuitton, {Waite Jr.},
  Kasprzak, Gell, Niemann, M{\"u}ller-Wodarg, Borggren, Fletcher, Patrick,
  Raaen, and Magee}}]{Cui2009}
Cui, J., R.~V. Yelle, V.~Vuitton, J.~H. {Waite Jr.}, W.~T. Kasprzak, D.~A.
  Gell, H.~B. Niemann, I.~C.~F. M{\"u}ller-Wodarg, N.~Borggren, G.~Fletcher,
  E.~Patrick, E.~Raaen, and B.~Magee (2009), {Analysis of Titan's neutral upper
  atmosphere from Cassini Ion Neutral Mass Spectrometer measurements},
  \textit{Icarus}, \textit{200}(2), 581 -- 615,
  \doi{10.1016/j.icarus.2008.12.005}.

\bibitem[{\textit{Davis et~al.}(1962)\textit{Davis, Rodewald, and
  Kurata}}]{Davis1962}
Davis, J.~A., N.~Rodewald, and F.~Kurata (1962), {Solid-liquid-vapor phase
  behavior of the methane-carbon dioxide system}, \textit{AIChE Journal},
  \textit{8}(4), 537--539, \doi{10.1002/aic.690080423}.

\bibitem[{\textit{Dietrich et~al.}(1975)\textit{Dietrich, Mackenzie, and
  Pawley}}]{Dietrich1975}
Dietrich, O., G.~Mackenzie, and G.~Pawley (1975), {The structural phase
  transition in solid DCN}, \textit{Journal of Physics C: Solid State Physics},
  \textit{8}, L98, \doi{10.1088/0022-3719/8/7/002}.

\bibitem[{\textit{Dubouloz et~al.}(1989)\textit{Dubouloz, Raulin, Lellouch, and
  Gautier}}]{Dubouloz1989}
Dubouloz, N., F.~Raulin, E.~Lellouch, and D.~Gautier (1989), {Titan's
  hypothesized ocean properties: The influence of surface temperature and
  atmospheric composition uncertainties}, \textit{Icarus}, \textit{82}(1), 81
  -- 96, \doi{10.1016/0019-1035(89)90025-0}.

\bibitem[{\textit{Dulmage and Lipscomb}(1951)}]{Dulmage1951}
Dulmage, W., and W.~Lipscomb (1951), {The crystal structures of hydrogen
  cyanide, HCN}, \textit{Acta Crystallographica}, \textit{4}, 330--334,
  \doi{10.1107/S0365110X51001070}.

\bibitem[{\textit{Etters and Kuchta}(1989)}]{Etters1989}
Etters, R., and B.~Kuchta (1989), {Static and dynamic properties of solid CO2
  at various temperatures and pressures}, \textit{Journal of Chemical Physics},
  \textit{90}, 4537, \doi{10.1063/1.456640}.

\bibitem[{\textit{Fleurant et~al.}(2008)\textit{Fleurant, Tucker, and
  Viles}}]{Fleurant2008}
Fleurant, C., G.~Tucker, and H.~Viles (2008), \textit{{Landscape Evolution:
  Denudation, Climate and Tectonics over Different Time and Space Scales}},
  vol. 296, chap. {A cockpit karst evolution model}, pp. 47 -- 62, Geological
  Society of London.

\bibitem[{\textit{Ford and Williams}(2007)}]{Ford2007}
Ford, D., and P.~Williams (2007), \textit{{Karst hydrogeology and
  geomorphology}}, 562 pp pp., Jon Wiley \& Sons Ltd, The Atrium, Southern
  Gate, Chichester, West Sussex, PO19 8SQ, England.

\bibitem[{\textit{French}(2007)}]{French2007}
French, H.~M. (2007), \textit{{The periglacial environment, Third Edition}},
  458 pp pp., Jon Wiley \& Sons Ltd, The Atrium, Southern Gate, Chichester,
  West Sussex PO19 8SQ, England.

\bibitem[{\textit{Frumkin}(1994)}]{Frumkin1994}
Frumkin, A. (1994), Hydrology and denudation rates of halite karst,
  \textit{Journal of Hydrology}, \textit{162}, 171--189,
  \doi{10.1016/0022-1694(94)90010-8}.

\bibitem[{\textit{Frumkin}(1996)}]{Frumkin1996}
Frumkin, A. (1996), \textit{{Salt tectonics}}, vol. 100, chap. {Uplift rate
  relative to base-levels of a salt diapir (Dead Sea Basin, Israel) as
  indicated by cave levels}, pp. 41--47, Geological Society Special
  Publication, London, \doi{10.1144/GSL.SP.1996.100.01.04}.

\bibitem[{\textit{Garasic}(2001)}]{Garasic2001}
Garasic, M. (2001), {New Speleohydrogeological Research of Crveno jezero (Red
  Lake) near Imotski in Dinaric Karst Area (Croatia, Europe) - International
  speleodiving expedition ``Crveno jezero 98''}, in \textit{13th International
  Congress of Speleology, 4th Speleological Congress of Latin Am\'erica and
  Caribbean, 26th Brazilian Congress of Speleology}, pp. 555 -- 559.

\bibitem[{\textit{{Garasic}}(2012)}]{Garasic2012}
{Garasic}, M. (2012), {Crveno Jezero - the biggest sinkhole in Dinaric Karst
  (Croatia)}, in \textit{EGU General Assembly Conference Abstracts},
  \textit{EGU General Assembly Conference Abstracts}, vol.~14, edited by
  A.~{Abbasi} and N.~{Giesen}, p. 7132.

\bibitem[{\textit{Glein and Shock}(2013)}]{Glein2013}
Glein, C.~R., and E.~L. Shock (2013), {A geochemical model of non-ideal
  solutions in the methane-ethane-propane-nitrogen-acetylene system on Titan},
  \textit{Geochimica et Cosmochimica Acta}, \textit{115}(0), 217 -- 240,
  \doi{10.1016/j.gca.2013.03.030}.

\bibitem[{\textit{Goudie and Wells}(1995)}]{Goudie1995}
Goudie, A.~S., and G.~L. Wells (1995), {The nature, distribution and formation
  of pans in arid zones}, \textit{Earth-Science Reviews}, \textit{38}, 1 -- 69,
  \doi{10.1016/0012-8252(94)00066-6}.

\bibitem[{\textit{Graves et~al.}(2008)\textit{Graves, McKay, Griffith, Ferri,
  and Fulchigoni}}]{Graves2008}
Graves, S. D.~B., C.~P. McKay, C.~A. Griffith, F.~Ferri, and M.~Fulchigoni
  (2008), {Rain and hail can reach the surface of Titan}, \textit{Planetary and
  Space Science}, \textit{56}, 346--357, \doi{10.1016/j.pss.2007.11.001}.

\bibitem[{\textit{Harrisson}(2012)}]{Harrisson2012}
Harrisson, K.~P. (2012), {Thermokarst processes in Titan's lakes: Comparison
  with terrestrial data}, in \textit{43rd Lunar and Planetary Science
  Conference}, pp. 2271--2272.

\bibitem[{\textit{{Hayes} et~al.}(2008)\textit{{Hayes}, {Aharonson},
  {Callahan}, {Elachi}, {Gim}, {Kirk}, {Lewis}, {Lopes}, {Lorenz}, {Lunine},
  {Mitchell}, {Mitri}, {Stofan}, and {Wall}}}]{Hayes2008}
{Hayes}, A., O.~{Aharonson}, P.~{Callahan}, C.~{Elachi}, Y.~{Gim}, R.~{Kirk},
  K.~{Lewis}, R.~{Lopes}, R.~{Lorenz}, J.~{Lunine}, K.~{Mitchell}, G.~{Mitri},
  E.~{Stofan}, and S.~{Wall} (2008), {Hydrocarbon lakes on Titan: Distribution
  and interaction with an isotropic porous regolith}, \textit{Geophysical
  Research Letters}, \textit{35}, L09,204, \doi{10.1029/2008GL033409}.

\bibitem[{\textit{Heintz and Bich}(2009)}]{Heintz2009}
Heintz, A., and E.~Bich (2009), {Thermodynamics in an icy world: The atmosphere
  and internal structure of Saturn's moon Titan}, \textit{Pure and Applied
  Chemistry}, \textit{81}(10), 1903--1920, \doi{10.1351/PAC-CON-08-10-04}.

\bibitem[{\textit{Hipondoka}(2005)}]{Hipondoka2005}
Hipondoka, M. H.~T. (2005), {The development and evolution of Etosha Pan,
  Namibia}, Ph.D. thesis, University of Wurzburg (Germany), 154pp.

\bibitem[{\textit{Jennings et~al.}(2009)\textit{Jennings, Flasar, Kunde,
  Samuelson, Pearl, Nixon, Carlson, Mamoutkine, Brasunas, Guandique,
  Achterberg, Bjoraker, Romani, Segura, Albright, Elliott, Tingley, Calcutt,
  Coustenis, and Courtin}}]{Jennings2009}
Jennings, D.~E., F.~M. Flasar, V.~G. Kunde, R.~E. Samuelson, J.~C. Pearl, C.~A.
  Nixon, R.~C. Carlson, A.~A. Mamoutkine, J.~C. Brasunas, E.~Guandique, R.~K.
  Achterberg, G.~L. Bjoraker, P.~N. Romani, M.~E. Segura, S.~A. Albright, M.~H.
  Elliott, J.~S. Tingley, S.~Calcutt, A.~Coustenis, and R.~Courtin (2009),
  {Titan's surface brightness temperatures}, \textit{The Astrophysical Journal
  Letters}, \textit{691}(2), L103--L105, \doi{10.1088/0004-637X/691/2/L103}.

\bibitem[{\textit{Kargel et~al.}(2007)\textit{Kargel, Furfaro, Hays, Lopes,
  Lunine, Mitchell, Wall, and {the Cassini RADAR Team}}}]{Kargel2007}
Kargel, J.~S., R.~Furfaro, C.~C. Hays, R.~M.~C. Lopes, J.~I. Lunine, K.~L.
  Mitchell, S.~D. Wall, and {the Cassini RADAR Team} (2007), {Titan's
  GOO-sphere: glacial, permafrost, evaporite, and other familiar processes
  involving exotic materials}, in \textit{38th Lunar and Planetary Science
  Conference}, pp. 1992--1993.

\bibitem[{\textit{Kirk and {Howington-Kraus}}(2008)}]{Kirk2008}
Kirk, R.~L., and E.~{Howington-Kraus} (2008), {Radargrammetry on three
  planets}, \textit{{The International Archives of the Photogrammetry, Remote
  Sensing and Spatial Information Sciences}}, \textit{37}(B4), 973 -- 980.

\bibitem[{\textit{{Kirk} et~al.}(2007)\textit{{Kirk}, {Howington-Kraus},
  Mitchell, Hensley, Stiles, and {the Cassini RADAR Team}}}]{Kirk2007}
{Kirk}, R.~L., E.~{Howington-Kraus}, K.~L. Mitchell, S.~Hensley, B.~W. Stiles,
  and {the Cassini RADAR Team} (2007), {First stereoscopic Radar images of
  Titan}, in \textit{38th Lunar and Planetary Science Conference}, pp.
  1427--1428.

\bibitem[{\textit{Kirk et~al.}(2012)\textit{Kirk, {Howington-Kraus}, Redding,
  Callahan, Hayes, {Le Gall}, Lopes, Lorenz, Lucas, Mitchell, Neish, Aharonson,
  Radebaugh, Stiles, Stofan, Wall, and Wood}}]{Kirk2012}
Kirk, R.~L., E.~{Howington-Kraus}, B.~Redding, P.~S. Callahan, A.~G. Hayes,
  A.~{Le Gall}, R.~M.~C. Lopes, R.~D. Lorenz, A.~Lucas, K.~L. Mitchell, C.~D.
  Neish, O.~Aharonson, J.~Radebaugh, B.~W. Stiles, E.~R. Stofan, S.~D. Wall,
  and C.~A. Wood (2012), {Topographic mapping of Titan: Latest results}, in
  \textit{43rd Lunar and Planetary Science Conference}, pp. 2759--2760.

\bibitem[{\textit{Klime\v{s} et~al.}(2011)\textit{Klime\v{s}, Bowler, and
  Michaelides}}]{Klimes2011}
Klime\v{s}, J., D.~R. Bowler, and A.~Michaelides (2011), {Van Der Waals density
  functionals applied to solids}, \textit{Physical Review B}, \textit{83},
  195,131, \doi{10.1103/PhysRevB.83.195131}.

\bibitem[{\textit{Krasnopolsky}(2009)}]{Krasnopolsky2009}
Krasnopolsky, V.~A. (2009), {A photochemical model of Titan's atmosphere and
  ionosphere}, \textit{Icarus}, \textit{201}(1), 226 -- 256,
  \doi{10.1016/j.icarus.2008.12.038}.

\bibitem[{\textit{Kuebler and McKinley}(1975)}]{Kuebler1975}
Kuebler, G.~P., and G.~McKinley (1975), \textit{Advances in Cryogenic
  Engineering}, vol.~21, chap. {Solubility of solid n-butane and n-pentane in
  liquid methane}, pp. 509--515, Springer US,
  \doi{10.1007/978-1-4757-0208-8\_60}.

\bibitem[{\textit{Kuebler and McKinley}(1995)}]{Kuebler1995}
Kuebler, G.~P., and G.~McKinley (1995), \textit{Advances in Cryogenic
  Engineering}, vol.~19, chap. {Solubility of solid benzene, toluene, n-hexane,
  n-heptane, in liquid methane}, pp. 320--326, Springer US,
  \doi{10.1007/978-1-4613-9847-9\_39}.

\bibitem[{\textit{Langmuir}(1997)}]{Langmuir1997}
Langmuir, D. (1997), \textit{Aqueous Environmental Geochemistry}, 618 pp. pp.,
  Prentice Hall, Inc., Upper Saddle River, New Jersey.

\bibitem[{\textit{Lara et~al.}(1996)\textit{Lara, Lellouch, {L\'opez-Moreno},
  and Rodrigo}}]{Lara1996}
Lara, L.~M., E.~Lellouch, J.~J. {L\'opez-Moreno}, and R.~Rodrigo (1996),
  {Vertical distribution of Titan's atmospheric neutral constituents},
  \textit{Journal of Geophysical Research}, \textit{101}(E10), 23,161--23,283,
  \doi{10.1029/96JE02036}.

\bibitem[{\textit{Lavvas et~al.}(2008{\natexlab{a}})\textit{Lavvas, Coustenis,
  and Vardavas}}]{Lavvas2008a}
Lavvas, P.~P., A.~Coustenis, and I.~M. Vardavas (2008{\natexlab{a}}), {Coupling
  photochemistry with haze formation in Titan's atmosphere, Part I: Model
  description}, \textit{Planetary and Space Science}, \textit{56}(1), 27 -- 66,
  \doi{10.1016/j.pss.2007.05.026}.

\bibitem[{\textit{Lavvas et~al.}(2008{\natexlab{b}})\textit{Lavvas, Coustenis,
  and Vardavas}}]{Lavvas2008b}
Lavvas, P.~P., A.~Coustenis, and I.~M. Vardavas (2008{\natexlab{b}}), {Coupling
  photochemistry with haze formation in Titan's atmosphere, Part II: Results
  and validation with Cassini/Huygens data}, \textit{Planetary and Space
  Science}, \textit{56}(1), 67 -- 99, \doi{10.1016/j.pss.2007.05.027}.

\bibitem[{\textit{Lee et~al.}(2010)\textit{Lee, Murray, Kong, Lundqvist, and
  Langreth}}]{Lee2010}
Lee, K., E.~D. Murray, L.~Kong, B.~I. Lundqvist, and D.~C. Langreth (2010),
  {Higher-accuracy Van der Waals density functional}, \textit{Physical Review
  B}, \textit{82}, 081,101, \doi{10.1103/PhysRevB.82.081101}.

\bibitem[{\textit{Leitner et~al.}(2014)\textit{Leitner, Singh, and
  Chevrier}}]{Leitner2014}
Leitner, M., S.~Singh, and V.~F. Chevrier (2014), {Solubility and detectability
  of acetonitrile in Titan lakes}, in \textit{45th Lunar and Planetary Science
  Conference}, p. 2658.

\bibitem[{\textit{Lide}(2010)}]{CRC}
Lide, D. (2010), \textit{{CRC Handbook of CHemistry and Physics, 90th edition
  (CD-ROM version 2010)}}, CRC Press/Taylor and Francis, Boca Raton, FL.

\bibitem[{\textit{Loerting et~al.}(2011)\textit{Loerting, Bauer, Kohl,
  Watschinger, Winkel, and Mayer}}]{Loerting2011}
Loerting, T., M.~Bauer, I.~Kohl, K.~Watschinger, K.~Winkel, and E.~Mayer
  (2011), {Cryoflotation: densities of amorphous and cristalline ices},
  \textit{The Journal of Physical Chemistry B}, \textit{115}, 14,167 -- 14,175,
  \doi{10.1021/jp204752w}.

\bibitem[{\textit{Lopes et~al.}(2007)\textit{Lopes, Mitchell, Wall, Mitri,
  Janssen, Ostro, Kirk, Hayes, Stofan, Lunine, Lorenz, Wood, Radebaugh,
  Paillou, Zebker, and Paganelli}}]{Lopes2007}
Lopes, R.~M.~C., K.~L. Mitchell, S.~D. Wall, G.~Mitri, M.~Janssen, S.~Ostro,
  R.~L. Kirk, A.~G. Hayes, E.~R. Stofan, J.~I. Lunine, R.~D. Lorenz, C.~Wood,
  J.~Radebaugh, P.~Paillou, H.~Zebker, and F.~Paganelli (2007), {The lakes and
  seas of Titan}, \textit{EOS, Transactions American Geophysical Union},
  \textit{88}(51), 569 -- 576, \doi{10.1029/2007EO510001}.

\bibitem[{\textit{Lora et~al.}(2011)\textit{Lora, Goodman, Russell, and
  Lunine}}]{Lora2011}
Lora, J.~M., P.~J. Goodman, J.~L. Russell, and J.~I. Lunine (2011), {Insolation
  in Titan's troposphere}, \textit{Icarus}, \textit{216}(1), 116 -- 119,
  \doi{10.1016/j.icarus.2011.08.017}.

\bibitem[{\textit{Lorenz and Lunine}(2002)}]{Lorenz2002}
Lorenz, R.~D., and J.~I. Lunine (2002), {Titan's snowline}, \textit{Icarus},
  \textit{158}(2), 557 -- 559, \doi{10.1006/icar.2002.6880}.

\bibitem[{\textit{Lorenz et~al.}(2008)\textit{Lorenz, Mitchell, Kirk, Hayes,
  Aharonson, Zebker, Paillou, Radebaugh, Lunine, Janssen, Wall, Lopes, Stiles,
  Ostro, Mitri, and Stofan}}]{Lorenz2008organic}
Lorenz, R.~D., K.~L. Mitchell, R.~L. Kirk, A.~G. Hayes, O.~Aharonson, H.~A.
  Zebker, P.~Paillou, J.~Radebaugh, J.~I. Lunine, M.~A. Janssen, S.~D. Wall,
  R.~M. Lopes, B.~Stiles, S.~Ostro, G.~Mitri, and E.~R. Stofan (2008), {Titan's
  inventory of organic surface materials}, \textit{Geophysical Research
  Letters}, \textit{35}, L02,206, \doi{10.1029/2007GL03118}.

\bibitem[{\textit{Lorenz et~al.}(2013)\textit{Lorenz, Stiles, Aharonson, Lucas,
  Hayes, Kirk, Zebker, Turtle, Neish, Stofan, Barnes, and {the Cassini RADAR
  Team}}}]{Lorenz2013}
Lorenz, R.~D., B.~W. Stiles, O.~Aharonson, A.~Lucas, A.~G. Hayes, R.~L. Kirk,
  H.~A. Zebker, E.~P. Turtle, C.~D. Neish, E.~R. Stofan, J.~W. Barnes, and {the
  Cassini RADAR Team} (2013), {A global topographic map of Titan},
  \textit{Icarus}, \textit{225}(1), 367 -- 377,
  \doi{10.1016/j.icarus.2013.04.002}.

\bibitem[{\textit{Lorenz et~al.}(2014)\textit{Lorenz, Kirk, Hayes, Anderson,
  Lunine, Tokano, Turtle, Malaska, Soderblom, Lucas, Karatekin, and
  Wall}}]{Lorenz2014}
Lorenz, R.~D., R.~L. Kirk, A.~G. Hayes, Y.~Z. Anderson, J.~I. Lunine,
  T.~Tokano, E.~P. Turtle, M.~J. Malaska, J.~M. Soderblom, A.~Lucas,
  O.~Karatekin, and S.~Wall (2014), {A Radar map of Titan seas: Tidal
  dissipation and ocean mixing through the throat of Kraken}, \textit{Icarus},
  \textit{237}, 9--15, \doi{10.1016/j.icarus.2014.04.005}.

\bibitem[{\textit{Lorenz}(1986)}]{Lorenz1986}
Lorenz, V. (1986), {On the growth of maars and diatremes and its relevance to
  the formation of tuff rings}, \textit{Bulletin of Volcanology}, \textit{48},
  265--274, \doi{10.1007/BF01081755}.

\bibitem[{\textit{Luks et~al.}(1981)\textit{Luks, Hottovy, and
  Kohn}}]{Luks1981}
Luks, K.~D., J.~D. Hottovy, and J.~P. Kohn (1981), {Three-phase
  solid-liquid-vapor equilibriums in the binary hydrocarbon systems
  methane-n-hexane and methane-benzene}, \textit{Journal of Chemical
  Engineering Data}, \textit{26}(4), 402--403, \doi{10.1021/je00026a016}.

\bibitem[{\textit{Lunine et~al.}(1983)\textit{Lunine, Stevenson, and
  Yung}}]{Lunine1983}
Lunine, J.~I., D.~J. Stevenson, and Y.~L. Yung (1983), {Ethane ocean on Titan},
  \textit{Science}, \textit{222}(4629), 1229 -- 1230,
  \doi{10.1126/science.222.4629.1229}.

\bibitem[{\textit{Lunine et~al.}(2008)\textit{Lunine, Elachi, Wall, Janssen,
  Allison, Anderson, Boehmer, Callahan, Encrenaz, Flamini, Franceschetti, Gim,
  Hamilton, Hensley, Johnson, Kelleher, Kirk, Lopes, Lorenz, Muhleman, Orosei,
  Ostro, Paganelli, Paillou, Picardi, Posa, Radebaugh, Roth, Seu, Shaffer,
  Soderblom, Stiles, Stofan, Vetrella, West, Wood, Wye, Zebker, Alberti,
  Karkoschka, Rizk, McFarlane, See, and Kazeminejad}}]{Lunine2008}
Lunine, J.~I., C.~Elachi, S.~D. Wall, M.~A. Janssen, M.~D. Allison,
  Y.~Anderson, R.~Boehmer, P.~Callahan, P.~Encrenaz, E.~Flamini,
  G.~Franceschetti, Y.~Gim, G.~Hamilton, S.~Hensley, W.~T.~K. Johnson,
  K.~Kelleher, R.~L. Kirk, R.~M. Lopes, R.~Lorenz, D.~O. Muhleman, R.~Orosei,
  S.~J. Ostro, F.~Paganelli, P.~Paillou, G.~Picardi, F.~Posa, J.~Radebaugh,
  L.~E. Roth, R.~Seu, S.~Shaffer, L.~A. Soderblom, B.~Stiles, E.~R. Stofan,
  S.~Vetrella, R.~West, C.~A. Wood, L.~Wye, H.~Zebker, G.~Alberti,
  E.~Karkoschka, B.~Rizk, E.~McFarlane, C.~See, and B.~Kazeminejad (2008),
  {Titan's diverse landscapes as evidenced by Cassini Radar's third and fourth
  looks at Titan}, \textit{Icarus}, \textit{195}(1), 415 -- 433,
  \doi{10.1016/j.icarus.2007.12.022}.

\bibitem[{\textit{{Luspay-Kuti} et~al.}(2012)\textit{{Luspay-Kuti}, Chevrier,
  Wasiak, Roe, Welivitiya, Cornet, Singh, and
  {Rivera-Valentin}}}]{LuspayKuti2012grl}
{Luspay-Kuti}, A., V.~F. Chevrier, F.~C. Wasiak, L.~A. Roe, W.~D. D.~P.
  Welivitiya, T.~Cornet, S.~Singh, and E.~G. {Rivera-Valentin} (2012),
  {Experimental simulations of CH$_4$ evaporation on Titan},
  \textit{Geophysical Research Letters}, \textit{39}(23), L23,808,
  \doi{10.1029/2012GL054003}.

\bibitem[{\textit{{Luspay-Kuti} et~al.}(2014)\textit{{Luspay-Kuti}, Chevrier,
  Cordier, {Rivera-Valentin}, Singh, Wagner, and Wasiak}}]{LuspayKuti2014}
{Luspay-Kuti}, A., V.~F. Chevrier, D.~Cordier, E.~G. {Rivera-Valentin},
  S.~Singh, A.~Wagner, and F.~Wasiak (2014), {Experimental constraints on the
  composition and dynamics of Titan's polar lakes}, \textit{Earth and Planetary
  Science Letters}, \textit{410}, 75--83, \doi{10.1016/j.epsl.2014.11.023}.

\bibitem[{\textit{Lyew-Ayee}(2010)}]{LyewAyee2010}
Lyew-Ayee, P. (2010), \textit{{Geomorphological Landscapes of the World}},
  chap. {The cockpit country of Jama\"ica: An island within an island}, pp. 69
  -- 77, Springer, Dordrecht Heidelberg London New York,
  \doi{10.1007/978-90-481-3055-9\_8}.

\bibitem[{\textit{MacKenzie et~al.}(2014)\textit{MacKenzie, Barnes, Sotin,
  Soderblom, {Le Mou\'elic}, Rodriguez, Baines, Buratti, Clark, Nicholson, and
  McCord}}]{MacKenzie2014}
MacKenzie, S.~M., J.~W. Barnes, C.~Sotin, J.~M. Soderblom, S.~{Le Mou\'elic},
  S.~Rodriguez, K.~H. Baines, B.~J. Buratti, R.~N. Clark, P.~D. Nicholson, and
  T.~B. McCord (2014), {Evidence of Titan's climate history from evaporite
  distribution}, \textit{Icarus}, (0), --,
  \doi{http://dx.doi.org/10.1016/j.icarus.2014.08.022}.

\bibitem[{\textit{Magee et~al.}(2009)\textit{Magee, Waite, Mandt, Westlake,
  Bell, and Gell}}]{Magee2009}
Magee, B.~A., J.~H. Waite, K.~E. Mandt, J.~Westlake, J.~Bell, and D.~A. Gell
  (2009), {INMS-derived composition of Titan's upper atmosphere: Analysis
  methods and model comparison}, \textit{Planetary and Space Science},
  \textit{57}(14-15), 1895 -- 1916, \doi{10.1016/j.pss.2009.06.016}.

\bibitem[{\textit{Malaska et~al.}(2010)\textit{Malaska, Radebaugh, Lorenz,
  Mitchell, Farr, and Stofan}}]{Malaska2010}
Malaska, M., J.~Radebaugh, R.~Lorenz, K.~Mitchell, T.~Farr, and E.~Stofan
  (2010), {Identification of karst-like terrain on Titan from valley analysis},
  in \textit{41st Lunar and Planetary Institute Science Conference}, pp. 1544
  -- 1545.

\bibitem[{\textit{Malaska et~al.}(2011)\textit{Malaska, Radebaugh, Mitchell,
  Lopes, Wall, and Lorenz}}]{Malaska2011dissolution}
Malaska, M., J.~Radebaugh, K.~Mitchell, R.~Lopes, S.~Wall, and R.~Lorenz
  (2011), {Surface dissolution model for Titan karst}, in \textit{First
  International Planetary Cave Research Workshop}, pp. 8018--8019.

\bibitem[{\textit{Malaska and Hodyss}(2014)}]{Malaska2014}
Malaska, M.~J., and R.~Hodyss (2014), {Dissolution of benzene, naphthalene, and
  biphenyl in a simulated Titan lake}, \textit{Icarus}, \textit{242}, 74 -- 81,
  \doi{10.1016/j.icarus.2014.07.022}.

\bibitem[{\textit{Mastrogiuseppe et~al.}(2014)\textit{Mastrogiuseppe, Poggiali,
  Hayes, Lorenz, Lunine, Picardi, Seu, Flamini, Mitri, Notarnicola, Paillou,
  and Zebker}}]{Mastrogiuseppe2014}
Mastrogiuseppe, M., V.~Poggiali, A.~Hayes, R.~Lorenz, J.~Lunine, G.~Picardi,
  R.~Seu, E.~Flamini, G.~Mitri, C.~Notarnicola, P.~Paillou, and H.~Zebker
  (2014), {The bathymetry of a Titan sea}, \textit{Geophysical Research
  Letters}, \textit{41}(5), 1432 -- 1437, \doi{10.1002/2013GL058618}.

\bibitem[{\textit{{McMullan} et~al.}(1992)\textit{{McMullan}, Kvick, and
  Popelier}}]{McMullan1992}
{McMullan}, R., A.~Kvick, and P.~Popelier (1992), {Structures of cubic and
  orthorhombic phases of acetylene by single-crystal neutron diffraction},
  \textit{Acta Crystallographica Section B: Structural Science}, \textit{48},
  726--731, \doi{10.1107/S0108768192004774}.

\bibitem[{\textit{Mihevc et~al.}(2010)\textit{Mihevc, Prelovsek, and {Zupan
  Hajna}}}]{Mihevc2010}
Mihevc, A., M.~Prelovsek, and N.~{Zupan Hajna} (Eds.) (2010),
  \textit{{Introduction to the Dinaric Karst}}, 72 pp. pp., Karst Research
  Institute at ZRC SAZU, Postojna.

\bibitem[{\textit{Miller et~al.}(2010)\textit{Miller, Pickford, and
  Senut}}]{Miller2010}
Miller, R.~M., M.~Pickford, and B.~Senut (2010), {The geology, palaeontology
  and evolution of the Etosha Pan, Namibia: Implications for terminal Kalahari
  deposition}, \textit{South African Journal of Geology}, \textit{113}(3), 307
  -- 334, \doi{10.2113/gssajg.113.307}.

\bibitem[{\textit{Mitchell}(2008)}]{Mitchell2008}
Mitchell, J.~L. (2008), {The drying of Titan's dunes: Titan's methane hydrology
  and its impact on atmospheric circulation}, \textit{Journal of Geophysical
  Research}, \textit{113}, E08,015, \doi{10.1029/2007JE003017}.

\bibitem[{\textit{Mitchell et~al.}(2009)\textit{Mitchell, Pierrehumbert,
  Frierson, and Caballero}}]{Mitchell2009}
Mitchell, J.~L., R.~T. Pierrehumbert, D.~M. Frierson, and R.~Caballero (2009),
  {The impact of methane thermodynamics on seasonal convection and circulation
  in a model Titan atmosphere}, \textit{Icarus}, \textit{203}(1), 250 -- 264,
  \doi{10.1016/j.icarus.2009.03.043}.

\bibitem[{\textit{Mitchell et~al.}(2011)\textit{Mitchell, \'Ad\'amkovics,
  Caballero, and Turtle}}]{Mitchell2011}
Mitchell, J.~L., M.~\'Ad\'amkovics, R.~Caballero, and E.~P. Turtle (2011),
  {Locally enhanced precipitation organized by planetary-scale waves on Titan},
  \textit{Nature Geoscience}, \textit{4}, 589 -- 592, \doi{10.1038/ngeo1219}.

\bibitem[{\textit{Mitchell and Malaska}(2011)}]{Mitchell2011karst}
Mitchell, K.~L., and M.~Malaska (2011), {Karst on Titan}, in \textit{First
  International Planetary Cave Research Workshop}, pp. 8021--8022.

\bibitem[{\textit{Mitchell et~al.}(2007)\textit{Mitchell, Kargel, Wood,
  Radebaugh, Lopes, Lunine, Stofan, Kirk, and {the Cassini RADAR
  Team}}}]{Mitchell2007}
Mitchell, K.~L., J.~S. Kargel, C.~A. Wood, J.~Radebaugh, R.~M.~C. Lopes, J.~I.
  Lunine, E.~R. Stofan, R.~L. Kirk, and {the Cassini RADAR Team} (2007),
  {Titan's crater lakes: Caldera VS karst ?}, in \textit{38th Lunar and
  Planetary Science}, pp. 2061--2064.

\bibitem[{\textit{Moore and Howard}(2010)}]{Moore2010}
Moore, J.~M., and A.~D. Howard (2010), {Are the basins of Titan's Hotei Regio
  and Tui Regio sites of former low latitude seas ?}, \textit{Geophysical
  Research Letters}, \textit{37}, L22,205, \doi{10.1029/2010GL045234}.

\bibitem[{\textit{Moore and Pappalardo}(2011)}]{Moore2011}
Moore, J.~M., and R.~T. Pappalardo (2011), {Titan: An exogenic world ?},
  \textit{Icarus}, \textit{212}(2), 790 -- 806,
  \doi{10.1016/j.icarus.2011.01.019}.

\bibitem[{\textit{Moriconi et~al.}(2010)\textit{Moriconi, Lunine, Adriani,
  D'Aversa, Negrao, Filacchione, and Coradini}}]{Moriconi2010}
Moriconi, M.~L., J.~I. Lunine, A.~Adriani, E.~D'Aversa, A.~Negrao,
  G.~Filacchione, and A.~Coradini (2010), {Characterization of Titan's Ontario
  Lacus region from Cassini/VIMS observations}, \textit{Icarus},
  \textit{210}(2), 823 -- 831, \doi{10.1016/j.icarus.2010.07.023}.

\bibitem[{\textit{Neish and Lorenz}(2012)}]{Neish2012}
Neish, C.~D., and R.~D. Lorenz (2012), {Titan's global crater population: A new
  assessment}, \textit{Planetary and Space Science}, \textit{60}(1), 26 -- 33,
  \doi{10.1016/j.pss.2011.02.016}.

\bibitem[{\textit{Neumann and Mann}(1969)}]{Neumann1969}
Neumann, A., and R.~Mann (1969), {Die L\"oslichkeit von festem Acetylen in
  fl\"ussigen Methan/\"Athylen-Mischungen}, \textit{Chemie Ingenieur Technik},
  \textit{41}(12), 708--711, \doi{10.1002/cite.330411204}.

\bibitem[{\textit{Poling et~al.}(2007)\textit{Poling, Prausnitz, and
  O'Connell}}]{Poling2007}
Poling, B.~E., J.~M. Prausnitz, and J.~P. O'Connell (2007), \textit{{The
  properties of gases and liquids, Fifth edition}}, McGraw-Hill Professional,
  Englewood Cliffs.

\bibitem[{\textit{Preston and Prausnitz}(1970)}]{Preston1970}
Preston, G.~T., and J.~M. Prausnitz (1970), {Thermodynamics of solid solubility
  in cryogenic solvents}, \textit{Industrial and Engineering Chemistry Process
  Design and Development}, \textit{9}(2), 264--271, \doi{10.1021/i260034a017}.

\bibitem[{\textit{Preston et~al.}(1971)\textit{Preston, Funk, and
  Prausnitz}}]{Preston1971}
Preston, G.~T., E.~W. Funk, and J.~M. Prausnitz (1971), {Solubilities of
  hydrocarbons and carbon dioxide in liquid methane and in liquid argon},
  \textit{Journal of Physical Chemistry}, \textit{75}(15), 2345--2352.

\bibitem[{\textit{Quirico et~al.}(2008)\textit{Quirico, Montagnac, Lees,
  McMillan, Szopa, Cernogora, Rouzaud, Simon, Bernard, Coll, Fray, Minard,
  Raulin, Reynard, and Schmitt}}]{Quirico2008}
Quirico, E., G.~Montagnac, V.~Lees, P.~F. McMillan, C.~Szopa, G.~Cernogora,
  J.-N. Rouzaud, P.~Simon, J.-M. Bernard, P.~Coll, N.~Fray, R.~D. Minard,
  F.~Raulin, B.~Reynard, and B.~Schmitt (2008), {New experimental constraints
  on the composition and structure of tholins}, \textit{Icarus},
  \textit{198}(1), 218 -- 231, \doi{10.1016/j.icarus.2008.07.012}.

\bibitem[{\textit{Rannou et~al.}(2003)\textit{Rannou, McKay, and
  Lorenz}}]{Rannou2003}
Rannou, P., C.~P. McKay, and R.~D. Lorenz (2003), {A model of Titan's haze of
  fractal aerosols constrained by multiple observations}, \textit{Planetary and
  Space Science}, \textit{51}(14-15), 963 -- 976,
  \doi{10.1016/j.pss.2003.05.008}.

\bibitem[{\textit{Rannou et~al.}(2006)\textit{Rannou, Montmessin, Hourdin, and
  Lebonnois}}]{Rannou2006}
Rannou, P., F.~Montmessin, F.~Hourdin, and S.~Lebonnois (2006), {The
  latitudinal distribution of clouds on Titan}, \textit{Science}, \textit{311},
  201 -- 205, \doi{10.1126/science.1118424}.

\bibitem[{\textit{Raulin}(1987)}]{Raulin1987}
Raulin, F. (1987), {Organic chemistry in the oceans of Titan}, \textit{Advances
  in Space Research}, \textit{7}(5), 571--581,
  \doi{10.1016/0273-1177(87)90358-9}.

\bibitem[{\textit{Refson and Pawley}(1986)}]{Refson1986}
Refson, K., and G.~Pawley (1986), {The Structure and Orientational Disorder in
  Solid n-Butane by Neutron Powder Diffraction}, \textit{Acta Crystallographica
  Section B}, \textit{42}, 402--410, \doi{10.1107/S010876818609804X}.

\bibitem[{\textit{Rodriguez et~al.}(2009)\textit{Rodriguez, {Le Mou\'elic},
  Rannou, Tobie, Baines, Barnes, Griffith, Hirtzig, Pitman, Sotin, Brown,
  Buratti, Clark, and Nicholson}}]{Rodriguez2009cloud}
Rodriguez, S., S.~{Le Mou\'elic}, P.~Rannou, G.~Tobie, K.~H. Baines, J.~W.
  Barnes, C.~A. Griffith, M.~Hirtzig, K.~M. Pitman, C.~Sotin, R.~H. Brown,
  B.~J. Buratti, R.~N. Clark, and P.~D. Nicholson (2009), {Global circulation
  as the main source of cloud activity on Titan}, \textit{Nature},
  \textit{459}, 678--682, \doi{10.1038/nature08014}.

\bibitem[{\textit{Rodriguez et~al.}(2011)\textit{Rodriguez, {Le Mou\'elic},
  Rannou, Sotin, Brown, Barnes, Griffith, Burgalat, Baines, Buratti, Clark, and
  Nicholson}}]{Rodriguez2011}
Rodriguez, S., S.~{Le Mou\'elic}, P.~Rannou, C.~Sotin, R.~H. Brown, J.~W.
  Barnes, C.~A. Griffith, J.~Burgalat, K.~H. Baines, B.~J. Buratti, R.~N.
  Clark, and P.~D. Nicholson (2011), {Titan's cloud seasonal activity from
  winter to spring with Cassini/VIMS}, \textit{Icarus}, \textit{216}(1), 89 --
  110, \doi{10.1016/j.icarus.2011.07.031}.

\bibitem[{\textit{Rodriguez et~al.}(2014)\textit{Rodriguez, Garcia, Lucas,
  App\'er\'e, Gall, Reffet, Corre, {Le Mou\'elic}, Cornet, du~Pont, Narteau,
  Bourgeois, Radebaugh, Arnold, Barnes, Stephan, Jaumann, Sotin, Brown, Lorenz,
  and Turtle}}]{Rodriguez2014}
Rodriguez, S., A.~Garcia, A.~Lucas, T.~App\'er\'e, A.~L. Gall, E.~Reffet, L.~L.
  Corre, S.~{Le Mou\'elic}, T.~Cornet, S.~C. du~Pont, C.~Narteau, O.~Bourgeois,
  J.~Radebaugh, K.~Arnold, J.~W. Barnes, K.~Stephan, R.~Jaumann, C.~Sotin,
  R.~H. Brown, R.~D. Lorenz, and E.~P. Turtle (2014), {Global mapping and
  characterization of Titan's dune fields with Cassini: Correlation between
  \{RADAR\} and \{VIMS\} observations}, \textit{Icarus}, \textit{230}, 168 --
  179, \doi{http://dx.doi.org/10.1016/j.icarus.2013.11.017}.

\bibitem[{\textit{Roe et~al.}(2002)\textit{Roe, {de Pater}, Macintosh, Gibbard,
  Max, and McKay}}]{Roe2002}
Roe, H.~G., I.~{de Pater}, B.~A. Macintosh, S.~G. Gibbard, C.~E. Max, and C.~P.
  McKay (2002), {Titan's atmosphere in late southern spring observed with
  adaptive optics on the W. M. Keck II 10-meter telescope}, \textit{Icarus},
  \textit{157}(1), 254 -- 258, \doi{10.1006/icar.2002.6831}.

\bibitem[{\textit{Schaller et~al.}(2006)\textit{Schaller, Brown, Roe, and
  Bouchez}}]{Schaller2006tempete}
Schaller, E.~L., M.~E. Brown, H.~G. Roe, and A.~H. Bouchez (2006), {A large
  cloud outburst at Titan's south pole}, \textit{Icarus}, \textit{182}(1), 224
  -- 229, \doi{10.1016/j.icarus.2005.12.021}.

\bibitem[{\textit{Schaller et~al.}(2009)\textit{Schaller, Roe, Schneider, and
  Brown}}]{Schaller2009}
Schaller, E.~L., H.~G. Roe, T.~Schneider, and M.~E. Brown (2009), {Storms in
  the tropics of Titan}, \textit{Nature}, \textit{460}(7257), 873--875,
  \doi{10.1038/nature08193}.

\bibitem[{\textit{Schneider et~al.}(2012)\textit{Schneider, Graves, Schaller,
  and Brown}}]{Schneider2012}
Schneider, T., S.~D.~B. Graves, E.~L. Schaller, and M.~E. Brown (2012), {Polar
  methane accumulation and rainstorms on Titan from simulations of the methane
  cycle}, \textit{Nature}, \textit{481}, 58--61, \doi{10.1038/nature10666}.

\bibitem[{\textit{Sharma and Byrne}(2010)}]{Sharma2010}
Sharma, P., and S.~Byrne (2010), Constraints on titan's topography through
  fractal analysis of shorelines, \textit{Icarus}, \textit{209}, 723 -- 737,
  \doi{10.1016/j.icarus.2010.04.023}.

\bibitem[{\textit{Sharma and Byrne}(2011)}]{Sharma2011}
Sharma, P., and S.~Byrne (2011), {Comparison of Titan's north polar lakes with
  terrestrial analogs}, \textit{Geophysical Research Letters}, \textit{38},
  L24,203, \doi{10.1029/2011GL049577}.

\bibitem[{\textit{Shaw and Thomas}(2000)}]{Shaw2000}
Shaw, P.~A., and D.~S.~G. Thomas (2000), \textit{{Arid zone geomorphology,
  Process, form and change in drylands, 2nd Edition}}, chap. {Pans, playa and
  salt lakes}, pp. 293 -- 317, {D. S. G. Thomas (eds.), John Wiley and Sons
  (England)}.

\bibitem[{\textit{Simon and Peters}(1980)}]{Simon1980}
Simon, A., and K.~Peters (1980), {Single-Crystal Refinement of the Structure of
  Carbon Dioxide}, \textit{Acta Crystallographica Section B}, \textit{36},
  2750--2751, \doi{10.1107/S0567740880009879}.

\bibitem[{\textit{Singh et~al.}(2014)\textit{Singh, Chevrier, Wagner, Leitner,
  Gainor, Roe, Cornet, and Combe}}]{Singh2014}
Singh, S., V.~F. Chevrier, A.~Wagner, M.~Leitner, M.~Gainor, L.~Roe, T.~Cornet,
  and J.-P. Combe (2014), {Solubility of acetylene in liquid hydrocarbons under
  Titan surface conditions}, in \textit{45th Lunar and Planetary Science
  Conference}, p. 2850.

\bibitem[{\textit{Sket}(2012)}]{Sket2012}
Sket, B. (2012), \textit{Encyclopedia of caves (2nd Ed.)}, chap. {Diversity
  patterns in the Dinaric Karst}, pp. 228 -- 238, Elsevier Inc., Amsterdam.

\bibitem[{\textit{Sotin et~al.}(2012)\textit{Sotin, Lawrence, Reinhardt,
  Barnes, Brown, Hayes, {Le Mou\'elic}, Rodriguez, Soderblom, Soderblom,
  Baines, Buratti, Clark, Jaumann, Nicholson, and Stephan}}]{Sotin2012}
Sotin, C., K.~J. Lawrence, B.~Reinhardt, J.~W. Barnes, R.~H. Brown, A.~G.
  Hayes, S.~{Le Mou\'elic}, S.~Rodriguez, J.~M. Soderblom, L.~A. Soderblom,
  K.~H. Baines, B.~J. Buratti, R.~N. Clark, R.~Jaumann, P.~D. Nicholson, and
  K.~Stephan (2012), {Observations of Titan's northern lakes at 5 microns:
  Implications for the organic cycle and geology}, \textit{Icarus},
  \textit{221}(2), 768 -- 786, \doi{10.1016/j.icarus.2012.08.017}.

\bibitem[{\textit{Spencer and Danner}(1972)}]{Spencer1972}
Spencer, C.~F., and R.~P. Danner (1972), {Improved equation for prediction of
  saturated liquid density}, \textit{Journal of Chemical Engineering Data},
  \textit{17}, 236--241, \doi{10.1021/je60053a012}.

\bibitem[{\textit{Stephan et~al.}(2009)\textit{Stephan, Jaumann, Karkoschka,
  Kirk, Barnes, Tomasko, Turtle, {Le Corre}, Langhans, {Le Mou\'elic}, Lorenz,
  and Perry}}]{Stephan2009}
Stephan, K., R.~Jaumann, E.~Karkoschka, R.~L. Kirk, J.~W. Barnes, M.~G.
  Tomasko, E.~P. Turtle, L.~{Le Corre}, M.~Langhans, S.~{Le Mou\'elic}, R.~D.
  Lorenz, and J.~Perry (2009), \textit{{Titan from Cassini-Huygens}}, chap.
  {Mapping products of Titan's surface}, pp. 489 -- 510, Springer, Dordrecht
  Heidelberg London New York, \doi{10.1007/978-1-4020-9215-2}.

\bibitem[{\textit{Stiles et~al.}(2009)\textit{Stiles, Hensley, Gim, Bates,
  Kirk, Hayes, Radebaugh, Lorenz, Mitchell, Callahan, Zebker, Johnson, Wall,
  Lunine, Wood, Janssen, Pelletier, West, and Veeramacheneni}}]{Stiles2009}
Stiles, B.~W., S.~Hensley, Y.~Gim, D.~M. Bates, R.~L. Kirk, A.~Hayes,
  J.~Radebaugh, R.~D. Lorenz, K.~L. Mitchell, P.~S. Callahan, H.~Zebker, W.~T.
  Johnson, S.~D. Wall, J.~I. Lunine, C.~A. Wood, M.~Janssen, F.~Pelletier,
  R.~D. West, and C.~Veeramacheneni (2009), {Determining Titan surface
  topography from Cassini SAR data}, \textit{Icarus}, \textit{202}(2), 584 --
  598, \doi{10.1016/j.icarus.2009.03.032}.

\bibitem[{\textit{Stofan et~al.}(2007)\textit{Stofan, Elachi, Lunine, Lorenz,
  Stiles, Mitchell, Ostro, Soderblom, Wood, Zebker, Wall, Janssen, Kirk, Lopes,
  Paganelli, Radebaugh, Wye, Anderson, Allison, Boehmer, Callahan, Encrenaz,
  Flamini, Francescetti, Gim, Hamilton, Hensley, Johnson, Kelleher, Muhleman,
  Paillou, Picardi, Posa, Roth, Seu, Shaffer, Vetrella, and West}}]{Stofan2007}
Stofan, E.~R., C.~Elachi, J.~I. Lunine, R.~D. Lorenz, B.~Stiles, K.~L.
  Mitchell, S.~Ostro, L.~Soderblom, C.~Wood, H.~Zebker, S.~Wall, M.~Janssen,
  R.~Kirk, R.~Lopes, F.~Paganelli, J.~Radebaugh, L.~Wye, Y.~Anderson,
  M.~Allison, R.~Boehmer, P.~Callahan, P.~Encrenaz, E.~Flamini,
  G.~Francescetti, Y.~Gim, G.~Hamilton, S.~Hensley, W.~T.~K. Johnson,
  K.~Kelleher, D.~Muhleman, P.~Paillou, G.~Picardi, F.~Posa, L.~Roth, R.~Seu,
  S.~Shaffer, S.~Vetrella, and R.~West (2007), The lakes of {Titan},
  \textit{Nature}, \textit{445}, 61 -- 64, \doi{10.1038/nature05438}.

\bibitem[{\textit{{Szczepaniec-Cieciak}
  et~al.}(1978)\textit{{Szczepaniec-Cieciak}, Dabrowska, Lagan, and
  Wojtaszek}}]{SzczepaniecCieciak1978}
{Szczepaniec-Cieciak}, E., B.~Dabrowska, J.~M. Lagan, and Z.~Wojtaszek (1978),
  {Estimation of the solubility of solidified substances in liquid methane by
  the Preston-Prausnitz method}, \textit{Cryogenics}, \textit{18}(10),
  591--600, \doi{10.1016/0011-2275(78)90186-8}.

\bibitem[{\textit{Tan et~al.}(2013)\textit{Tan, Kargel, and Marion}}]{Tan2013}
Tan, S.~P., J.~S. Kargel, and G.~M. Marion (2013), {Titan's atmosphere and
  surface liquid: New calculation using Statistical Associating Fluid Theory},
  \textit{Icarus}, \textit{222}(1), 53 -- 72,
  \doi{10.1016/j.icarus.2012.10.032}.

\bibitem[{\textit{Tewelde et~al.}(2013)\textit{Tewelde, Perron, Ford, Miller,
  and Black}}]{Tewelde2013}
Tewelde, Y., J.~T. Perron, P.~Ford, S.~Miller, and B.~Black (2013), Estimates
  of fluvial erosion on titan from sinuosity of lake shorelines,
  \textit{Journal of Geophysical Research: Planets}, \textit{118}, 1--15,
  \doi{10.1002/jgre.20153}.

\bibitem[{\textit{Tobie et~al.}(2006)\textit{Tobie, Lunine, and
  Sotin}}]{Tobie2006}
Tobie, G., J.~I. Lunine, and C.~Sotin (2006), {Episodic outgassing as the
  origin of atmospheric methane on Titan}, \textit{Nature}, \textit{440}(7080),
  61 -- 64, \doi{10.1038/nature04497}.

\bibitem[{\textit{Tokano}(2009)}]{Tokano2009}
Tokano, T. (2009), {Impact of seas/lakes on polar meteorology of Titan:
  Simulation by a coupled GCM-Sea model}, \textit{Icarus}, \textit{204}(2), 619
  -- 636, \doi{10.1016/j.icarus.2009.07.032}.

\bibitem[{\textit{Toublanc et~al.}(1995)\textit{Toublanc, Parisot, Brillet,
  Gautier, Raulin, and McKay}}]{Toublanc1995}
Toublanc, D., J.~P. Parisot, J.~Brillet, D.~Gautier, F.~Raulin, and C.~P. McKay
  (1995), {Photochemical modeling of Titan's atmosphere}, \textit{Icarus},
  \textit{113}(1), 2 -- 26, \doi{10.1006/icar.1995.1002}.

\bibitem[{\textit{Truesdell and Jones}(1974)}]{Truesdell1974}
Truesdell, A., and B.~Jones (1974), {WATEQ, a computer program for calculating
  chemical equilibria of natural waters}, \textit{Journal of Research of the
  United States Geological Survey}, \textit{2}, 233--248.

\bibitem[{\textit{Tucker et~al.}(2001)\textit{Tucker, Lancaster, Gasparini, and
  Bras}}]{Tucker2001}
Tucker, G.~E., S.~T. Lancaster, N.~M. Gasparini, and R.~L. Bras (2001),
  \textit{{Landscape Erosion and Evolution Modeling}}, chap. {The
  Channel-Hillslope Integrated Landscape Development (CHILD) Model}, pp. 349 --
  388, {Springer US}.

\bibitem[{\textit{Turtle et~al.}(2011{\natexlab{a}})\textit{Turtle, Perry,
  Hayes, Lorenz, Barnes, McEwen, West, Del~Genio, Barbara, Lunine, Schaller,
  Ray, Lopes, and Stofan}}]{Turtle2011rains}
Turtle, E.~P., J.~E. Perry, A.~G. Hayes, R.~D. Lorenz, J.~W. Barnes, A.~S.
  McEwen, R.~A. West, A.~D. Del~Genio, J.~M. Barbara, J.~I. Lunine, E.~L.
  Schaller, T.~L. Ray, R.~M.~C. Lopes, and E.~R. Stofan (2011{\natexlab{a}}),
  {Rapid and extensive surface changes near Titan's equator: Evidence of April
  showers}, \textit{Science}, \textit{331}, 1414--1417,
  \doi{10.1126/science.1201063}.

\bibitem[{\textit{Turtle et~al.}(2011{\natexlab{b}})\textit{Turtle, {Del
  Genio}, Barbara, Perry, Schaller, {McEwen}, West, and Ray}}]{Turtle2011grl}
Turtle, E.~P., A.~D. {Del Genio}, J.~M. Barbara, J.~E. Perry, E.~L. Schaller,
  A.~S. {McEwen}, R.~A. West, and T.~L. Ray (2011{\natexlab{b}}), {Seasonal
  changes in Titan's meteorology}, \textit{Geophysical Research Letters},
  \textit{38}, L03,203, \doi{10.1029/2010GL046266}.

\bibitem[{\textit{Vinatier et~al.}(2010)\textit{Vinatier, B\'ezard, Nixon,
  Mamoutkine, Carlson, Jennings, Guandique, Teanby, Bjoraker, Flasar, and
  Kunde}}]{Vinatier2010}
Vinatier, S., B.~B\'ezard, C.~A. Nixon, A.~Mamoutkine, R.~C. Carlson, D.~E.
  Jennings, E.~A. Guandique, N.~A. Teanby, G.~L. Bjoraker, F.~M. Flasar, and
  V.~G. Kunde (2010), {Analysis of Cassini/CIRS limb spectra of Titan acquired
  during the nominal mission: I. Hydrocarbons, nitriles and CO2 vertical mixing
  ratio profiles}, \textit{Icarus}, \textit{205}(2), 559 -- 570,
  \doi{10.1016/j.icarus.2009.08.013}.

\bibitem[{\textit{Vixie et~al.}(2012)\textit{Vixie, Barnes, Jackson, and
  Wilson}}]{Vixie2012lac}
Vixie, G., J.~W. Barnes, B.~Jackson, and P.~Wilson (2012), {Temperate lakes
  discovered on Titan}, in \textit{43rd Lunar and Planetary Science
  Conference}, p. 2766.

\bibitem[{\textit{Vlahovic et~al.}(2002)\textit{Vlahovic, Tisljar, Velic, and
  Maticec}}]{Vlahovic2002}
Vlahovic, I., J.~Tisljar, I.~Velic, and D.~Maticec (2002), {The Karst Dinarides
  are composed of relics of a single Mesozoic platform: Facts and
  consequences}, \textit{Geologia Croatica}, \textit{55}(2), 171--183.

\bibitem[{\textit{{von Szalghary}}(1972)}]{Szalghary1972}
{von Szalghary}, W.-D. (1972), {L\"oslichkeit von festem benzol in fl\"ussigen
  Kohlenwasserstoffen}, \textit{{K\"altetechnik-Klimatisierung}}, \textit{24},
  145--149.

\bibitem[{\textit{Waltham}(2008)}]{Waltham2008}
Waltham, T. (2008), {Fengcong, fenglin, cone karst and tower karst},
  \textit{Cave and Karst Science}, \textit{35}, 77--88.

\bibitem[{\textit{Wang and Li}(2009)}]{Wang2009}
Wang, P., and Q.~Li (2009), \textit{Encyclopedia of paleoclimatology and
  ancient environments}, chap. {Monsoons: Pre-Quaternary}, pp. 583--589,
  Springer, Netherlands.

\bibitem[{\textit{Wasiak et~al.}(2013)\textit{Wasiak, Androes, Blackburn,
  Tullis, Dixon, and Chevrier}}]{Wasiak2013}
Wasiak, F.~C., D.~Androes, D.~G. Blackburn, J.~A. Tullis, J.~Dixon, and V.~F.
  Chevrier (2013), {A geological characterization of Ligeia Mare in the
  northern polar region of Titan}, \textit{Planetary and Space Science},
  \textit{84}(0), 141 -- 147, \doi{10.1016/j.pss.2013.05.007}.

\bibitem[{\textit{White}(1984)}]{White1984}
White, W.~B. (1984), \textit{{Groundwater as a geomorphic agent}}, chap. {Rate
  processes: chemical kinetics and karst landform development}, pp. 227--248,
  Allen and Unwin, Inc, Boston.

\bibitem[{\textit{{White}}(2012)}]{White2012}
{White}, W.~B. (2012), \textit{Encyclopedia of caves (2nd Ed.)}, chap.
  {Hydrogeology of karst aquifers}, pp. 383 -- 391, Elsevier Academic Press,
  Amsterdam.

\bibitem[{\textit{White}(2013)}]{White2013}
White, W.~M. (2013), \textit{Geochemistry}, 672 pp., John Wiley \& Sons.

\bibitem[{\textit{Wilson and Atreya}(2004)}]{Wilson2004}
Wilson, E.~H., and S.~K. Atreya (2004), {Current state of modeling the
  photochemistry of Titan's mutually dependent atmosphere and ionosphere},
  \textit{Journal of Geophysical Research}, \textit{109}, E06,002,
  \doi{10.1029/2003JE002181}.

\bibitem[{\textit{Wood et~al.}(2007)\textit{Wood, Mitchell, Lopes, Radebaugh,
  Stofan, Lunine, and {the Cassini RADAR Team}}}]{Wood2007}
Wood, C.~A., K.~L. Mitchell, R.~M.~C. Lopes, J.~Radebaugh, E.~Stofan,
  J.~Lunine, and {the Cassini RADAR Team} (2007), {Volcanic calderas in the
  North Polar Region of Titan}, in \textit{38th Lunar and Planetary Science
  Conference}, pp. 1454--1455.

\bibitem[{\textit{Xuewen and Weihai}(2006)}]{Xuewen2006}
Xuewen, Z., and C.~Weihai (2006), {Tiankengs in the karst of China},
  \textit{Speleogenesis and Evolution of Karst Aquifers}, \textit{4}(1), 1--18.

\bibitem[{\textit{Yaws}(1996)}]{Yaws1996}
Yaws, C.~L. (1996), {Appendix D - Critical Properties and Acentric Factor for
  Inorganic Compounds and Elements}, in \textit{Inorganic Compounds and
  Elements}, \textit{Handbook of Thermodynamic Diagrams}, vol.~4, pp. 351 --
  356, Gulf Professional Publishing, \doi{10.1016/B978-0-88415-860-8.50036-2}.

\bibitem[{\textit{Yung et~al.}(1984)\textit{Yung, Allen, and Pinto}}]{Yung1984}
Yung, Y.~L., M.~Allen, and J.~P. Pinto (1984), {Photochemistry of the
  atmosphere of Titan: Comparison between model and observations}, \textit{The
  Astrophysical Journal}, \textit{55}, 465--506, \doi{10.1086/190963}.

\bibitem[{\textit{{Zupan Hajna}}(2012)}]{ZupanHajna2012}
{Zupan Hajna}, N. (2012), \textit{Encyclopedia of caves (2nd Ed.)}, chap.
  {Dinaric Karst: Geography and geology}, pp. 195 -- 203, Elsevier Academic
  Press, Amsterdam.

\end{thebibliography}

\end{document}